\documentclass[12pt]{article}
\pdfoutput=1

\usepackage{amsmath,amssymb,amscd}
\usepackage{listings}
\usepackage{caption}
\usepackage{dsfont}
\usepackage{slashed}
\usepackage{color}
\usepackage{ulem}
\usepackage[pdftex]{graphicx}
\usepackage{epstopdf}
\usepackage{subfigure}
\usepackage{epsfig}
\usepackage{listings}
\usepackage{caption}
\usepackage{cite}
\usepackage{multirow}
\usepackage{booktabs}

\newcommand{\beq}{\begin{equation}}
\newcommand{\eeq}{\end{equation}}
\newcommand{\nbea}{\begin{align*}}
\newcommand{\neea}{\end{align*}}
\newcommand{\nbeq}{\begin{equation*}}
\newcommand{\neeq}{\end{equation*}}

\newcommand{\intinf}{\int\limits_{-\infty}^{+\infty}}
\newcommand{\lat}{{\it Fermi}-LAT}
\usepackage{multirow}
\usepackage{array}
\newcolumntype{M}[1]{>{\centering\arraybackslash}m{#1}}
\newcolumntype{N}{@{}m{0pt}@{}}

\begin{document}

\pagestyle{plain}

\baselineskip=21pt

\begin{center}

{\large {\bf Robust Constraint on Lorentz Violation Using {\bf\it Fermi}-LAT Gamma-Ray Burst Data}}

\vskip 0.1in

{\bf John Ellis}\textsuperscript{a,b},~
{\bf Rostislav Konoplich}\textsuperscript{c,d},~
{\bf Nikolaos E. Mavromatos}\textsuperscript{a,e},~\\
{\bf Linh Nguyen}\textsuperscript{d},~
{\bf Alexander S. Sakharov}\textsuperscript{c,d,f},~
{\bf Edward K. Sarkisyan-Grinbaum}\textsuperscript{f,g},~\\

\vskip 0.1in

{\small {\it

\textsuperscript{a}Theoretical Particle Physics and Cosmology Group, Physics Department, \\
King's College London, Strand, London WC2R 2LS, United Kingdom\\
\vspace{0.25cm}
\textsuperscript{b}National Institute of Chemical Physics \& Biophysics, R{\" a}vala 10, 10143 Tallinn, Estonia; \\
Theoretical Physics Department, CERN, CH-1211 Gen\`eve 23, Switzerland\\
\vspace{0.25cm}
\textsuperscript{c}Department of Physics, New York University\\
726 Broadway, New York, NY 10003, United States of America\\
\vspace{0.25cm}
\textsuperscript{d}Physics Department, Manhattan College\\
{\mbox 4513 Manhattan College Parkway, Riverdale, NY 10471, United States of America}\\
\vspace{0.25cm}
\textsuperscript{e} Currently also at: Department of Theoretical Physics and IFIC, University of Valencia - CSIC, \\
{\mbox  Valencia, E-46100, Spain }\\
\vspace{0.25cm}
\textsuperscript{f}Experimental Physics Department, CERN, CH-1211 Gen\`eve 23, Switzerland \\
\vspace{0.25cm}
\textsuperscript{g}Department of Physics, The University of Texas at Arlington\\
{\mbox 502 Yates Street, Box 19059, Arlington, TX 76019, United States of America} \\
}
}

\vskip 0.1in

{\bf Abstract}

\end{center}

\baselineskip=18pt \noindent



Models of quantum gravity suggest that the vacuum should be regarded as a medium with quantum
structure that may have non-trivial effects on photon propagation,
including the violation of Lorentz invariance. {\it Fermi} Large Area Telescope (LAT) observations
of gamma-ray bursts (GRBs) are sensitive probes of
Lorentz invariance, via studies of energy-dependent timing shifts in their rapidly-varying photon emissions.
In this paper we analyze the {\it Fermi}-LAT measurements of high-energy gamma rays from GRBs
with known redshifts, allowing for the possibility of energy-dependent variations in emission times
at the sources as well as a possible non-trivial refractive index {\it in vacuo} for photons. We use
statistical estimators based on the irregularity, kurtosis and skewness of bursts that are relatively
bright in the 100~MeV to multi-GeV energy band to constrain possible dispersion effects during propagation.
We find that the energy scale characterizing a linear energy dependence of the refractive index should exceed
a few $\times 10^{17}$~GeV, and we estimate the sensitivity attainable with additional future sources
to be detected by {\it Fermi}-LAT.

\vskip 3mm

Keywords: Lorentz invariance; Gamma-ray burst; Quantum gravity
\vskip 5mm
\leftline{KCL-PH-TH/2018-28, CERN-TH/2018-138,  IFIC/17-62}
\leftline{March 2019}


\section{Introduction and Summary}

The idea that the space-time vacuum should
be regarded as a non-trivial medium - baptized `space-time foam' by Wheeler~\cite{Wheeler} - is based on very general intuition.
This intuition arises from the feature of quantum mechanics that on time scales $\Delta t$ any physical system must exhibit
virtual energy fluctuations $\Delta E$ with magnitudes $\Delta E \sim \hslash/\Delta t$. Wheeler~\cite{Wheeler} argued that on time
scales $\Delta t \sim 1/M_P$, where $M_P \sim 10^{19}$~GeV is the Planck mass: $M_P = \sqrt{\hslash c/G_N}$ and
$G_N$ is the Newton constant of classical gravity, there would appear quantum-gravitational fluctuations in
the space-time continuum with $\Delta E \sim M_P$, resulting in a `foamy'
structure on short time scales $\Delta t \sim \hslash/M_P$. This observation led
Wheeler to argue that space-time would no longer appear smooth at
distance scales $\Delta x \sim \hslash/M_P$, and that it might {exhibit} both non-topological irregularities and topological fluctuations.

This intuitive picture suggests the appearance of a refractive index $\eta$ for particles such as photons propagating through
`empty' space, corresponding to a phase velocity $v_{\rm ph} = p/E = c/\eta$~\cite{dfoam}. It can be argued
on general grounds that photons should not travel faster than $c$~\footnote{From now on, we use natural
units in which $\hslash, c \equiv 1$.}, because otherwise they
would emit gravitational {\v C}erenkov radiation and lose energy unacceptably quickly~\cite{KT}~\footnote{This
argument assumes that gravitational waves do not propagate superluminally, which is consistent with the recent near-coincident
observations of gravitational waves and photons from the merger of two neutron stars~\cite{NS-NS}.}.
Hence, the photon refractive  index $\eta \ge 1$, corresponding to
{\it subluminal} propagation of energetic photons, as predicted in simple models~\cite{dfoam,loop,pospelov,dsr}.
One would also expect that the refractive index
should {\it increase} with energy, because gravitational interactions are proportional to some {\it negative} power of $M_P$, in general,
and therefore should increase as some {\it positive} power of the energy~\footnote{This is a characteristic signature of space-time foam,
which is to be contrasted with the refractive index of an ordinary material medium that generally decreases for photons with
shorter wavelengths. The effect is analogous to boats and ships navigating in stormy seas.
Ships that are considerably longer than the distances between the peaks of the waves (corresponding to long-wavelength (low-energy) photons),
can pass straight through the waves, whereas small boats (corresponding to short-wavelength (high-energy) photons), ascend
and then descend each wave, and progress significantly more slowly than large ships.}.
Models~\cite{dfoam,loop,pospelov,dsr} suggest that the
photon group velocity might deviate from that of light linearly in photon energy E:
\begin{equation}\label{mdr1}
v_g \sim \; 1 - \frac{E}{M_1} ,
\end{equation}
where one might expect that $M_1 = {\cal O}(M_P)$. However, the Lorentz-violating (LV) scale
$M_1$ would depend on unknown parameters of the microscopic theory, including
the string scale, which $\ne M_P$, in general. In the D-foam model discussed below~\cite{dfoam}, $M_1$
would depend also on couplings to D-particles, which would depend on the particle species,
and on the local density of D-particles~\footnote{If the density of D-particles is not uniform,
but depends on the cosmological epoch~\cite{emnewuncert},
$M_1$ could depend also on the redshift, but we do not discuss here this potential complication.}. Moreover,  other energy
dependences: $\eta - 1 \sim (E/M_n)^n$ should also be considered, such as the case
$n = 2$~\cite{mitsou, EMNSS}~\footnote{For a general discussion, see~\cite{pfifer}.}.

The proposal that there might be observable
effects on the propagation of particles such as photons~\cite{AEMNS} was made originally
in the context of concrete models of space-time foam motivated by (non-critical) string/brane theories~\cite{dfoam,mitsou}.
These models go beyond conventional local quantum field theories, and contain the necessary ingredients
for discussing the interaction between a propagating matter particle and a quantum-gravitational `environment'.
The former is described as an (open) string excitation representing some observable (Standard-Model-like) particle
of matter or radiation, moving through a (3+1)-dimensional brane Universe~\cite{polchinski}.
The `environment' of quantum-gravitational fluctuations is provided in this context by ensembles of
quantum space-time defects described as D-particles~\cite{dfoam}, which move in the higher-dimensional bulk.
They consist of branes that are compactified in such a way that, from the point of view of a low-energy four-dimensional
observer living on the (3+1)-dimensional brane Universe, they  look  approximately point-like~\cite{emnewuncert}.
When D-particles cross this D-brane world, they are perceived in our Universe
as space-time events localized at specific locations $x$ and specific times $t$, which we call `D-foam'.

There are non-trivial interactions between bosonic open strings and such D-particles, consisting of splitting of the open string and
emission of other open string excitations stretching between the D-particle and the brane world~\cite{emnewuncert}.
These interactions must respect the gauge symmetries on the brane world, such as the U(1) of electromagnetism.
The consequent charge conservation implies that `D-foam' appears transparent to charged bosons,
but not to neutral ones such as photons and gravitons. Moreover, in the absence of low-energy supersymmetry,
fermions such as right-handed neutrinos would have suppressed interactions with the D-foam~\cite{synchr}. Conventional
left-handed neutrinos are  doublets under the electroweak SU(2) group of the Standard Model,
so their interactions with the D-particles are further suppressed.

We have studied the possible observable consequences for photons propagating through our D-brane Universe
in several previous papers~\cite{AEMNS,dfoam,mitsou,emnewuncert,EMNS,EMNSS}.
In general, if a photon encounters a D-particle, its interaction with it may resemble that of a photon
propagating through a transparent material medium such as glass. In that case it may interact with the
electrons that it contains, via absorption and subsequent re-emission, with the net
effect of slowing down the photon. Thus, light travelling through glass has a
refractive index $\eta > 1$. Moreover, the value of $\eta$ varies with the colour of the light, i.e., the energy of the
associated photon. Similarly, we expect in general that light travelling through the
quantum-gravitational vacuum would acquire an energy-dependent
refractive index $\eta > 1$, that we may model via interactions with D-particles~\cite{emnewuncert}.
On the other hand, because of the absence of interactions between charged particles and D-foam
the deviation of the refractive index of the electron from unity would be suppressed~\cite{synchr,emnewuncert},
as required phenomenologically.

Many other models of Lorentz violation have been proposed.
These include purely phenomenological models motivated by
aspects of cosmic-ray physics~\cite{mestres} and other considerations~\cite{pheno}.
A more theoretical suggestion -- the `Standard Model Extension' -- is the possibility that Lorentz invariance is broken
spontaneously~\cite{sme,pospelov,Kost_LV_operators}, and it has been
argued in the context of some models of loop quantum gravity~\cite{loop}
that the vacuum might exhibit non-trivial optical properties. Moreover,
quantum field theories of the Lifshitz type~\cite{lifsh}, in which the space
and time coordinates scale differently, have attracted renewed interest
in the context of quantum gravity. In Lifshitz theories
Lorentz invariance can be violated at high energies, but is
restored in the low-energy limit. Another approach is that of doubly
(or deformed) special relativity~\cite{dsr}, in which Lorentz invariance
is fundamentally deformed, rather than violated~\footnote{See \cite{dsrdoubts} for a discussion of the definition of
measurable momenta in such models.}.

How can one probe such ideas, in particular the modification (\ref{mdr1}) of
photon propagation {\it in vacuo}? It was suggested in~\cite{AEMNS} that variable astrophysical sources
would provide the most sensitive probes of such Lorentz violation, in view of their large distances and
and non-trivial time structures, as well as their emissions of high-energy photons. Examples of such sources
that were suggested in~\cite{AEMNS} include pulsars, active galactic
nuclei and gamma-ray bursts (GRBs). The {\it Fermi}-LAT sample of the latter is the subject of the analysis
in this paper~\footnote{In field-theoretical
models such as the Standard Model Extension~\cite{sme,pospelov,Kost_LV_operators} in which Lorentz invariance is broken
spontaneously, there are also birefringence effects. Probes of the rotation of the polarization of light from distant astrophysical
sources~\cite{bireAngle,GRB_Polal} constrain this effect very strongly~\cite{Stec_Polar,INTEGRAL_Polar,strict_Polar,Lin_Pol,Kislat_Polar}.
However, birefringence is not expected in the models of space-time foam studied here.}.

There have been many previous analyses of such variable astrophysical emissions.
The first systematic study of possible Lorentz violation using
the light curves of a number GRBs with emissions in the sub-MeV energy range distributed over
a range of red shifts was presented in~\cite{mitsou}~\footnote{We recall, however,
that the effective quantum-gravity scales would depend on the density of D-brane defects
in D-foam models~\cite{emnewuncert}, which could vary with the cosmological epoch.}.
The study was extended subsequently in~\cite{EMNS,EMNSS},
incorporating the light curves of substantially larger samples of GRBs,
applying advanced time series analysis techniques (wavelets), and scrutinizing
the possible systematic uncertainties inherent to such kind of analyses.  Studies
exhibiting similar levels of sensitivity have also been performed elsewhere~\cite{Lamon,Bolmont}.
Recent searches for Lorentz violation using samples
of sub-MeV light curves of short GRBs~\cite{SWIFT_SHORT_LV}
detected by the {\it Swift} satellite have not reported an improvement of sensitivity, compared
with the analysis of \cite{EMNSS}.  On the other hand, observations by~\lat~\cite{Fermi-LAT} of high-energy emission from GRBs where
the energies of some individual gamma rays exceeded 10~GeV made possible
a substantial increase in the sensitivity to $M_1$, approaching the Planck scale.
For example, an analysis of time differences in the arrival times of individual gamma rays from
the single source GRB080916C suggested a limit on $M_1$ that was
about 2 orders of magnitude~\cite{Fermi_080916} stronger than that in~\cite{EMNSS}.
Another source GRB090510A detected by~\lat~was used to give
a {trans-Planckian} lower limit on this scale of quantum gravity~\cite{Fermi_090510,NonFermi_090510,NonFermi_090510_develop}.
Moreover, it was argued in~\cite{Nemiroff} that the assumption of a particular ``rhythm'' in the arrival of multi-GeV gamma rays from
GRB090510A could even push the lower limit on the quantum gravity scale
up to two orders higher than the Planck scale. Another assumption was made in~\cite{ch1,ch2,ch3,ch4}, where it was suggested that the source frame
time offset of the individual highest-energy gamma rays in emissions of \lat~objects should coincide to
very high precision with the time offset of the peak emission of the sub-MeV energy light curves of  the objects. This
analysis led to a claim of a signal for the violation of Lorentz invariance, rather than a lower bound,
corresponding to a quantum-gravity scale $M_1 \sim10^{17}$~GeV.

However, reports of sensitivities to Lorentz-violating effects with $M_1 \sim M_P$ and, {\it a fortiori}, claims of signals, are beset
with systematic uncertainties associated with our ignorance of the energy dependence of the times at which
photons are emitted at the source. In particular, the literature also contains considerable
discussion of the possibility that some higher-energy photons may be emitted later than prompt lower-energy photons, see~\cite{HE_GRB}, powered
by a relativistic blast wave in the circum-GRB medium. However, we do not enter this discussion here.

Instead, our aim is to develop statistical techniques that minimize the impact of such source effects,
which is the central point of our paper~\footnote{Some attempts to combine such source effects with propagation effects due to a
potential quantum gravity medium, based on a particular ``magnetic-jet'' model for GRB emission~\cite{mj}, can be found in \cite{fits}.}.
In the current study, we consider three distinct statistical measures of GRB emissions that mitigate source effects, which we use in an
analysis of {\it Fermi}-LAT data in an attempt to obtain the most robust constraints on Lorentz violation associated with
modified dispersion relations of photons during their propagation in a quantum foamy space-time medium.
Although our analysis is motivated by one particular framework for Lorentz violation, the statistical techniques developed and the
results obtained here are applicable to a wide class of such models.

The structure of the article is as follows: in Section \ref{secPulse} we
review the basic features of propagation of a pulse in a dispersive quantum-gravity medium,
while in Section \ref{sec:DataInUse} we present the Fermi-LAT data to be used in our analysis. In Section
\ref{sec:recov}, we discuss methods to recover properties of source timing that will play an important r\^ole in our attempts to extract
robust constraints on Lorentz violation. In Section \ref{sec:estim},  we embark on our main part of the analysis,
by describing the various statistical measures of GRB emissions that we use to mitigate source effects.
The first, the {\it Irregularity Estimator} is
based on the observation that dispersion due to propagation through the space-time medium would tend
to `dilute' any burst-like feature, leading asymptotically to a distribution that shows no time-dependent features above the background.
One may then constrain the Lorentz violation parameters $M_n$ by minimizing this dilution.
The second statistical measure, the {\it Kurtosis Estimator} is based on the related observation
that the kurtosis, namely the height of a distribution relative to its standard deviation,
would also be reduced by the effects of propagation through the space-time medium.
Finally, the {\it Skewness Estimator} exploits the fact that an energy-dependent reduction in photon velocity
would increase the skewness, or asymmetry, of burst-like features in the emissions.
Uncertainties in these estimators are discussed in Section~\ref{sec:stab}, and in Section~\ref{sec:consol} we apply these methods
to the ensemble of {\it Fermi}-LAT data on emissions from GRBs with
bright emissions in the 100~MeV to multi-GeV energy band. Our analysis
leads to a lower limit on $M_1$ in the range
$2.4$ to $8.4 \times 10^{17}$~GeV,
which we consider to be the most robust  constraint to date on this type of Lorentz violation induced by a dispersive quantum-gravitational medium.
A brief discussion of these results, the associated uncertainties and ways to improve them, is presented in Section~\ref{sec:concl}.
Finally conclusions and outlook are presented in Section~\ref{sec:concl2}.

\section{Pulse Propagation in a Quantum-Gravity Medium}
\label{secPulse}

In this Section we review basic features of the deformation of the envelope of an electromagnetic wave packet during its
propagation in a quantum-gravity dispersive medium~\cite{mitsou} that leads to the refractive index effect (\ref{mdr1}).

The basic solution of the wave equation is a plane wave of the form
\beq
u(x,t) = A(k)e^{ikx-i\omega (k)t} \, ,
\eeq
where $k$ is the momentum and $\omega$ the frequency, and the superposition principle leads to the general solution
\beq
u(x,t) = \frac{1}{\sqrt{2\pi}}\intinf A(k)e^{ikx-i\omega (k)t}dk. \,
\eeq
Conversely, the amplitude $A(k)$ can be expressed as
\beq
A(k) = \frac{1}{\sqrt{2\pi}}\intinf u(x,0)e^{-ikx}dx \, .
\eeq
Here, for simplicity, we consider a normalized Gaussian wave packet, with variance $a^2$:
\beq
u(x,0) = \frac{1}{a\sqrt{2\pi}} e^{-\frac{x^2}{2a^2}} \, .
\label{gaussPack0}
\eeq
The amplitude in momentum space of such a wave is
\beq
\label{amplGauss1}
A(k)=\frac{1}{\sqrt{2\pi}}\intinf u(x,0)e^{-ikx}dx = \frac{1}{\sqrt{2\pi}}e^{-\frac{a^2k^2}{2}} \, .
\eeq
If the distribution $A(k)$ is sharply peaked around some value $k_0$, the group velocity of a traveling pulse is given by
\beq
v_g=\frac{d\omega}{dk}\vert_{k_0} \, .
\label{vGr}
\eeq
Provided that the quantum-gravity-induced deviation of the propagation velocity from the speed of light would then imply
that the dispersion relation has the form
\beq
\omega^2=k^2(1+2\beta_n k^n) \, ,
\label{dispRel1}
\eeq
for $\beta_n k^n<<1$, the group velocity is given
\beq
v_g \approx 1+(n+1)\beta_n k_0^n \, ,
\label{vg}
\eeq
which yields a correction of the form (\ref{mdr1}) if $\beta_1$ is negative and $n = 1$.

In the case of a Gaussian packet formed by superposing traveling plane waves of momentum $k_0$
\beq
u(x,0) = \frac{1}{a\sqrt{2\pi}} e^{i k_0 x} e^{-\frac{x^2}{2a^2}} \, ,
\label{gaussPackk0}
\eeq
that locates at $x=0$ at $t=0$ and propagates along the $x$ direction, the corresponding
Fourier amplitude is given by
\beq
A(k)=\frac{1}{\sqrt{2\pi}}e^{-\frac{a^2}{2}(k-k_0)^2} \, .
\label{pulseAt0}
\eeq
At a later time $t > 0$, the pulse (\ref{pulseAt0}) will evolve as
\beq
u(x,t)=\frac{1}{\sqrt{2\pi}}\intinf A(k)e^{i(kx-\omega t)}dk=\frac{1}{2\pi}\intinf
e^{-\frac{a^2}{2}(k-k_0)^2}e^{i(kx-\omega t)}dk\, ,
\label{pulseT1}
\eeq
which for $n=1$ readily reduces to
\beq
u(x,t)=\frac{1}{2\pi}e^{i(xk_0-t\omega_0)}\intinf
e^{-\left(\frac{a^2}{2}+i\beta_1 t\right)(k-k_0)^2}e^{i(x-v_gt)(k-k_0)}dk \,
\label{pulseT2}
\eeq
where $\omega_0=k_0(1+\beta_1k_0)$.
Evaluating the integral in (\ref{pulseT2}), one arrives at the following final expression for the amplitude of the wave packet
\beq
u(x,t)=\frac{1}{2\sqrt{\pi}}\frac{e^{i(xk_0-t\omega_0)}}{\left(\frac{a^2}{2}+i\beta_1 t\right)^{1/2}}
\exp\left[-\frac{(x-v_gt)^2}{4\left(\frac{a^2}{2}+i\beta_1 t\right)}\right] \, .
\label{pulseT3}
\eeq
Eventually, one also obtains the intensity of the wave packet:
\beq
I(x,t,v_g)=|u(x,t)|^2= \frac{1}{2\pi a^2}\frac{1}{\left(1+4\frac{\beta_1^2t^2}{a^4}\right)^{1/2}}
\exp\left[-\frac{(x-v_gt)^2}{a^2\left(1+4\frac{\beta_1^2t^2}{a^4}\right)}\right] \, .
\label{gaussPack1}
\eeq
It is easy to see from (\ref{gaussPack1}) that, as time evolves, the peak of the amplitude gets shifted to $x+v_gt$
and the amplitude of the envelope reduces. The packet becomes wider in
such a way that an initially narrower packet spreads much faster compared to one that has a larger initial width $a$.

\begin{figure}
\centering
\includegraphics[scale=0.50]{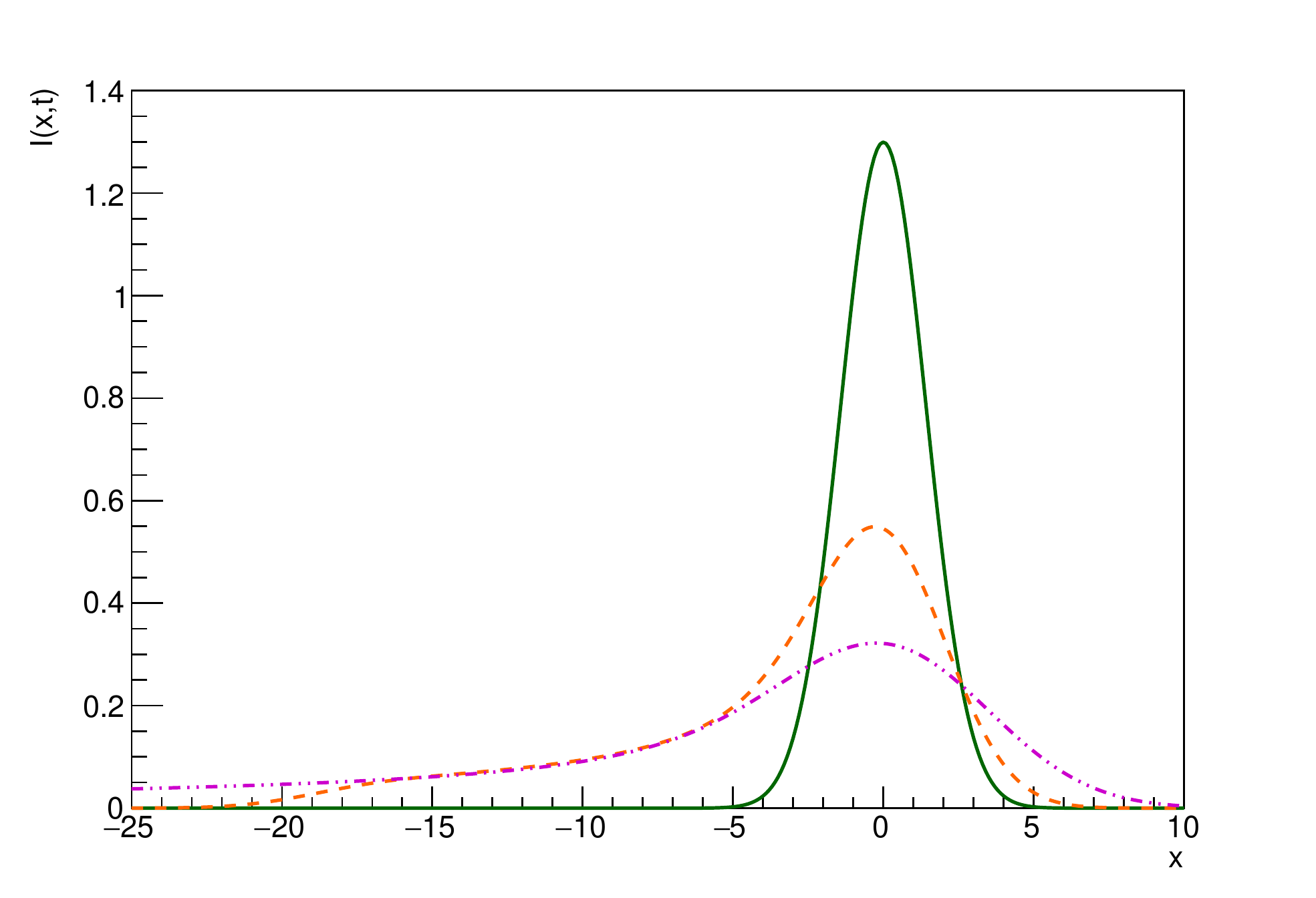}
\vspace{-0.4cm}
\caption{\it Time evolution of the profile of a Gaussian wave packet injected
into a dispersive medium characterized by (\ref{dispRel1}), convoluted at two different
points in time $t_2>t_1>0$ with a power-law energy spectrum~(\ref{spectrum1}).
The (green) solid line represents the profile
of the wave envelope at the injection time ($t=0$). The (orange) dashed and (magenta) dash-dotted lines represent
the profiles modified by the propagation effects at times $t_1$ and  $t_2$, respectively.
The energy range spans two orders of magnitude in dimensionless units, so as to reproduce a typical spectral range
of the high-energy GRB emissions measured by {\it Fermi}-LAT used in the analysis.}
\label{fig:packet}
\end{figure}

Let us now assume that a signal with Gaussian profile and some spectral content $\Phi (k)$ is located
within a band spanning a certain range from $k_1$ to $k_2$.
In this case, since the group velocity in a dispersive medium (\ref{vg}) depends on the wave number $k$, the overall
intensity should be calculated as the convolution
\beq
{\cal I}(x,t) = \int_{k_1}^{k_2}I(x,t,v_g(k))\Phi (k)dk \, .
\label{convSpectrum1}
\eeq
The {\it Fermi}-LAT high-energy emission spectra of GRBs can be well approximated by a power-law
model~\cite{LAT-Cathalog} with positive spectral index $\alpha$~\footnote{Despite the variety of spectral models
(see for example~\cite{LAT-Cathalog} and references therein), a simple power-law serves as an approximation
to the convolution (\ref{convSpectrum1}) that is sufficient to reveal the main features of the deformation of a Gaussian profile propagating
through a quantum-gravitational medium.}:
\beq
\label{spectrum1}
\Phi (E) \propto E^{-\alpha} = \Phi_0k^{-\alpha} \, ,
\eeq
and results of the numerical convolution (\ref{convSpectrum1}) at two points in time, using to the profile
(\ref{gaussPack1}) and the spectral model (\ref{spectrum1}) are presented in Fig.~\ref{fig:packet}.

As can be seen in Fig.~\ref{fig:packet}, the burst-like feature of a GRB intensity profile modelled by a Gaussian
envelope is deformed during propagation in a quantum-gravity dispersive medium of the type considered in this work.
Three features of this deformation can be distinguished, which we exploit subsequently in our
analysis.

(i) As the signal of the GRB propagates, irregularities in its intensity profile, superposed upon a background, get diluted,
causing any pulsing intensity profile to approach an almost featureless the background like time profile at large times.
One can invert this possible quantum-gravity propagation effect, converting
the timings and energies of photons arriving in high-energy emissions from distant GRBs
detected by {\it Fermi}-LAT back to the intensity distribution that would have been injected at the source, compensating the signal timings
for the propagation delays (see Section~\ref{sec:DataInUse} for details).
One can then estimate the amount of Lorentz violation affecting the propagation of the signal by calculating the amount of compensation that
maximizes the irregularities (spikes) of the GRB intensity distribution injected
at the source, denoted by $t=0$ in Fig.~\ref{fig:packet}.
The qualitative and quantitative picture of smoothing of irregularities in the initial intensity distribution
of a burst-like signal by Lorentz violation during its propagation holds for any shapes of  initial spikes, which may be quite irregular, unlike the Gaussian
example shown in Fig.~\ref{fig:packet}.

(ii) One such effect on the intensity profile is that the heights of peaks in the signal are reduced during its propagation in a dispersive medium.
Therefore, one can quantify the relative sizes of the peaks of the compensated intensity distribution and estimate the amount
of Lorentz violation by calculating the amount of compensation that would maximize the peaking of the distribution at the time of emission.
In making this estimate, we use in Section~\ref{sec:kurt} a method that does not depend on the particular shapes
of the peaks.

(iii)  The intensity time profile becomes more asymmetric with time.
The degree of asymmetry of a time profile can be estimated by comparing it with
a symmetric reference distribution, which we take for convenience to be
a normal distribution. One can use as an estimate of the effect of propagation through
a quantum-gravitational medium the amount of Lorentz violation
that maximizes the symmetry of the initial distribution.
We note that the estimator we use in Section~\ref{sec:skewness} to quantify this kind of deformation of the signal
is not restricted by any assumption on the shape of the initial peaks in the intensity distribution.

In later Sections we present in detail the methods used to estimate the deformation features outlined above.

\section{{\it Fermi}-LAT Data}
\label{sec:DataInUse}
The data analyzed in this work are taken from the {\it Fermi}-LAT Pass 8 transient event class
{\tt P8R2{\_}TRANSIENT010}~\cite{fermiLATdata}.
The background contamination in this set of data is calibrated to the best-fit power-law
parametrization of the Isotropic Diffuse Gamma-Ray Background (IGRB) emission from~\cite{Abdo}.
The loosest selection criteria for this {\tt TRANSIENT} class are designed for short-duration events,
such as GRBs, that benefit from increased photon statistics while tolerating a higher
background fraction and the broader point spread function (PSF) of LAT. This class has a background
rate that is equal to the IGRB reference spectrum and requires the presence
of a signal in both the tracker and the calorimeter.

The required data set is extracted from the publicly available archive~\cite{FermiLATarch} at
Fermi Science Support Center.

The Fermi mission provides a suite of tools, called the Fermi
Science Tools, for the analysis of LAT data, and the tool to perform selection cuts is called {\tt gtselect}.
A lower energy limit of 100 MeV on photon energies is chosen to reject photons with poorly reconstructed
energies and directions. No maximum energy cut is applied, since photon energies can reach
a few tens of GeV. We select photons reconstructed within a circular region of interest (ROI)
centered on the best available GRB localization with
a radius corresponding to the 95\% containment radius of the transient LAT PSF
estimated for a 100 MeV photon.
The GRB directions used to specify the center of the ROI  are obtained by follow-up ground-based
observations, and can be assumed for our purposes to coincide with the true direction of the GRB.

\begin{figure}
\centering
\includegraphics[width=0.5\textwidth]{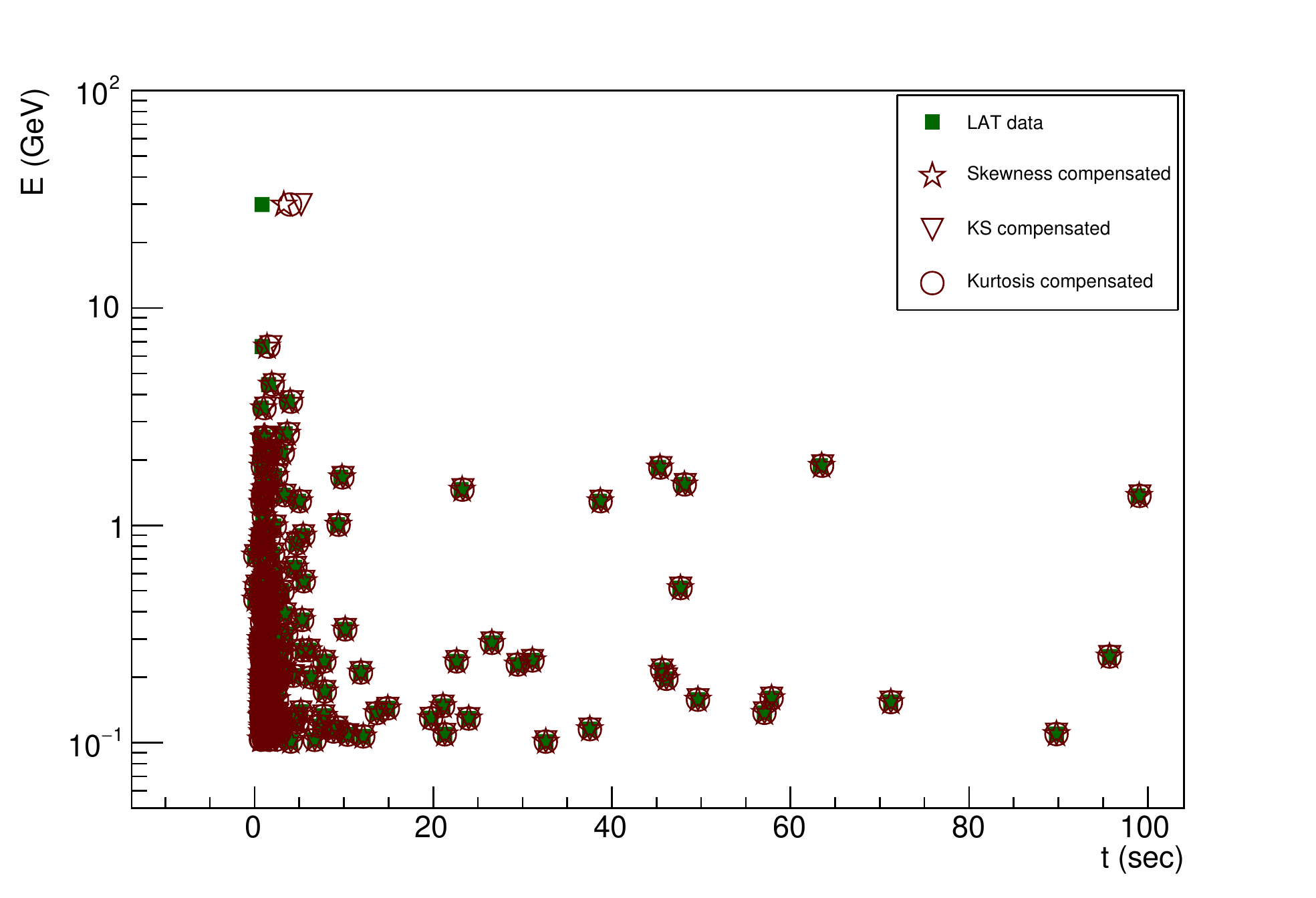}\hspace{0cm}\includegraphics[width=0.5\textwidth]{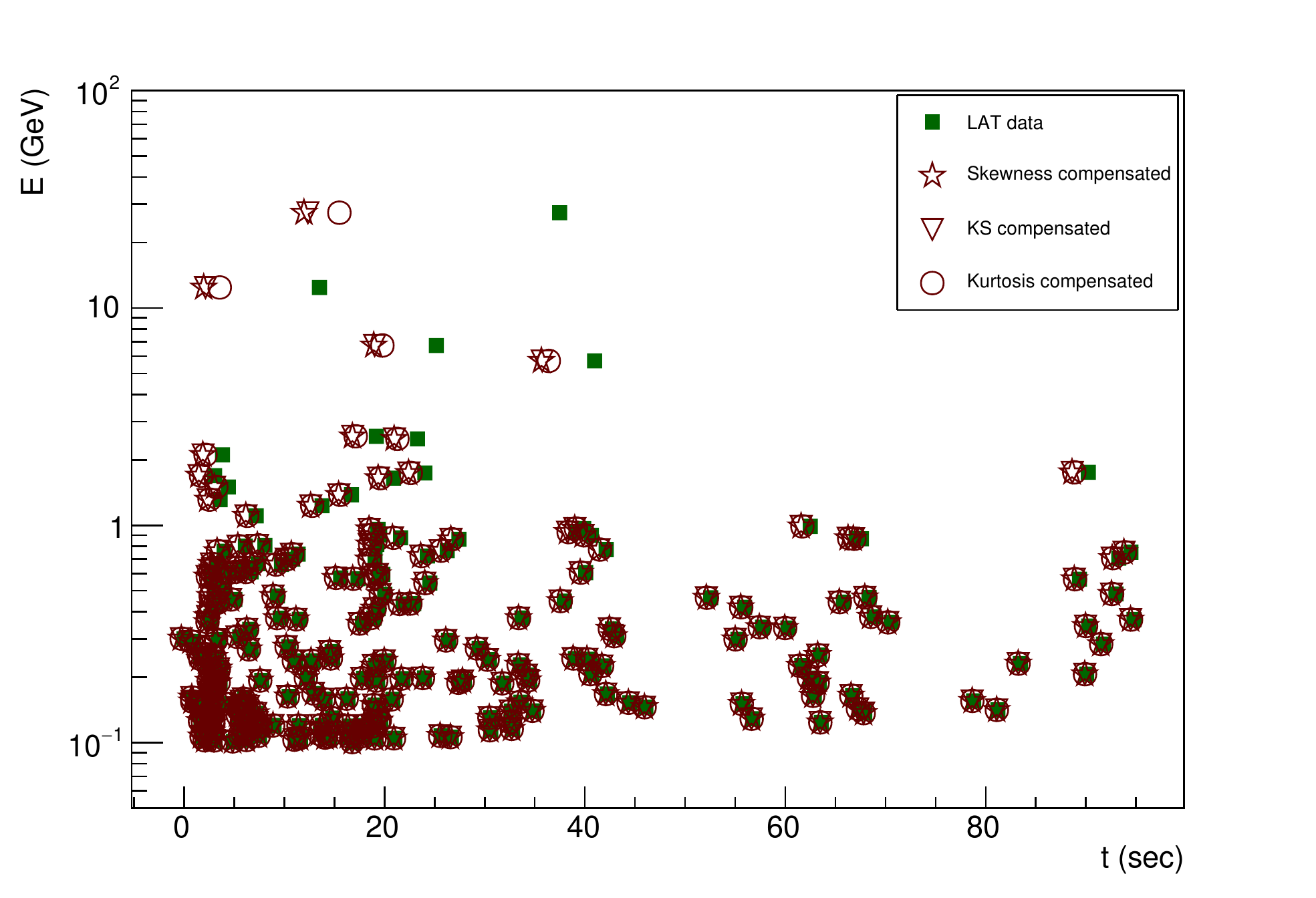}
\vspace{-0.4cm}
\caption{\it Energies vs. arrival times of \lat~photons that passed the transient off-line selection (solid green squares),
as outlined in Section~\ref{sec:DataInUse}, which are consistent with the direction of GRB090510A (left panel) and GRB080916C (right panel).
The earliest arrival time is set to zero in each case.
The detector frame delays indicated in the left panel by the open triangles, stars and circles  are calculated assuming a dependence $F(E)=E$,
with the irregularity estimator, kurtosis estimator and skewness estimators, respectively, (see the text in details) for compensation parameters
with $\tau  = -0.081$~s/GeV, $\tau  = -0.099$~s/GeV
and $\tau  = -0.104$~s/GeV. The same is shown in right panel
for $\tau  = 0.930$~s/GeV, $\tau  = 0.935$~s/GeV
and $\tau  = 0.801$~s/GeV.}
\label{fig:DATA_090510}
\end{figure}

The data for the 8 GRBs with measured
red shifts and relatively high numbers of photons with
energies above 100 MeV detected by {\it Fermi}-LAT that are used in our analysis are presented in Table~1.
Fig.~\ref{fig:DATA_090510} shows scatter plots of the photon energies versus arrival
times for two GRBs in this data sample. The data for all the GRBs in Table~1 are processed
similarly, using the various estimation procedures
described below.
\begin{table}[]
\centering
\label{tab:grbs}
\begin{tabular}{lllllll} \toprule
    \; GRB  & \, $z_{\rm src}$ & $N$ & $T_{\rm HE}^{\gamma}$ & $E_{\rm HE}^{\gamma}$ &
   \quad $\tau\pm\sigma_{\tau}$ &   $\sigma_{\tau}$ (Bias corr.)  \\
       &   &   & \ [s]  & [GeV] & \quad [s$\cdot {\rm GeV}^{-1}$] & \quad [s$\cdot {\rm GeV}^{-1}$] \\ \hline\hline
    080916C  & 4.350 & 220 & 16.5 & 13.2 & $0.892\pm 0.053$ & \quad \quad 0.096 \\
    090510A  & 0.903  & 222 & 0.8  & 31.3 & $-0.099\pm 0.014$ & \quad \quad 0.023 \\
    090902B  & 1.822  & 329 & 81.8  & 33.4 & $1.655\pm 0.088$ & \quad \quad 0.139 \\
    090926A  & 2.1062  & 310 & 24.8  & 19.6 & $0.534\pm 0.054$  &  \quad \quad 0.104 \\ 
    110731A  & 2.830   & 80 & 5.0   & 3.2 & $4.54\pm 1.12$ & \quad \quad 1.692 \\
    130427A  & 0.34  & 584 & 243.0 & 95 & $0.652\pm 0.107$ & \quad \quad 0.618  \\
    160509A  & 1.60  & 33 & 77.0  & 52 & $0.946\pm 0.054$ & \quad \quad 0.122   \\
    170214A  & 2.53  & 298 & 105  & 7.8 & $-3.68\pm 1.16$  & \quad \quad 3.084  \\ \hline\hline
\end{tabular}
\caption{List of \lat\ GRBs included in our analysis.
The notations used are: $z_{src}$  - red shift of the source;
$N$ - number of photons arrived from the source;
$T_{\rm HE}^{\gamma}$ - arrival of the most energetic photon detected by \lat ;
$E_{\rm HE}^{\gamma}$ - energy of the most energetic photon detected by \lat ;
$\tau\pm\sigma_{\tau}$ - the mean value and $1\sigma$ uncertainty of the overall
distribution of the correct values of the compensation parameter
as described in Sections~\ref{sec:recov} and \ref{sec:stab}; $\sigma_{\tau}$ (Bias corr.) -
$1\sigma$ uncertainty in the overall distribution of the values of the compensation parameters
corrected for the bias systematic, as described in Section~\ref{sec:stab}. The data were extracted from~\cite{FermiLATarch}.}
\end{table}

\section{Recovery of the Source Timing Properties}
\label{sec:recov}

If $\beta_n$ in (\ref{dispRel1}) is set to
\beq
\label{betanQG}
\beta_n=-\frac{1}{2}(M_{QGn}^{-1})^n \, ,
\eeq
where $M_{QGn}$ represents the scale at which Lorentz-violating quantum gravity effects set in,
the group velocity (\ref{vg}) acquires the form
\beq
\label{vg1}
v_g(E)=\left[ 1-\frac{n+1}{2}\left( \frac{E}{M_{QGn}}\right)^n\right] \, .
\eeq
The differential relation between time and red shift in the standard cosmological
$\Lambda$CDM model with dark energy and dark matter contributions $\Omega_{\Lambda} = 0.7$ and  $\Omega_M = 0.3$
respectively, is given by
\beq
\label{tz}
dt=-H_0^{-1}\frac{dz}{(1+z)h(z)} \, ,
\eeq
where $h(z)=\sqrt{\Omega_{\Lambda}+\Omega_M(1+z)^3}$ and $H_0=68$~km/s/Mpc is the present Hubble expansion rate (see for example~\cite{PDG}).
Thus, the difference in {proper} distance covered by two photons emitted  at red shift $z_{\rm src}$ with velocity difference $\Delta v_g$ is
\beq
\label{distz1}
\Delta L = H_0^{-1}\int\limits_{0}^{z_{\rm src}}\frac{\Delta v_g dz}{h(z)} \, .
\eeq
It follows from~(\ref{vg1}) that the velocity difference of two photons of energies $E_2>E_1$ is given by
\beq
\label{dvg}
\Delta v_g(E_1,E_2) = \frac{n+1}{2}\frac{E_2^n-E_1^n}{M_{QGn}^n} \, .
\eeq
Therefore, the difference in the arrival times of these photons is
\beq
\label{dtz}
\Delta t=\frac{n+1}{2}\frac{H_0^{-1}}{M_{QGn}^n}(E_2^n-E_1^n)\int\limits_0^{z_{\rm src}}\frac{(1+z)^ndz}{h(z)} =
a_{QGn}F_n(E_1,E_2)K_n(z_{\rm src}) \, ,
\eeq
{where the factors in (\ref{dtz}) are $a_{QGn}=\frac{n+1}{2}\frac{H_0^{-1}}{M_{QGn}^n}$, $F_n(E_1,E_2)=(E_2^n-E_1^n)$ and
$K_n(z_{\rm src})=\int\limits_0^{z_{\rm src}}\frac{(1+z)^ndz}{h(z)}$}.
In the following, the earliest arrival time  of a photon from a given GRB is always set to zero~\footnote{For reference, we
recall that the observed time difference is not simply $\Delta t $, but it is stretched by an additional factor $(1 + z_{\rm src})$,
due to the Universe's expansion (see (\ref{lag1}) below).}.

As discussed above, the expectation that photons with lower energies (longer wavelengths)
may be delayed less than photons with higher energies (shorter wavelengths)
implies that the temporal pattern of photons arriving from a given GRB
should be modified compared to the pattern when emitted by the source.
However, in order to elucidate the possible magnitude of the quantum gravity dispersion effect, we need
statistical estimators that enable us to discriminate between source and propagation effects.

For the purpose of our analysis, we allow the arrival time of every detected photon
to incorporate an {\it a priori} unknown source-related time-lag $b_{\rm sf}$ as well as
the energy-dependent time delay (\ref{dtz}) accumulated in the course of propagation because of
quantum gravity dispersion~\cite{EMNSS}. In the case of linear energy dependence, $n=1$ (\ref{mdr1}),
one finds the following expression for the arrival times of
individual photons:
\beq
t_{\rm obs}(E_i) = b_{\rm sf}(E_i)(1+z_{\rm src})+\tau(z_{\rm src})E_i \, ,
\label{lag1}
\eeq
where
\beq
\label{tauK1}
\tau(z_{\rm src}) = a_{\rm QG1}K_1(z_{\rm src})
\eeq
is a ``compensation parameter" that quantifies the amount of the linearly energy-dependent
propagation effect encoded in the signal from a given source. ``Compensation" is understood here
in the sense of recovery of the original pattern of the intrinsic emission times by removing a possible propagation
effect related to quantum-gravity dispersion. In practice, instead of the source frame intrinsic
timings $b_{\rm sf}(E_i)$ we use the detector frame intrinsic timings given by
\beq
b_{\rm df}(E_i,\tau) = t_{\rm obs}(E_i) - \tau(z_{\rm src})E_i \, .
\label{rec1}
\eeq
The correct value of the compensation parameter
applied in~(\ref{rec1}) recovers the intrinsic pattern of the timings $b_{\rm df}(E_i)$
before being dispersed by quantum-gravity effects.
Because of unknown details of the source activity due to our imperfect knowledge of the radiation mechanism
of GRBs, as well as of potential stochasticity during the burst,
the intrinsic source distribution $b_{\rm df}(E_i)$ is expected
to be different for different GRBs. This is the main challenge for inferring a common
propagation effect from samples of high-energy gamma rays emitted by different GRBs.

As we have demonstrated in Section~\ref{secPulse}, a Gaussian emission envelope would be deformed during its
propagation through a dispersive quantum-gravity medium.
We may assume that there would be similar deformations in the shape of an emission
envelope of arbitrary profile with burst-like features in its temporal intensity distribution.
Therefore, one may estimate the compensation parameter using statistical quantifications of
the deformations in the intensity profile of an envelope propagating in such a medium.

For a given source, the data are represented
by $N$ points, each one associated with the arrival time and energy of a photon reconstructed by the
\lat~(for two examples of sources used in our analysis, see Fig.~\ref{fig:DATA_090510}).
Describing the pattern of intrinsic timings of individual photons in the detector frame
by a probability distribution function ${\cal F}(b_{\rm df}(E_i,\tau ))$,
the shape of the intensity distribution is given by
\beq
{\cal I}(b_{\rm df}(E_i,\tau )) = W_i{\cal F}(b_{\rm df}(E_i,\tau )) \, ,
\label{intens}
\eeq
where
\beq
\label{weight}
W_i=\frac{E_i}{E_{\rm min}}
\eeq
is the energy weight of a given photon with energy $E_i$ normalized
by the energy $E_{\rm min}$ of the softest photon measured by the {\it Fermi}-LAT within the
sample from a given GRB. The compensation parameter $\tau $ in~(\ref{intens}) is
defined in such a way that the profile ${\cal I}(b_{\rm df}(E_i,\tau  = 0))$ coincides with that measured in the
detector, after the deformation by propagation effects.

This deformation can be characterized by non-parametric estimators whose optimization, using
appropriate criteria, allows one to estimate the correct value of the
compensation parameter.
In practice, we calculate the estimators for trial values
of the compensation parameter $\tau ^j$ applied in~(\ref{rec1}), where $j$ indicates one of a set of
values taken either from a regular grid or random values distributed uniformly over a
pre-defined range of values of $\tau $.
The correct value of $\tau ^j$ should generate an intensity distribution ${\cal I}(b_{\rm df}(E_i,\tau ^j))$ that
satisfies in the best possible way the criteria for recovery of the genuine pattern of the detector-frame intrinsic timing
for a given estimation technique, as we discuss in the next Section.

\section{Estimation Techniques \label{sec:estim}}

We present three estimators in this Section, one sensitive to each type of deformation
of a wave envelope with burst-like features described in Section~\ref{secPulse}. They are used subsequently in our analysis of the Fermi-LAT
data presented in Section~\ref{sec:DataInUse}. As we demonstrate below, these techniques lead to robust constraints on the
photon refractive index potentially induced by Lorentz-violating quantum-gravity effects.

\subsection{Irregularity Estimator}
\label{sec:ks}

We have shown in Section~\ref{secPulse} that a burst-like signal with a power-law spectrum,
as modelled by a Gaussian shape, gets ``diluted"
with time as it propagates in a dispersive medium. In general, a qualitatively similar result is expected
for a signal possessing any kind of  burst-like activity {superposed on a ``regular" background distribution}.
Thus we expect that during dispersive
propagation any irregular signal with burst-like features degenerates in shape, approaching
{this background distribution~\footnote{In general, the shape of the background distribution is suggested
by the physics of non-variable part of the GRB's engine.}.
Conversely, application of the procedure for compensating
quantum gravity propagation effects described above should
recover the intrinsic irregularities at the source. The degree of irregularity
can be estimated by comparing a compensated intensity distribution with a reference one, the latter being  an
{\it a priori} featureless {(smooth)} distribution with the same statistical strength.
It is clear that {, in the absence of any insight into the physics of the non-variable part of the
high-energy emission by the GRB engine and hence
any assumptions about the shape of the background,}
the best reference featureless distribution is the uniform one.
Since the shape of the intensity PDF at the source is unknown, we make the
comparison on the basis of non-parametric statistics.
This ensures that the analysis is as independent as possible of assumptions on the {shape of the
irregularities an hence the dynamics of the variability of of GRB engines in the high-energy band}.

We use the Kolmogorov-Smirnov (KS) statistic (see, for example, ref.~\cite{statBOOK1}) to estimate
the degree of irregularity of an intensity distribution. We define the KS difference between two
distribution functions $\Xi_T(t)$ and $U_T(t)$ as
\beq
D(\tau ) = \underset{t_0 < t < t_{N-1}}{\max}|\Xi_T(\tau ,t)-U_T(\tau ,t)| \, ,
\label{KSD1}
\eeq
where $t_0$ and $t_{N-1}$, which themselves are functions of $\tau $, represent the timings of the first and the
last photon within a compensated distribution, respectively.

Following (\ref{intens}), the \lat\ list of photons is converted as follows
into the distribution function $\Xi_T(\tau ,t)$. First, for a given source, we associate every photon with its energy weight~(\ref{weight}).
Then, the function $\Xi_T(\tau ,t)$ is constructed as the fraction of those energy weights $W_i$ whose
associated photons arrived within the time range $[t_0;T]$, where $T\le t$.
The reference distribution function $U_T(\tau ,t)$ is generated as a set of $N_U={\rm ceil}(\sum_NW_i)$
times generated randomly and distributed uniformly over the range $[t_0;t_{N-1}]$. The function $U_T(\tau ,t)$ is then defined as the
fraction of generated times within the range $[t_0;T]$, where $T\le t$.

Following the rigorous KS procedure~\cite{statBOOK1}, the distribution of the KS statistic can be calculated in the
case of the null hypothesis, which, in our case, is the set of $N_U$ uniformly distributed timings. The distribution of the KS statistic gives the significance
of any non-zero observed value of $D(\tau )$. The function that enters into the calculations
of the significance, $Q_{\rm KS}(\lambda)$, where $\lambda\propto D(\tau )$, is monotonic
with limiting values $Q_{\rm KS}(0)=1$ and $Q_{\rm KS}(\infty )=0$. Regardless
of the exact form of $Q_{\rm KS}(\lambda)$, the most significant incompatibility
between a compensated intensity distribution and the null hypothesis is achieved for the value
of $\tau $ in (\ref{rec1}) that maximizes the difference (\ref{KSD1}). In turn, the most significant
deviation of the data from a featureless signal (uniform distribution) unambiguously implies that
the data are maximally irregular, so that burst-like features of the signal are recovered optimally.

\begin{figure}
\centering
\includegraphics[width=0.5\textwidth]{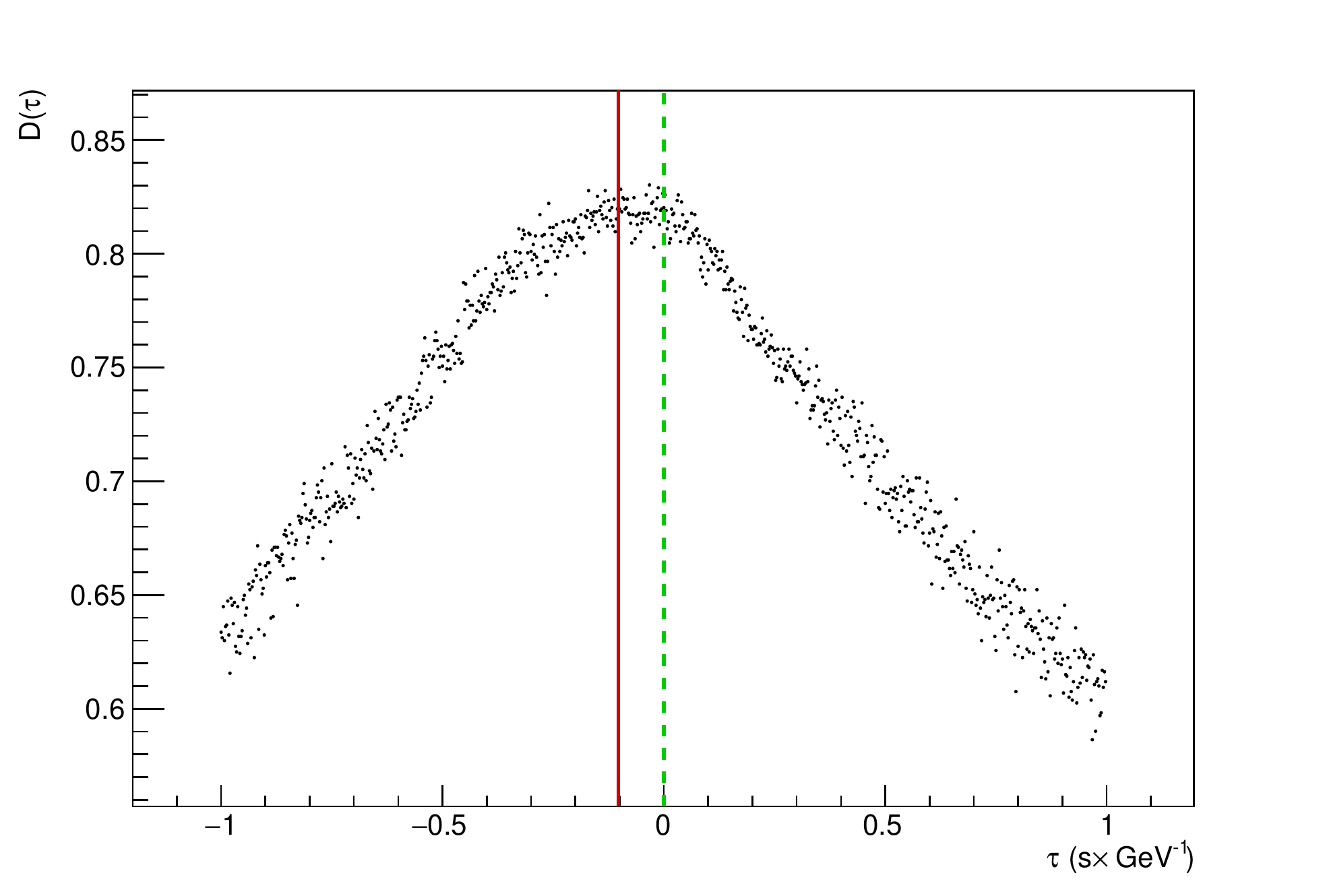}\hspace{0cm}\includegraphics[width=0.5\textwidth]{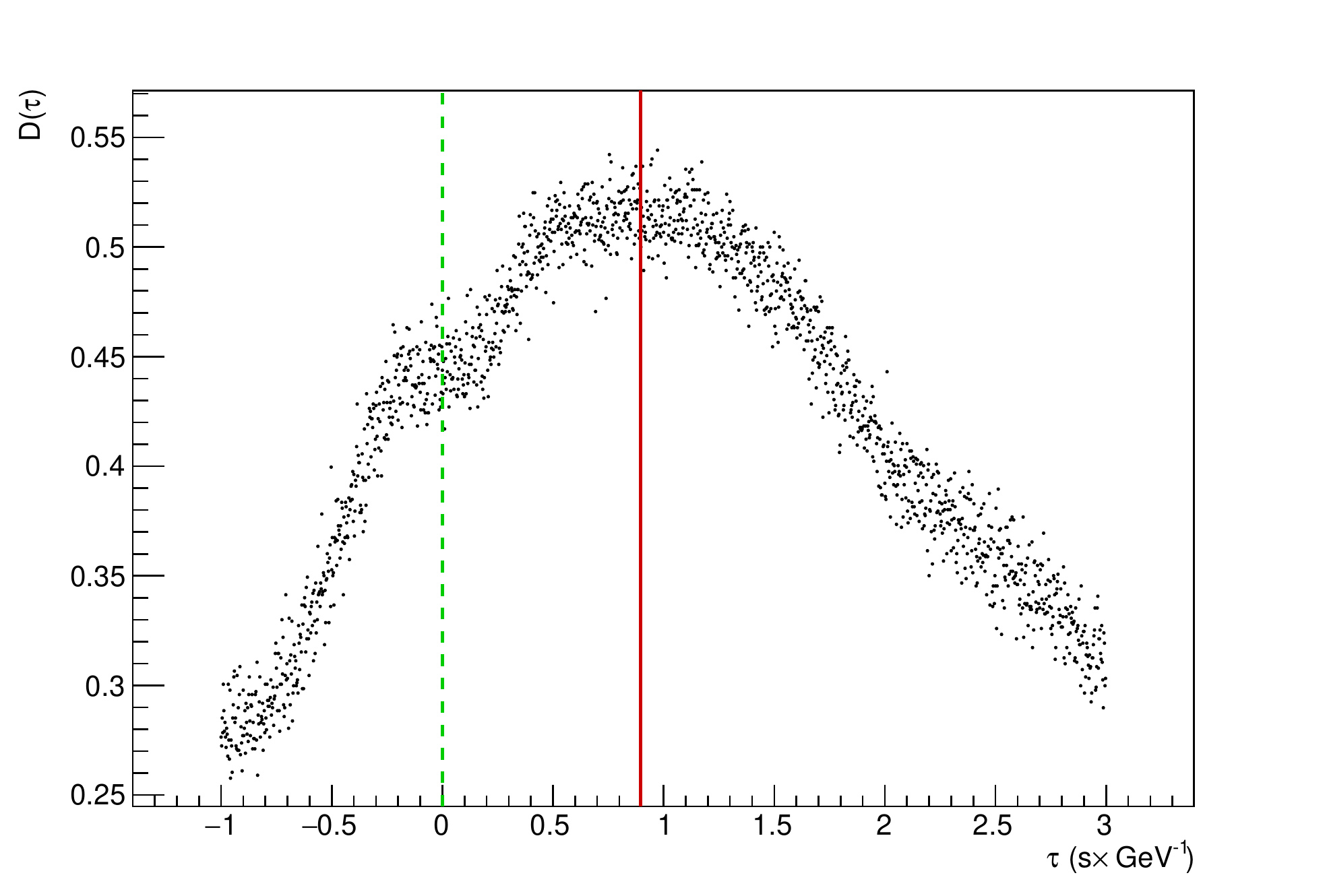}
\vspace{-0.4cm}
\caption{\it Values of $D(\tau )$ for a selected discrete set of trial values $\tau ^j$ of the compensation parameter,
where $j$ runs over a grid of several hundred values. The plots show the $D(\tau )$ dependences for the signal
from  GRB090510A (left panel) and GRB080916C (right panel). The positions of the maxima are
marked by the vertical (red) solid lines, $\tau  =- 0.103$~s/GeV (left panel)
and $\tau  = 0.90$~s/GeV (right panel), as estimated by
calculating the weighted average of the $\tau_\gamma$ values for which $D(\tau_\gamma)$ exceeds 95\% of its peak value.
The (green) dashed lines mark the positions of the maxima in the absence of any propagation effect.
}
\label{fig:KSD_090510}
\end{figure}

Examples of the $D(\tau )$ distributions for two GRBs (GRB090510A and GRB080916C) are presented in Fig.~\ref{fig:KSD_090510}.
The values of $\tau $ at the maxima are the best estimates of the values
of the compensation parameter that recover the initial irregularities
of the intensity distribution at the source. Since the $D(\tau )$ curves in Fig.~\ref{fig:KSD_090510}
exhibit significant fluctuations, in particular around the peaks, we use three methods to localize the
maxima of the $D(\tau )$ distributions, which are described in more detail in Section~\ref{sec:stab} below.
As an example, the positions of the maxima shown in Fig.~\ref{fig:KSD_090510} were obtained by averaging the distribution
of $\tau_\gamma$ values for which $D(\tau_\gamma)$ exceeds 95\% of its peak value.
The corresponding detector-frame intrinsic times derived from the
positions of the maxima of the $D(\tau )$ curves are plotted in Fig.~\ref{fig:DATA_090510} as open triangles, which
can be compared with the detected arrival times (green squares).

\subsection{Kurtosis Estimator}
\label{sec:kurt}

As already discussed in Section~\ref{secPulse}, as a signal becomes more ``diluted" during its propagation in a dispersive medium,
the peaks  in its burst-like features degrade. Therefore, the relative sizes of the peaks can serve as another measure of signal deformation
by quantum gravity effects. To quantify this effect on the intensity distribution we use a measure of  {\it kurtosis},
which provides information on the height of the peak of a distribution relative to the value of its standard deviation.
For the intensity defined in~(\ref{intens}), the kurtosis formula
for a compensated distribution is
\beq
\label{kurt1}
{\cal K(\tau )} = N_W\frac{\sum\limits_{i=0}^{N-1}((b_{\rm df}(E_i,\tau )-\overline{b_{\rm df}(\tau )})W_i)^4}
{\left(\sum\limits_{i=0}^{N-1}((b_{\rm df}(E_i,\tau )-\overline{b_{\rm df}(\tau )})W_i)^2\right)^2}-3,
\eeq
where, as above, every photon is associated with a weight $W_i$ given by (\ref{weight}), while $\overline{b_{\rm df}(\tau )}$
stands for average of intrinsic time
of a given signal in the detector frame, and $N_W$ is a normalization
factor~\footnote{The absolute value of the normalization factor is unimportant, since we study only the variation in ${\cal K}$
for different values of $\tau $.}.
Expression~(\ref{kurt1}) gives the excess of the kurtosis of the intensity distribution relative
to that of a normal distribution.
{Whatever the shape of the time profile at the source one expects that the
kurtosis of the intensity distribution always changes in a certain way in the course of propagation.}
{Namely,} a burst-like signal evolves towards a {\it platykurtic} (flattened) type
of intensity distribution~(\ref{intens}) as it propagates in a dispersive medium~\footnote{It is convenient to calculate the kurtosis~(\ref{kurt1}) of
a distribution relative to that of a normal distribution. However, this does not entail any assumption about the actual initial
shape of the time profile of a GRB.}.
{Therefore,} we use the compensation procedure described in Section~\ref{sec:recov}, see (\ref{rec1}),
to return the shape of the intensity distribution towards the maximally {\it leptokurtic} (peaky)
type~\footnote{We note that in other applications (see for example~\cite{KurtPeak}) the kurtosis
is used as a measure of the tail of a distribution,
rather than the shape of the peak}. {In other words, the estimator is based on the value of the compensation
parameter that maximizes the kurtosis, without any assumption on the shape of
the injected GRB time profile.}

\begin{figure}
\centering
\includegraphics[width=0.5\textwidth]{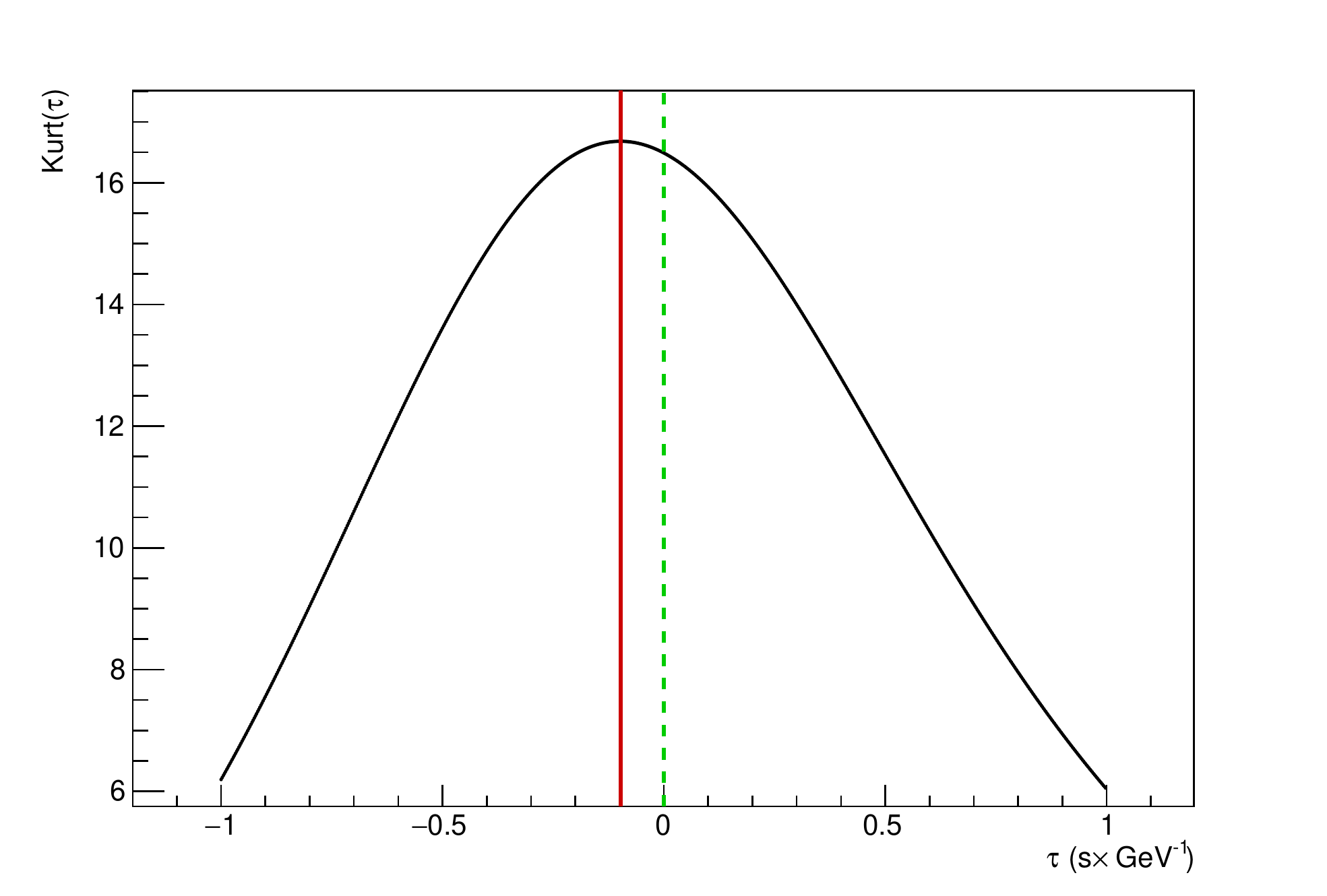}\hspace{0cm}\includegraphics[width=0.5\textwidth]{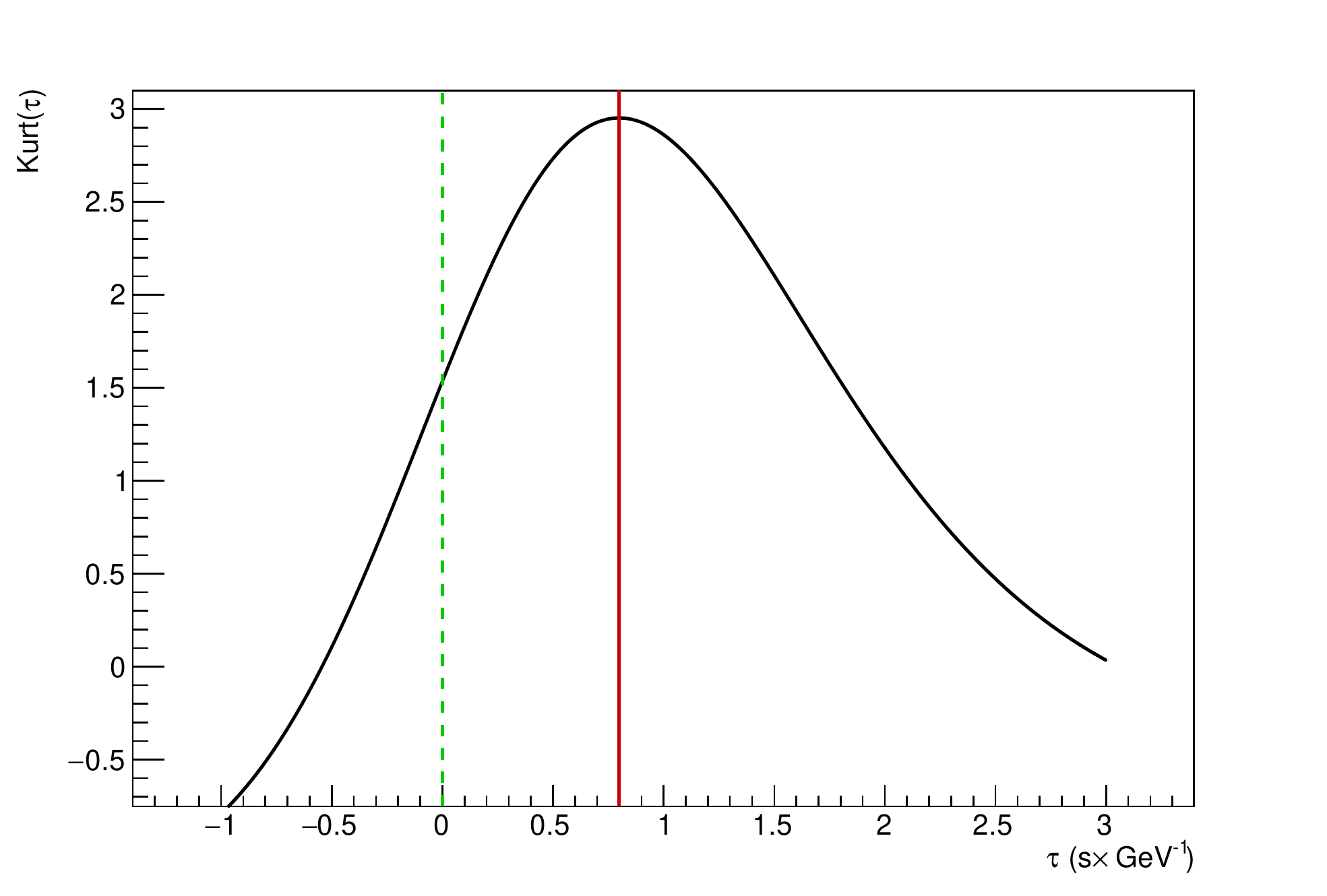}
\vspace{-0.4cm}
\caption{\it Curves of the kurtosis ${\cal K(\tau )}$ as functions of the compensation parameter $\tau $,
as calculated for a set of discrete values of the compensation parameter $\tau ^j$,
with $j$ running over a grid of several hundred values. The calculations are for
GRB090510A (left panel) and GRB080916C (right panel). The positions of the maxima are
marked by the vertical (red) solid lines, with the values $\tau  =- 0.098$~s/GeV (left panel)
and $\tau  = 0.80$~s/GeV (right panel).
The (green) dashed lines mark the positions of the maxima in the absence of any propagation effect.}
\label{fig:KURT}
\end{figure}

Examples of the ${\cal K(\tau )}$ curves calculated at different points $\tau ^j$ in a grid of values of the
compensation parameter are shown in Fig.~\ref{fig:KURT}. The values of $\tau $ that reshape the intensity
distributions to the most leptokurtic type are considered as those that best recover the original signal at the source.
The estimates are made for the same sources as in case of the irregularity estimator examples, namely for
GRB090510A and GRB080916C. The optimal values of $\tau $ are quite similar in both cases.

\subsection{Skewness Estimator}
\label{sec:skewness}

Skewness is a measure of the degree to which a distribution is asymmetrical. In Section~\ref{secPulse}
we used a symmetric distribution, namely a Gaussian, to model a burst-like feature at a GRB source, see the
solid line in Fig.~\ref{fig:packet}. The dashed and dashed dotted lines in Fig.~\ref{fig:packet} are
asymmetric distributions showing how this Gaussian envelope evolves when propagating trough a dispersive
mediim with a power-law energy spectrum.
The asymmetry may be measured using {\it skewness}
(see for example~\cite{statBOOK1}), which takes the following form
for the intensity distribution defined in~(\ref{intens}):
\beq
\label{skew1}
{\cal S(\tau )}=\sqrt{N_W}
\frac{\sum\limits_{i=0}^{N-1}((b_{\rm df}(E_i,\tau )-\overline{b_{\rm df}(\tau )})W_i)^3}
{\left(\sum\limits_{i=0}^{N-1}((b_{\rm df}(E_i,\tau )-\overline{b_{\rm df}(\tau )})W_i)^2\right)^{3/2}} \, ,
\eeq
where a weight $W_i$ given by (\ref{weight}) is assigned to every data point,
$\overline{b_{\rm df}(\tau )}$ stands for the average of the detector-frame intrinsic timing
of a given signal, and $N_W$ is a normalization parameter~\footnote{As previously,
since we only compare values of ${\cal S}$
for different values of $\tau $, the absolute normalization is not important.}.
Mathematically, the skewness is the ratio of the third moment of a distribution to its second moment raised
to the power $3/2$. The dispersed distribution shown in~Fig.~\ref{fig:packet} is described as negatively
skewed (or skewed to the left).
In general, dispersion of the form (\ref{mdr1}) causes the skewness of a signal with a burst-like feature
to become more negative.
{Therefore, whatever the shape of the time profile at the source and its degree of symmetry,
one expects that the skewness of the intensity distribution always changes towards more negative values in
course of propagation of the signal due to dispersion.
Conversely, the compensation procedure of Section~\ref{sec:recov} tends
to increase the skewness of the intensity distribution towards positive values, and we consider
the optimal value of the compensation parameter in~(\ref{rec1}) to be that maximizing the skewness.}

\begin{figure}
\centering
\includegraphics[width=0.5\textwidth]{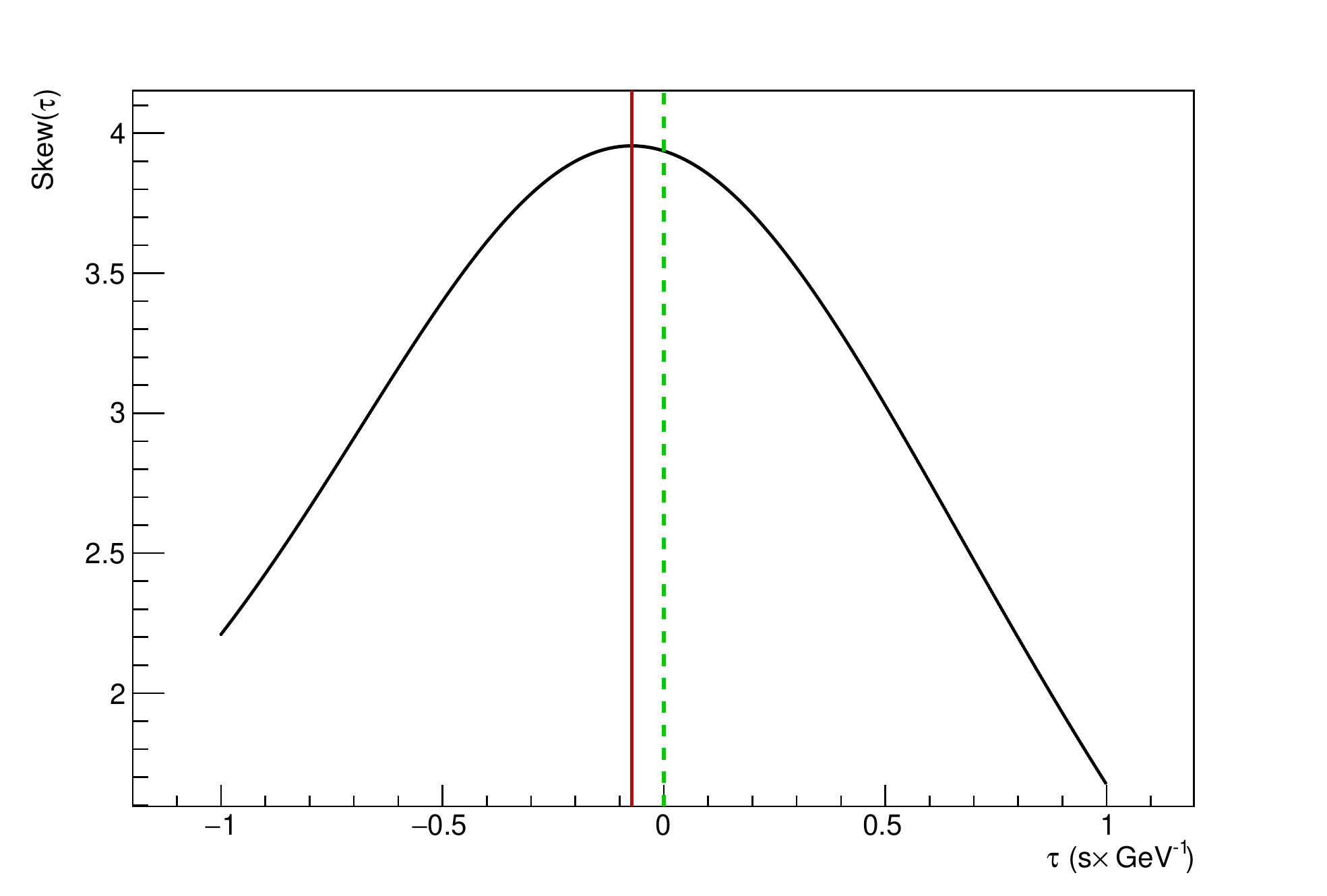}\hspace{0cm}\includegraphics[width=0.5\textwidth]{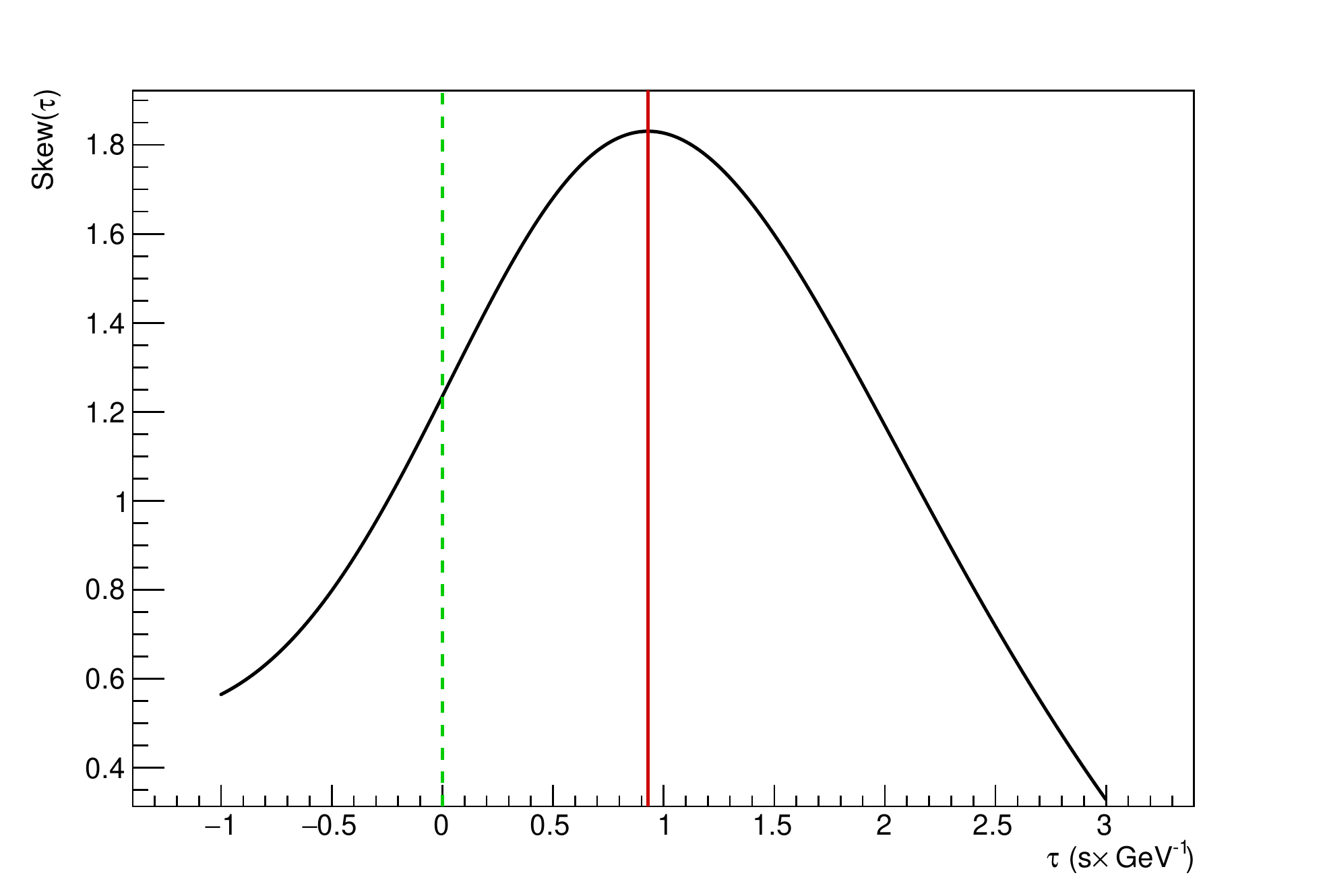}
\vspace{-0.4cm}
\caption{\it The skewness ${\cal S(\tau )}$ calculated for a set of discrete values of the compensation parameter $\tau ^j$
values of the compensation parameter $\tau $ with $j$ running over a grid of several hundred points, for
GRB090510A (left panel) and GRB080916C (right panel). The position of the maximum is
marked by the vertical (red) solid line: $\tau  =- 0.073$~s/GeV (left panel)
and $\tau  = 0.93$~s/GeV (right panel).
The (green) dashed lines mark the positions of the maxima in the absence of any propagation effect.}
\label{fig:SKEV}
\end{figure}

Examples of the values of ${\cal S(\tau )}$ calculated for different points $\tau ^j$ in a grid of values for the
compensation parameter are shown in Fig.~\ref{fig:SKEV}. The values of $\tau $ maximizing the skewness of the intensity
distributions are considered to be those that best recover the original signal at the source.
The estimates are made for the same GRBs as in cases of the irregularity and kutosis estimators discussed previously, namely
GRB090510A and GRB080916C, and we find values of $\tau $ that are similar to those found previously
{The skewness estimator we utilize here is fully non-parametric, and does not rely on any assumption
about the shape of the time profile at the source.}

\section{Uncertainties in the Estimators}
\label{sec:stab}

In this Section we quantify the stability of the estimators described above with respect to the performance of \lat~\cite{Fermi-LAT} and
account for the bias-induced systematic uncertainty in the overall measurement of the compensation parameter.

We comment first that the shortest timing shift in our studies
is expected to be at the level of the smallest estimated $|\tau |\gtrsim 0.1$~s/GeV multiplied by the lower energy cut,
100~MeV, which amounts to about $1$~ms. Since the time resolution of the instrument is better than $10\mu$s,
we may assume that our results are insensitive to the timing accuracy.
However, the  evolution of the timing patterns during the propagation of the signals depends upon the energies of individual photons.
Thus, the energy resolution of the instrument can influence the stability of the estimation of the correct value of the
compensation parameter.

\begin{figure}
\centering
\includegraphics[width=0.5\textwidth]{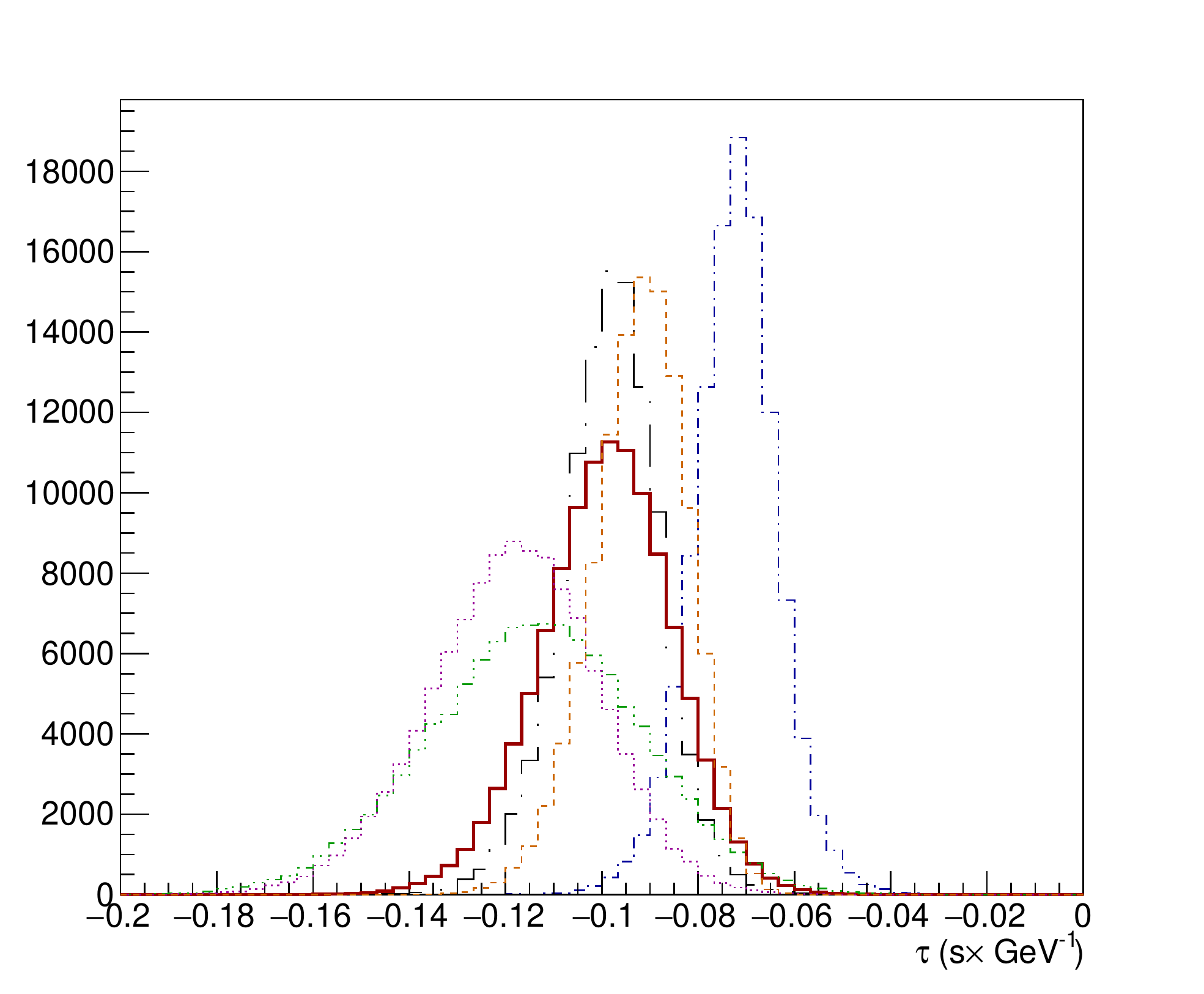}\hspace{0cm}\includegraphics[width=0.5\textwidth]{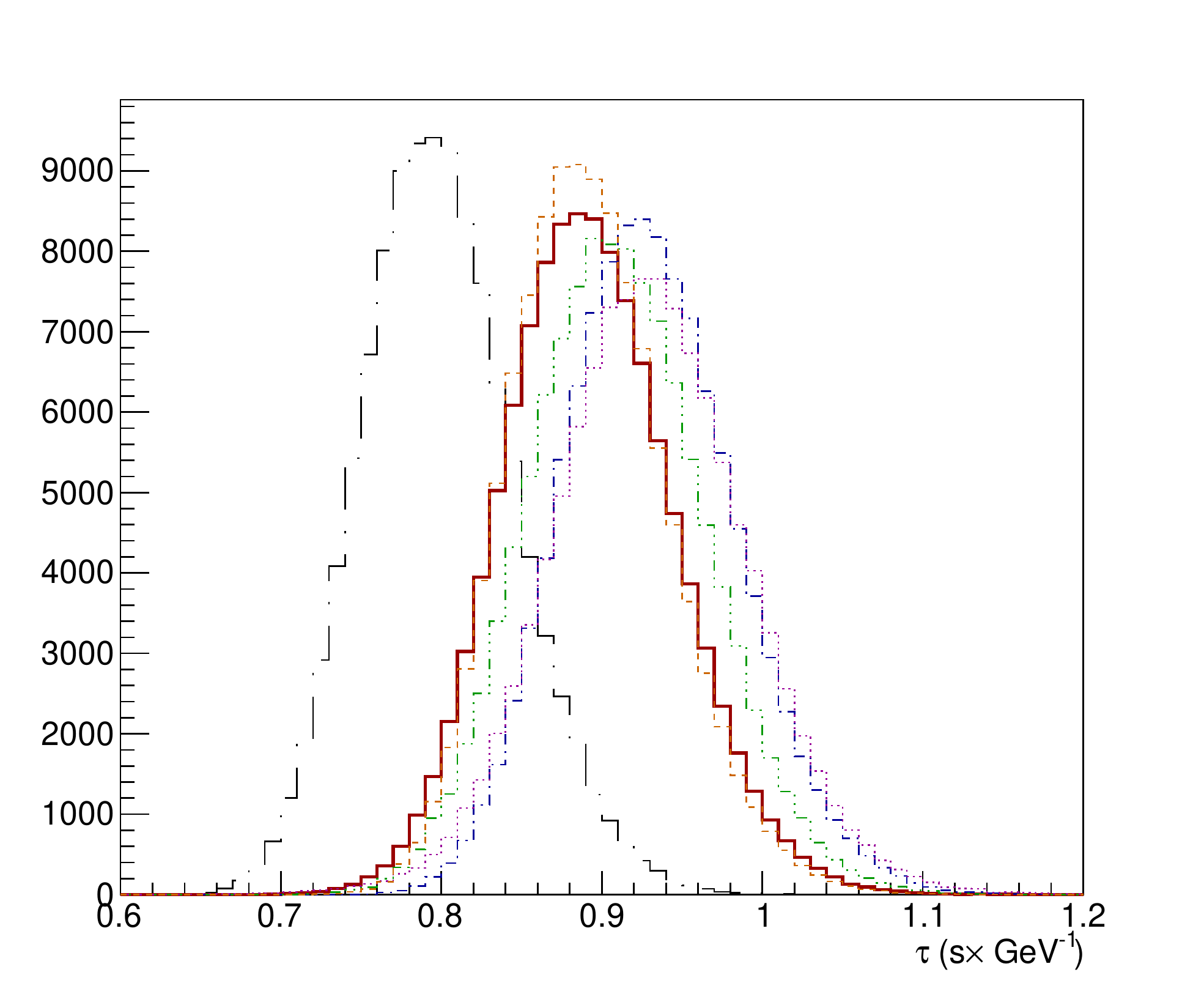}
\includegraphics[width=0.5\textwidth]{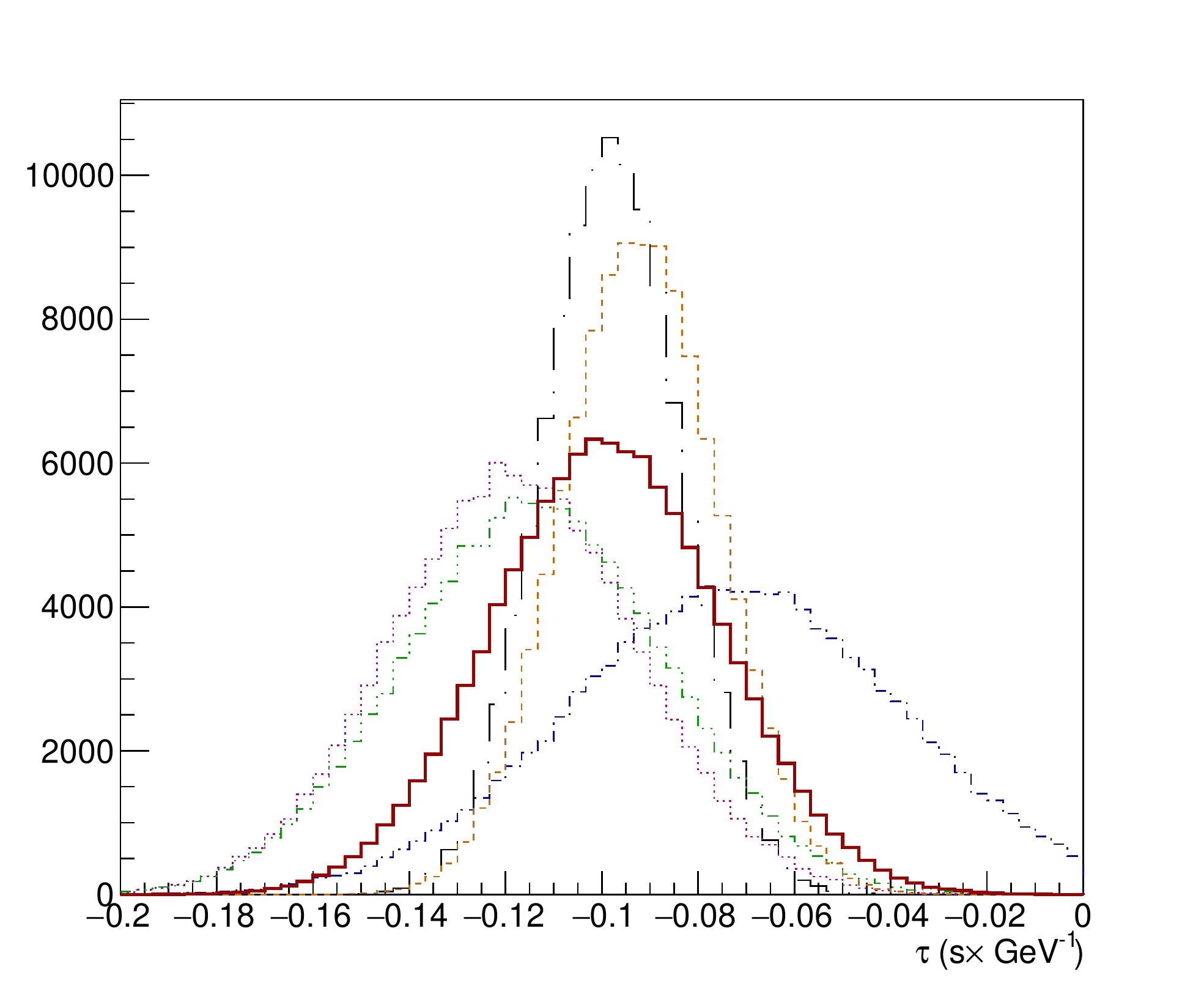}\hspace{0cm}\includegraphics[width=0.5\textwidth]{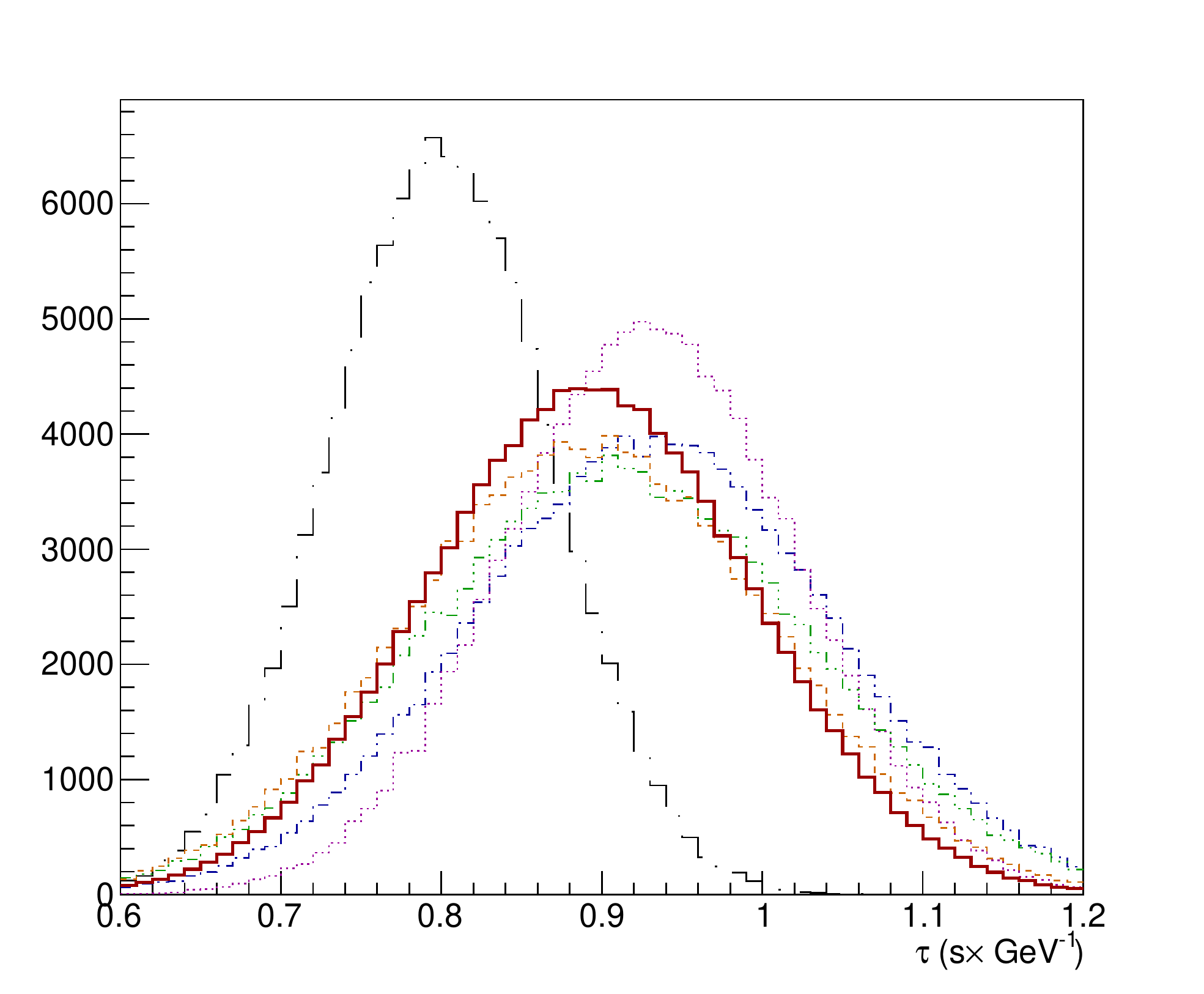}
\vspace{-0.4cm}
\caption{\it The upper panels show distributions of the correct values of the compensation parameter estimated for 112 thousand toy models
of GRB090510A (left panel) and GRB080916C (right panel).
The (magenta) dotted lines show the results of applying of the irregularity estimator using simply the middle values of line segments at 95\%
of the heights of the KS difference curves. The positions of the maxima of the irregularity estimators found
by averaging of the peaks of the KS difference curves are shown by (green) short-dashed three-dotted lines.
The results of using the kernel density estimator (KDE) to estimate the positions of the KS difference curves of the irregularity estimator are shown
by (orange) dashed lines. The (black) long-dashed dotted lines show the distribution of the results of the kurtosis
estimator applied to the toy models. The results of the skewness estimators are shown by
(blue) short-dashed dotted lines. Finally, the distributions called the overall distributions in the text are depicted by the (red) solid lines.
The plots in the lower panels shows the results corrected for the systematic bias explained in the text.}
\label{fig:TOYS}
\end{figure}

In the following we apply the estimators discussed in the previous Section to toy data sets generated by smearing the
energies of the individual photons using a model resolution function, so as to assess
the instability of the estimated compensation parameters.
For this purpose we use one of the {\tt P8R2\_V6} energy resolution performance plots from~\cite{FERMI_Eres}, and parameterize empirically the
energy resolution for 68\% half-width containment of the reconstructed incoming photon energy as:
\beq
\label{Eres}
\frac{\Delta E}{E} = 0.7234 - 0.4393x + 0.1133x^2 - 0.01459x^3 + 0.0008579x^4 \, ,
\eeq
where $x \equiv \log_{10}(E/{\rm MeV})$. Inaccuracy of the energy measurements superimposes
an instability into the estimations of the correct value of the compensation parameter.
We assign to every photon within an emission episode an energy generated randomly using a
normal distribution defined by the mean value of its observed energy $E$ and the
standard deviation derived from (\ref{Eres}). To maximize confidence in the accuracy
of estimates, we have analyzed $\sim 100$~thousand toys for every individual source.
Examples of distributions of the corresponding values of the compensation parameters obtained for the toy data sets
using the different estimators are shown in Fig.~\ref{fig:TOYS}.
For every source we apply five different estimation procedures based on the estimators described in the previous Sections.

Three distributions in Fig.~\ref{fig:TOYS} are obtained from the irregularity estimator described in Section~\ref{sec:ks},
using different methods to analyze the KS difference curve. The issue is that KS difference curves like those in Fig.~\ref{fig:KSD_090510}
are quite irregular, in particular near the peak. This is due to the fact that, in order to avoid an unwanted systematic, we utilize
different uniformly-distributed reference samples for every choice of $\tau $. These irregularities
can introduce an ambiguity in the estimation of the position of the maximum of the $D(\tau )$ function derived as in (\ref{KSD1}).
We utilize three methods to estimate the positions of the maxima of $D(\tau )$ curve.
The first is simply to define the position of a maximum
as the centre of a segment formed by a horizontal line cutting the $D(\tau )$ curve at a
certain fraction of the total height of the $D(\tau )$ curve~\footnote{Since the degrees of irregularities of $D(\tau )$ depend on the
statistics and spectral content of the GRB, we vary the cut fraction between 4\% and 18\% for different
sources.}. The resulting distributions for two GRBs are shown by (magenta) dotted lines in Fig.~\ref{fig:TOYS}.
Another method is to define the maximum by the weighted average of the
top part of the $D(\tau )$ curve after being cut by the same horizontal line, which is shown by the (green) short-dashed
three-dotted lines in Fig.~\ref{fig:TOYS}. Finally, we also used the kernel density estimation (KDE) technique~\cite{tkde}, which
provides an estimate of $D(\tau )$ within its whole support. The
maxima of the KDE curves are then used to estimate the correct values of the compensation parameter, with
the results shown by (orange) dashed lines in Fig.~\ref{fig:TOYS}.

Unlike the KS difference curves, the kurtosis curves (see Fig.~\ref{fig:KURT}) and the skewness curves (see Fig.~\ref{fig:SKEV}) are quite regular over
the whole support, so that the positions of the maxima for ${\cal K(\tau )}$ and ${\cal S(\tau )}$
are unambiguous. The resulting distributions for kurtosis and skewness are presented in Fig.~\ref{fig:TOYS} by (black) long-dashed and (blue)
short-dashed-dotted lines, respectively.

Another measure of the robustness of an estimator is its bias. In our case, this is
expressed by the deviation of the estimate of the correct value $\tau ^{\rm recov}$
of the compensation parameter from its true value $\tau ^{\rm true}$:
\beq
\label{bias}
B(\tau ) = \overline\tau ^{\rm recov} -\tau ^{\rm true},
\eeq
where the average $\overline\tau ^{\rm recov}$ stands for the expected recovered value over a large number
of repeated experiments.

We study the bias by analyzing a variety of realizations of the emission from a given source generated by
sampling the timing and energy distributions with respect to their distributions in the data.
The method of generating realizations used here is akin to the ``flux randomization" procedure initially prescribed in~\cite{biasMC} and
later applied in~\cite{GBMlagLAT} for simulations of \lat\ GRB light curves. Following the general idea of~\cite{biasMC},
we simulate realizations of a given signal in such a way that the average temporal and energetic characteristics of each realization
are equal, at some level of accuracy, to a specific timing-energy distribution obtained from the data.
The timing-energy distributions for different realizations are obtained as random numbers along two axes, distributed
according to the cell-contents of two-dimensional progenitor histograms with $N^2$ cells.
(We recall that $N$ is the number of events arriving at \lat~from a given source.)
The total energy of the photons composing a generated realization is set to be equal
to the total energy of the original signal with a certain accuracy.
The progenitor histograms are constructed from the detected
patterns with timings compensated for the propagation effect~Fig.~\ref{fig:BIAS_REALIZATIONS}.
The amount of the compensation is defined by the ``correct" value of
the compensation parameter found in any given estimation procedure. Examples of such realizations
generated from the data from GRB090510A and GRB080916C compensated for the results of the kurtosis estimation procedure are
presented in~Fig.~\ref{fig:BIAS_REALIZATIONS}.

The realizations produced by the compensated data are regarded as being unaffected by the dispersion effect.
In general, since the compensation values are different for the five estimation procedures used, one should study five different sets
of realizations of the detected emission episode for every source. However, in practice, in order to reduce CPU time,
for each GRB we use only one progenitor histogram compensated with respect to the result of one particular estimation technique, and
offsets of the compensations for other estimators are taken into account in the calculations of the final
uncertainties attributed to the bias corrections.
Once the progenitor histogram is obtained we produce a number of realizations with a common degree of dispersion
corresponding to a particular injected value of $\tau ^{\rm true}$. The total energy of
the generated photons is required to be the same, to within 15\%, as that measured in the data.

We then apply our estimation procedures to the set of realizations with a given injected dispersion signal, and calculate the average
over the set of toy realizations of the estimated correct
value of the compensation parameter. This average represents the expected recovered value of the compensation
parameter $\overline\tau ^{\rm recov}$ in (\ref{bias}).
Several reference values of $\tau ^{\rm true}$, each injected into separately-generated sets of realizations,
are tested for every estimation procedure.
Parameters of straight line fits to $\overline\tau ^{\rm recov}$ versus $\tau ^{\rm true}$,
like that ones shown in Fig.~\ref{fig:BIAS_KURT} and Fig.~\ref{fig:BIAS_KSav}, are used for the determination
of the uncertainties related to bias in the estimates of the correct values of the compensation parameters obtained using
our estimation procedures. The required value of  $B(\tau )$ given by (\ref{bias}) is given
by the difference between a given value of $\tau $  and
the value of the linear fitted function calculated at the same $\tau $.
Finally, the bias calculated in this way for every estimation procedure is included in the uncertainties of the
estimates of the compensation parameter that we present. The impacts of the bias corrections are illustrated in
the lower row of Fig.~\ref{fig:TOYS}, to be compared with the upper row, where no bias correction has been applied.

\begin{figure}
\centering
\includegraphics[width=0.5\textwidth]{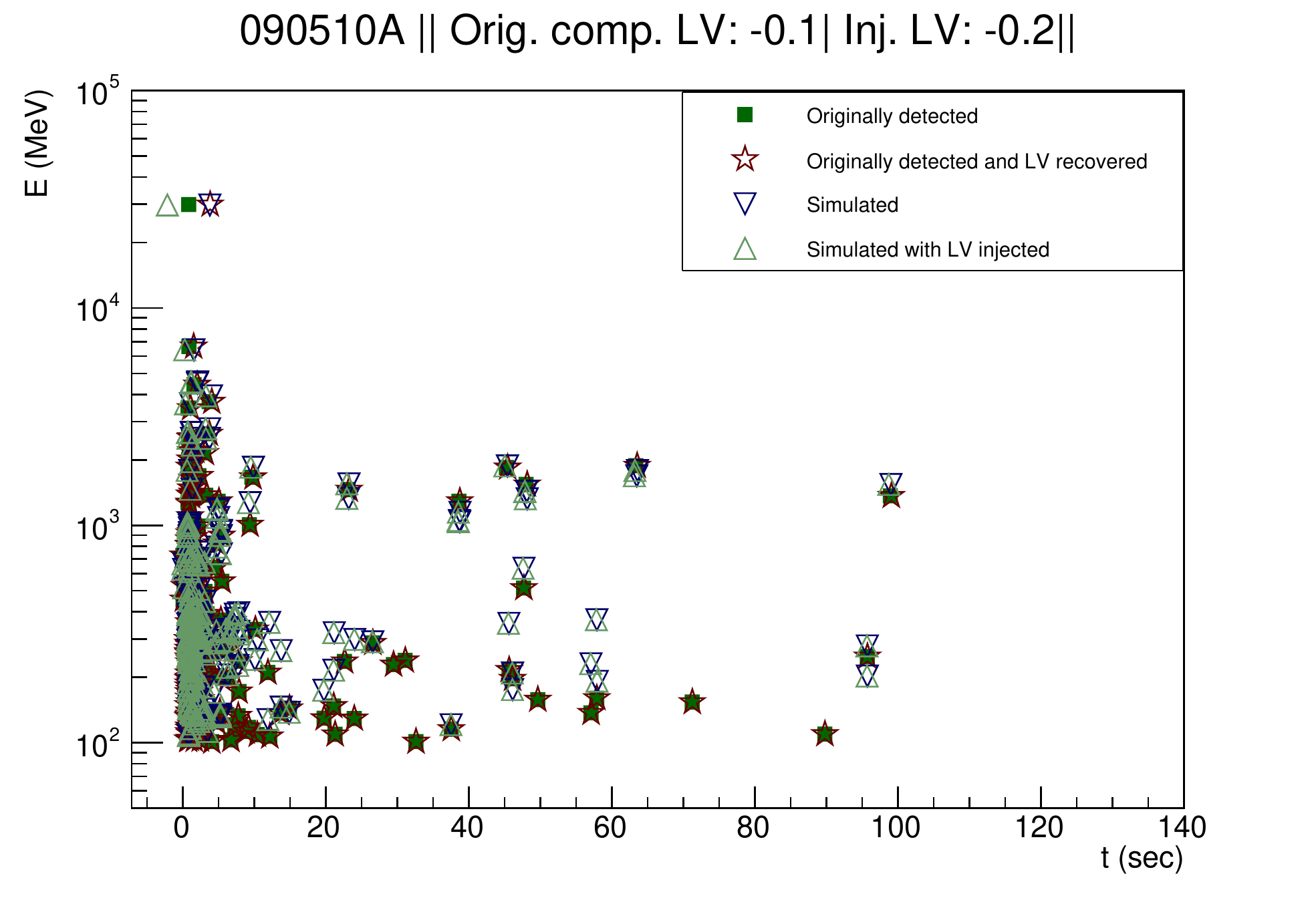}\hspace{0cm}\includegraphics[width=0.5\textwidth]{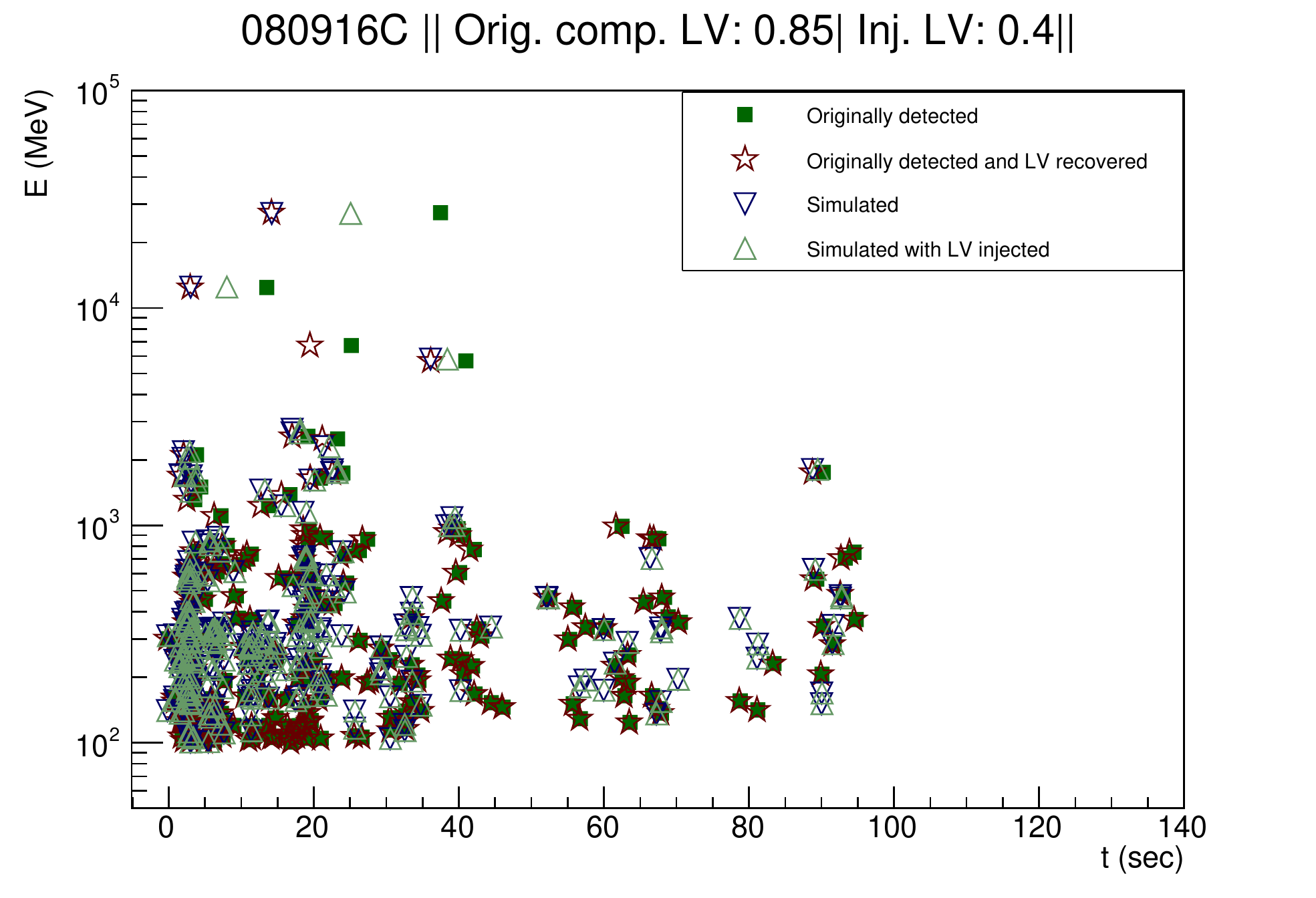}
\vspace{-0.4cm}
\caption{\it Examples of one particular realization with injected quantum-gravity effects for the
objects GRB090510A (left panel) and GRB080916C (right panel). The solid squares
represent the data recorded by {\it Fermi}-LAT, the stars represent the progenitor histogram obtained
with the quantum-gravity compensation estimated using the kurtosis estimator, and the downward-pointing triangles represent
a random simulation of the two-dimensional distribution constrained by the cell pattern of the
progenitor histogram. The upward-pointing triangles are obtained by injecting $\tau ^{\rm true}=-0.2$ for
GRB090510A and $\tau ^{\rm true}=0.4$ for GRB080916C into the simulated realization.}
\label{fig:BIAS_REALIZATIONS}
\end{figure}

\begin{figure}
\centering
\includegraphics[width=0.5\textwidth]{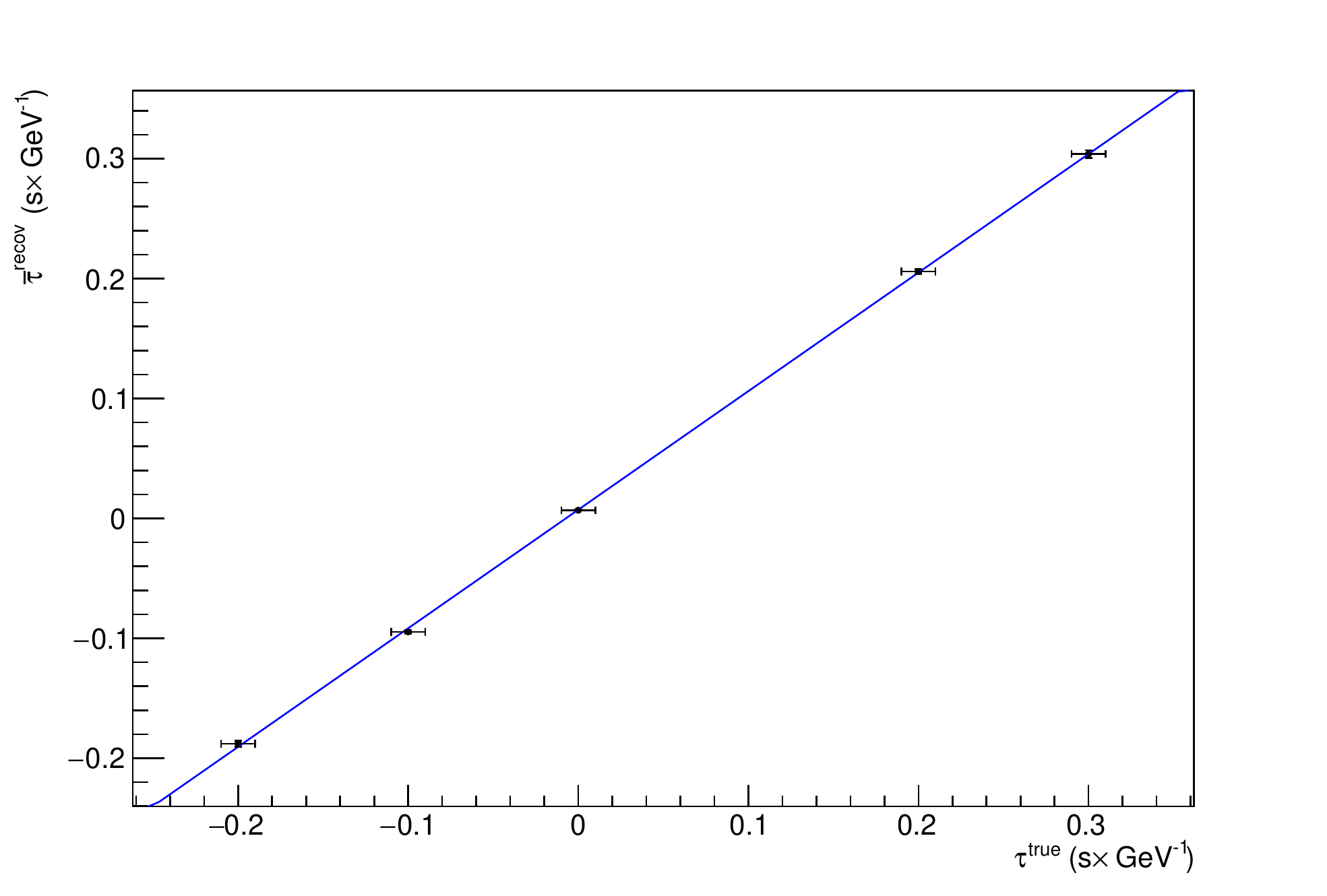}\hspace{0cm}\includegraphics[width=0.5\textwidth]{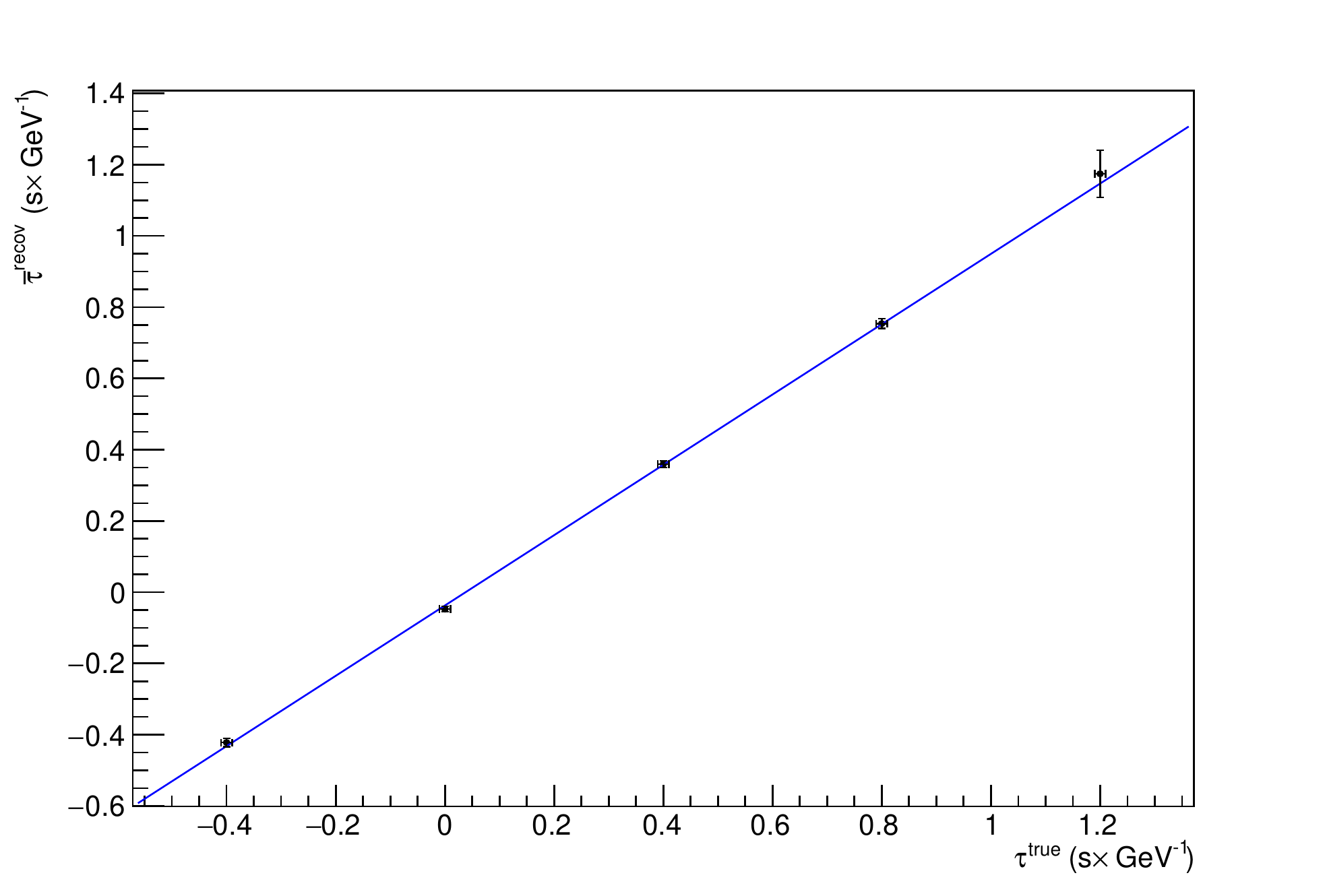}
\vspace{-0.4cm}
\caption{\it The results of studies of bias in the kustosis estimator for the objects GRB090510A (left panel) and  GRB080916C (right panel).
The amounts of injected and reconstructed Lorentz-violating signals are shown on the
horizontal and vertical axes, respectively. The injections have been made into 15 realizations of sources seeded by data with timings
compensated for the value of the dispersion effect estimated by the kurtosis estimation procedure.
The horizontal errors are 1$\sigma$ uncertainties in the kurtosis estimation procedure indicated by the data.}
\label{fig:BIAS_KURT}
\end{figure}

\begin{figure}
\centering
\includegraphics[width=0.5\textwidth]{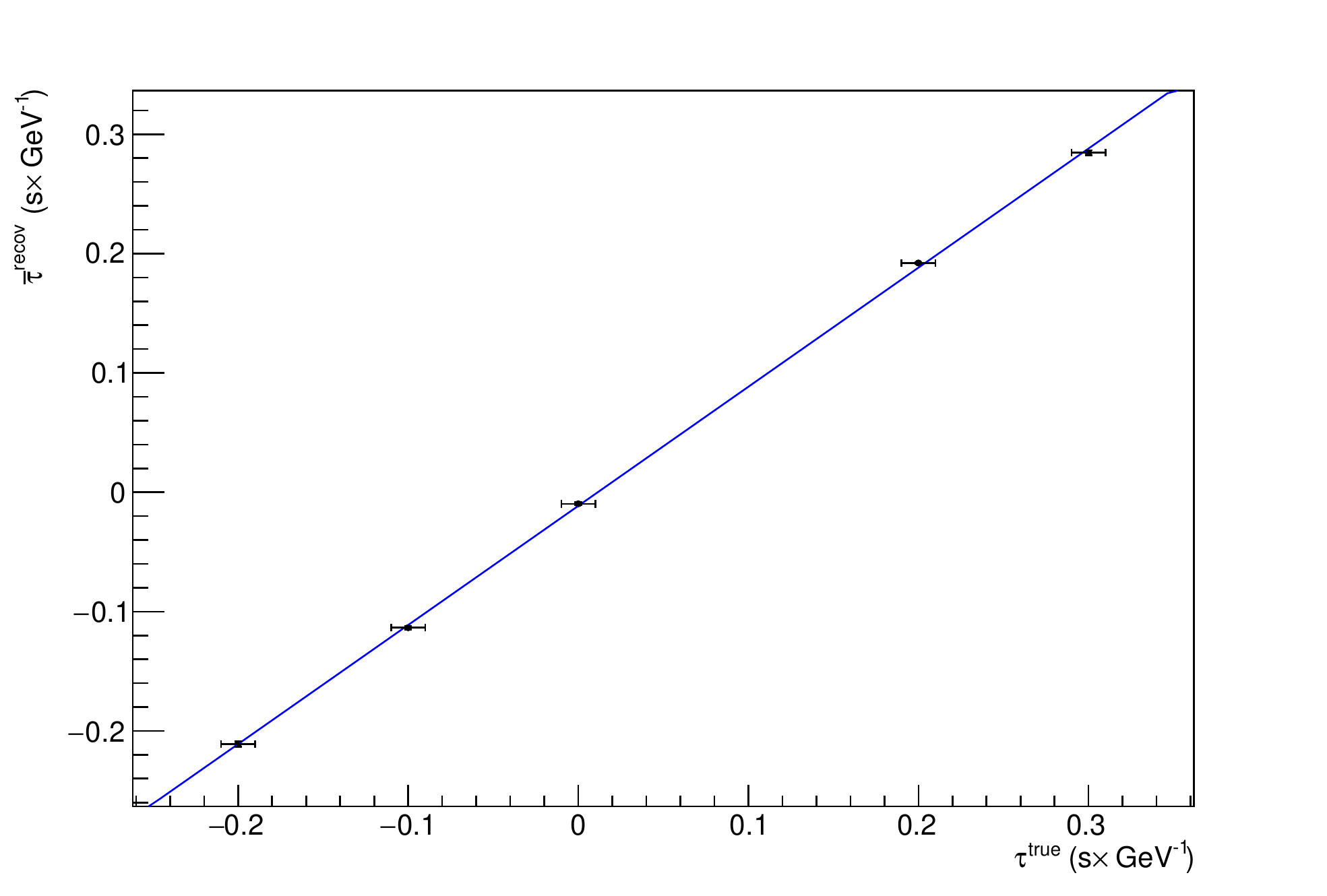}\hspace{0cm}\includegraphics[width=0.5\textwidth]{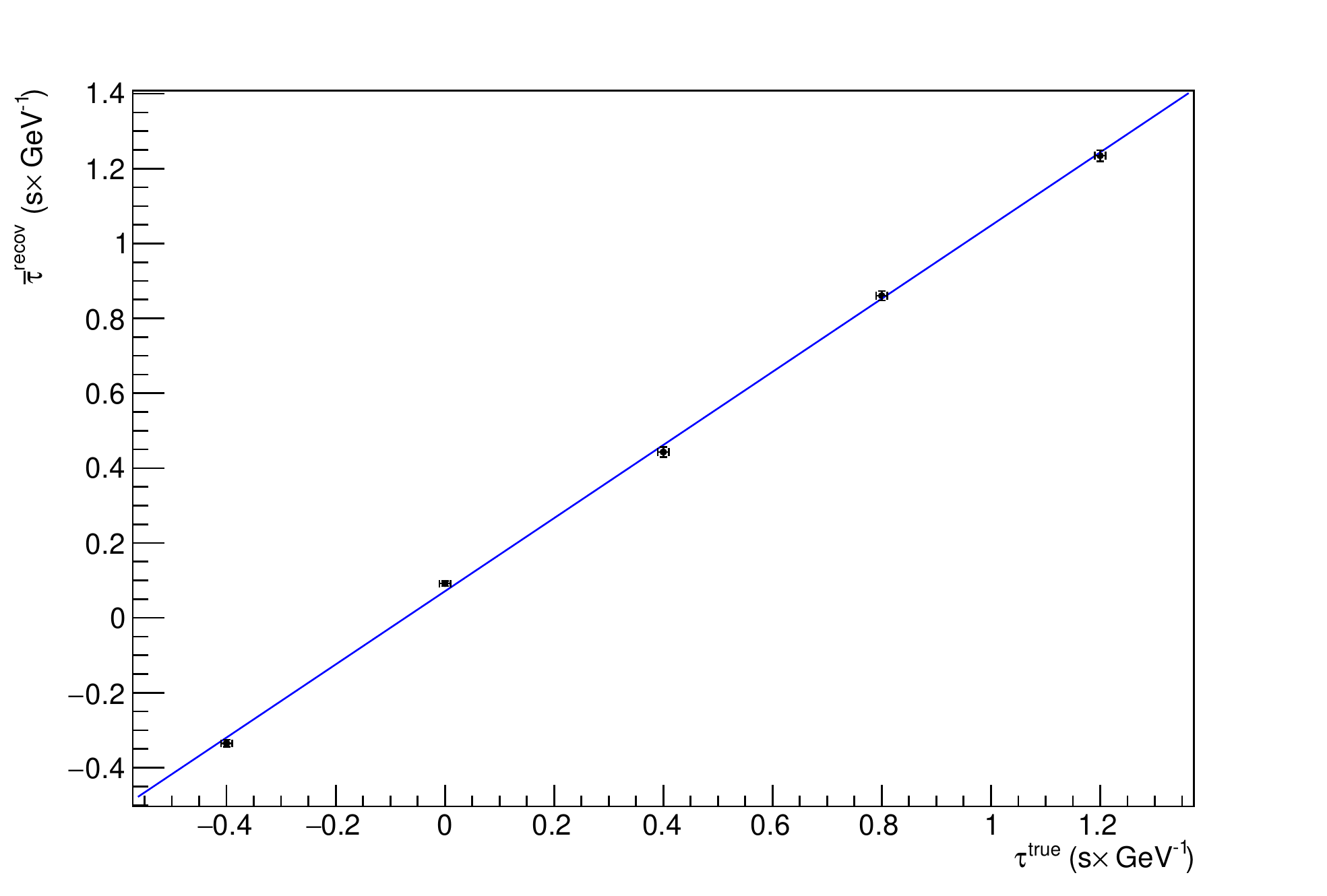}
\vspace{-0.4cm}
\caption{\it Same as Fig.~\ref{fig:BIAS_KURT}, but for the irregularity estimator with its maximum values calculated
by averaging of the tops of the KS difference curves.}
\label{fig:BIAS_KSav}
\end{figure}

Almost identical work flows were used to estimate bias uncertainties for all estimators in the data from all the sources analyzed.
We use for illustration results for the progenitor histogram for GRB080916C, shown in Fig.~\ref{fig:BIAS_REALIZATIONS} by stars,
compensated for the result obtained using the kutosis estimation procedure, namely $\tau  = 0.85$~s/GeV
(the pattern originally detected  is shown by the solid squares in Fig.~\ref{fig:BIAS_REALIZATIONS}). For the bias study, we use five reference values
of $\tau ^{\rm true}$, each injected into a separate set of 15
realizations of the detected emission generated from the progenitor histogram.
Thus, a total of 75 realizations has been generated for the bias study of GRB080916C.
One of the realizations (modified with $\tau ^{\rm true}=0.4$~s/GeV) is shown in Fig.~\ref{fig:BIAS_REALIZATIONS} by downward-
(upward-)pointing triangles. For every realization we apply the kurtosis estimation procedure
to a set of 16 thousand toys generated with the energy smearing procedure described earlier, to obtain the optimal
values of the compensation parameters~\footnote{It is enough to quantify the bias with 1$\sigma$ precision,
and a set of 15 realizations each with 16 thousand toys provides sufficient precision using the available CPU time capacity.}.
The final distribution of the recovered value of the compensation parameter
$\overline\tau ^{\rm recov}$ for a particular amount of injected dispersion $\tau ^{\rm true}$ is obtained as the
average over all 15 realizations within a single set. The result of the processing for this specific object using the kurtosis estimator
is shown in the right panel of~Fig.~\ref{fig:BIAS_KURT}. The difference between the outcome of the kurtosis estimator for the data of GRB090816C,
$\tau_{\gamma 0} = 0.8$~s/GeV, and the value of the function obtained from the straight line fit
with errors related to the fit added in quadrature yields an uncertainty of
$\pm 0.069$~s/GeV, to be compared with  $\pm 0.048$~s/GeV when the data are used directly. The irregularity estimator is more
affected by bias, as seen in the right panel of Fig.~\ref{fig:BIAS_KSav}.
The estimators are least biased
in the case of GBR090510A (see the left panels of Fig.~\ref{fig:BIAS_KURT} and Fig.~\ref{fig:BIAS_KSav}).

As can be seen from the examples in  Fig.~\ref{fig:TOYS}, the precisions of the different estimation
techniques are quite similar to each other, although differences appear at the 1$\sigma$ level,
in particular when the bias is not taken into account (upper row of Fig.~\ref{fig:TOYS}).
We attribute these differences to an unidentified systematic that is probably related to the fact that
different estimators deal with different kinds of deformation of the signal envelope.
To be conservative, instead of giving a preference to any particular estimator, we simply average
the results of the five estimation procedures for each energy smearing toy.
In this way, we obtain the overall distributions shown as the solid lines in the upper panels of Fig.~\ref{fig:TOYS}.
The results obtained from the bias-corrected distributions shown in the lower panels of Fig.~\ref{fig:TOYS} as
solid lines are used for combination studies in the next Section.

\section{Consolidated Distribution and the Robust Limit
\label{sec:consol}}

Our final goal is to infer the common degree of quantum-gravity-induced dispersion that is most compatible
with the estimates obtained for all the sources we have analyzed.
We note that the relation (\ref{tauK1}) implies that correct value of the compensation
parameter $\tau (z)$ obtained for sources at different red shifts
can be adjusted to the value at a reference red shift $z_0$ via the scaling
\beq
\label{scaleK}
\tau (z_0) = \tau (z)\frac{K_1(z_0)}{K_1(z)} \, .
\eeq
We apply this adjustment to every toy model generated by energy smearing for each source. For simplicity,
we choose $z_0 = 0.8944$, which corresponds to $K_1(z_0)=1$. In this case the compensation parameter can be
converted trivially into the main parameter of interest, namely, the scale of violation of Lorentz invariance
\beq
\label{mQGtau}
M_1=\frac{H_0^{-1}}{\tau (z_0)} \, .
\eeq
We have transformed the overall distributions, point by point, into a combined distribution
of values (\ref{scaleK}), which we call the K-reduced distribution. The overall distributions of all the sources entered in the analysis
together with their K-reduced versions are presented in Fig.~\ref{fig:COMB1}.

\begin{figure}
\centering
\includegraphics[width=0.5\textwidth]{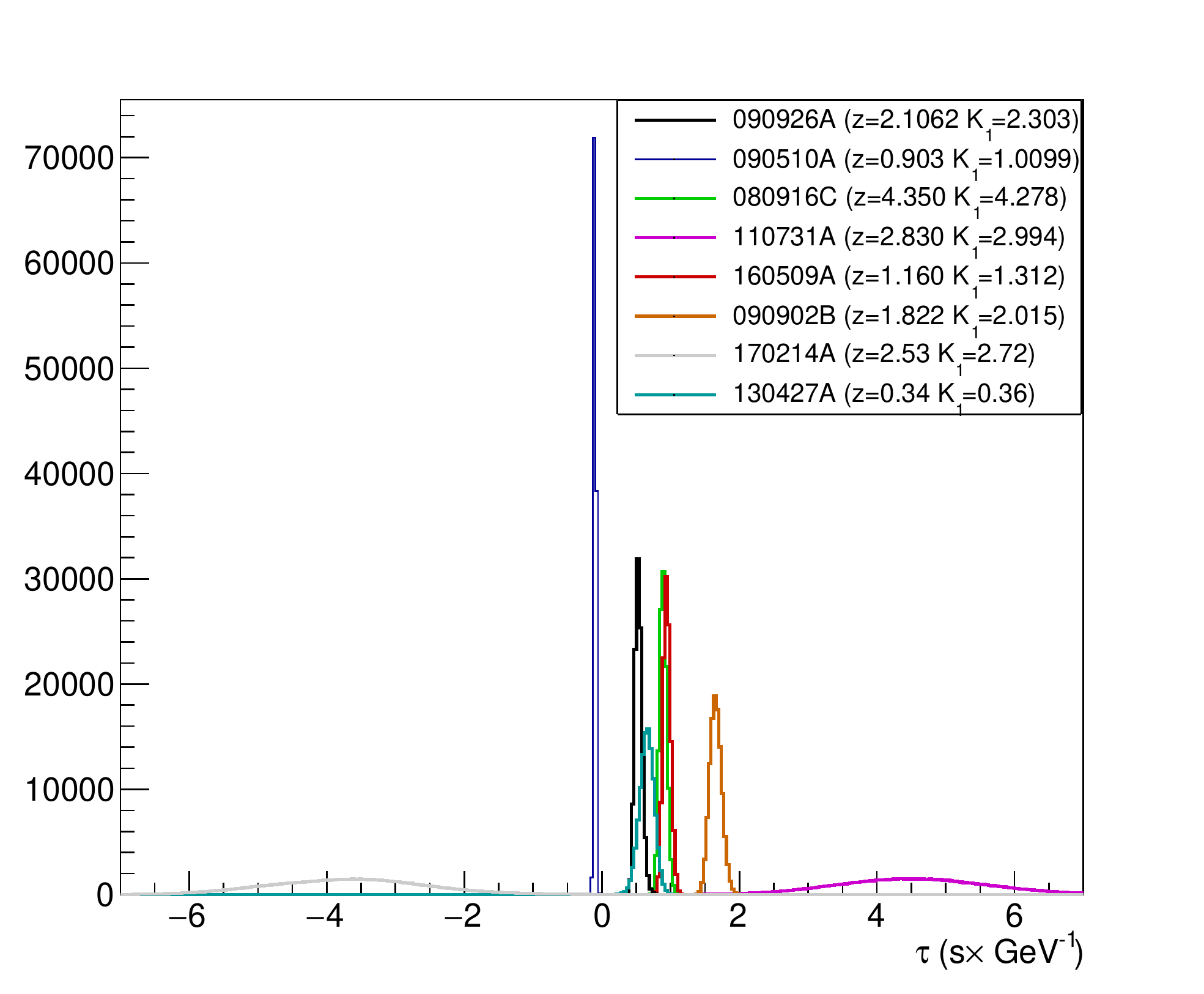}\hspace{0cm}\includegraphics[width=0.5\textwidth]{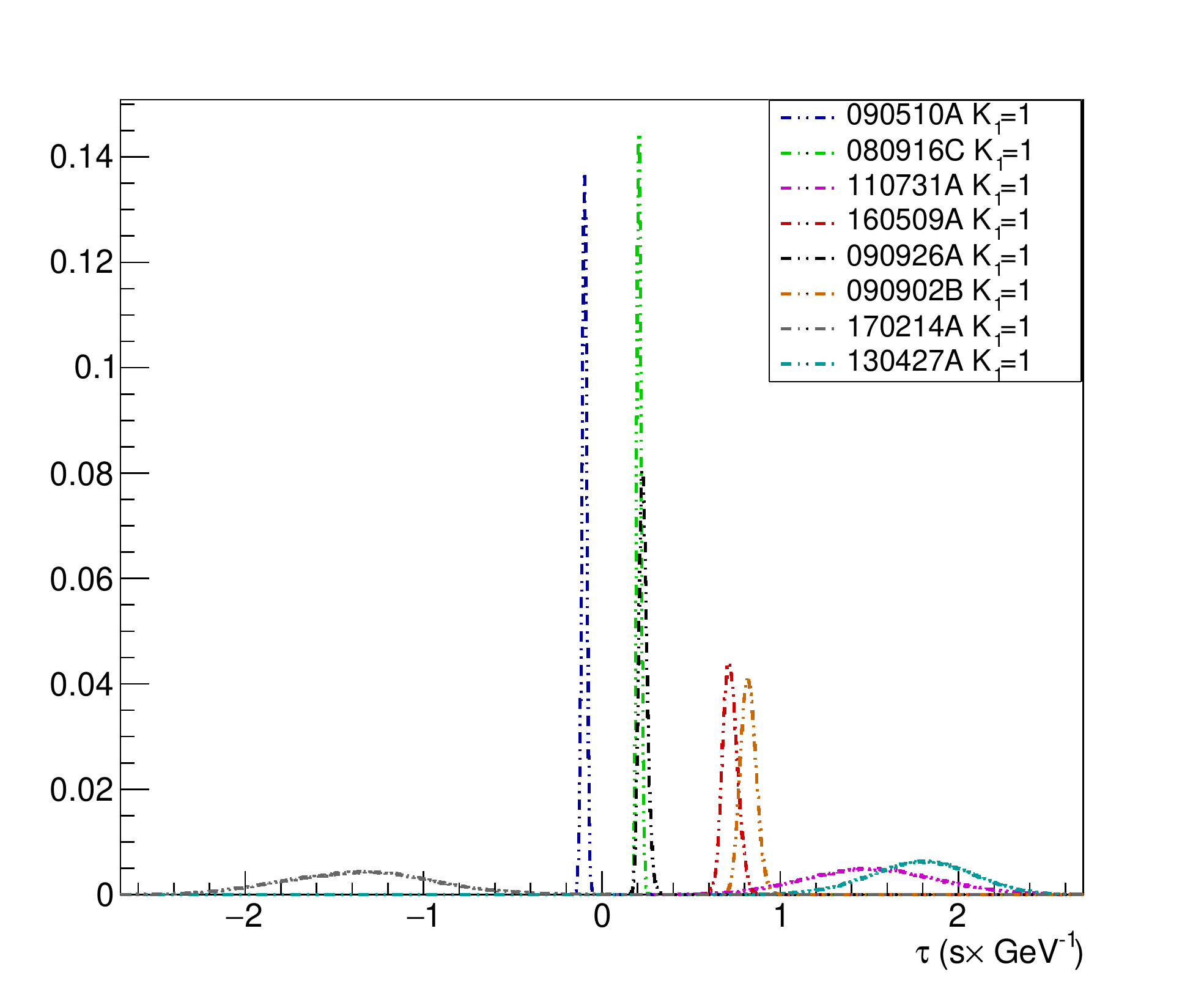}
\includegraphics[width=0.5\textwidth]{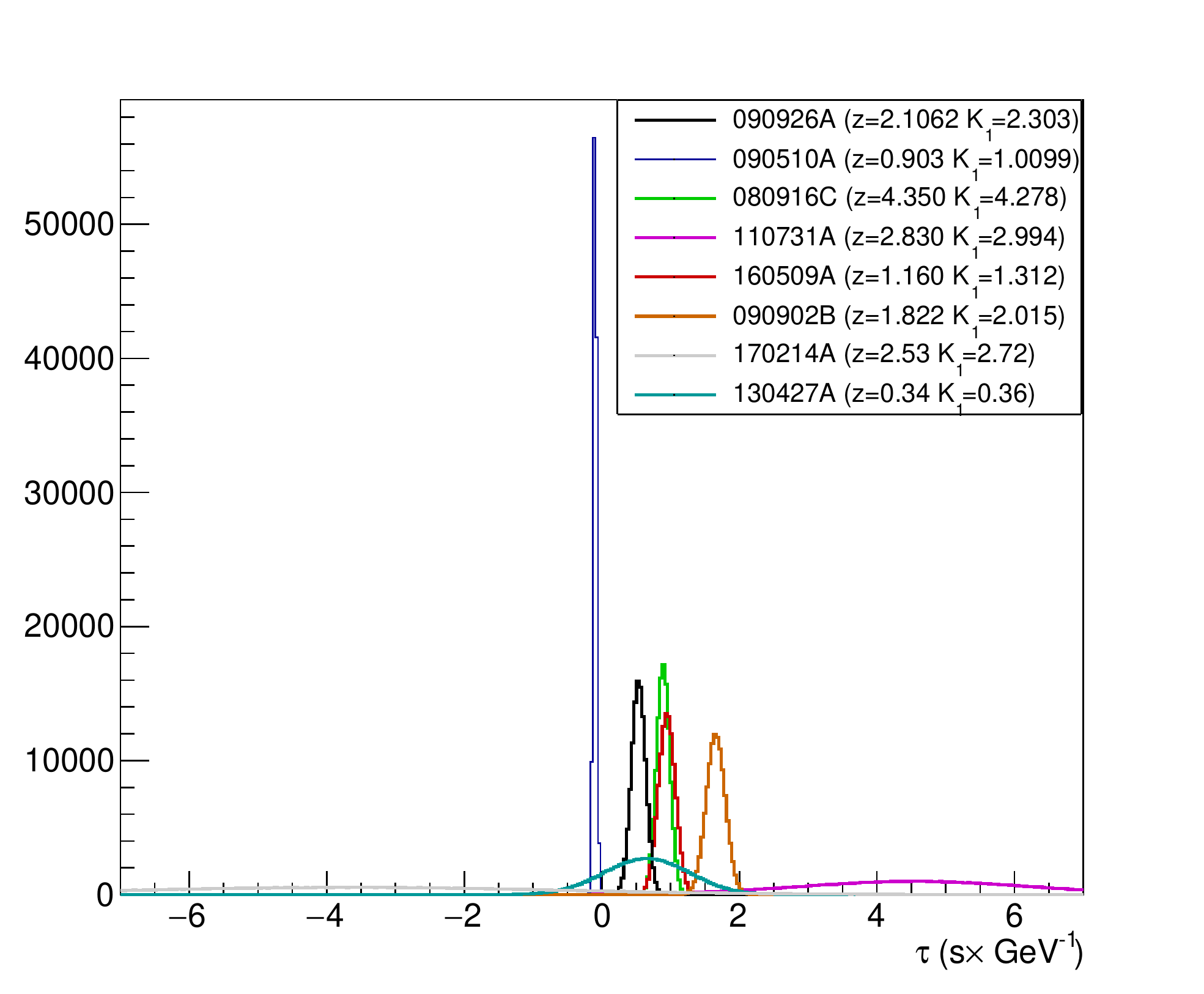}\hspace{0cm}\includegraphics[width=0.5\textwidth]{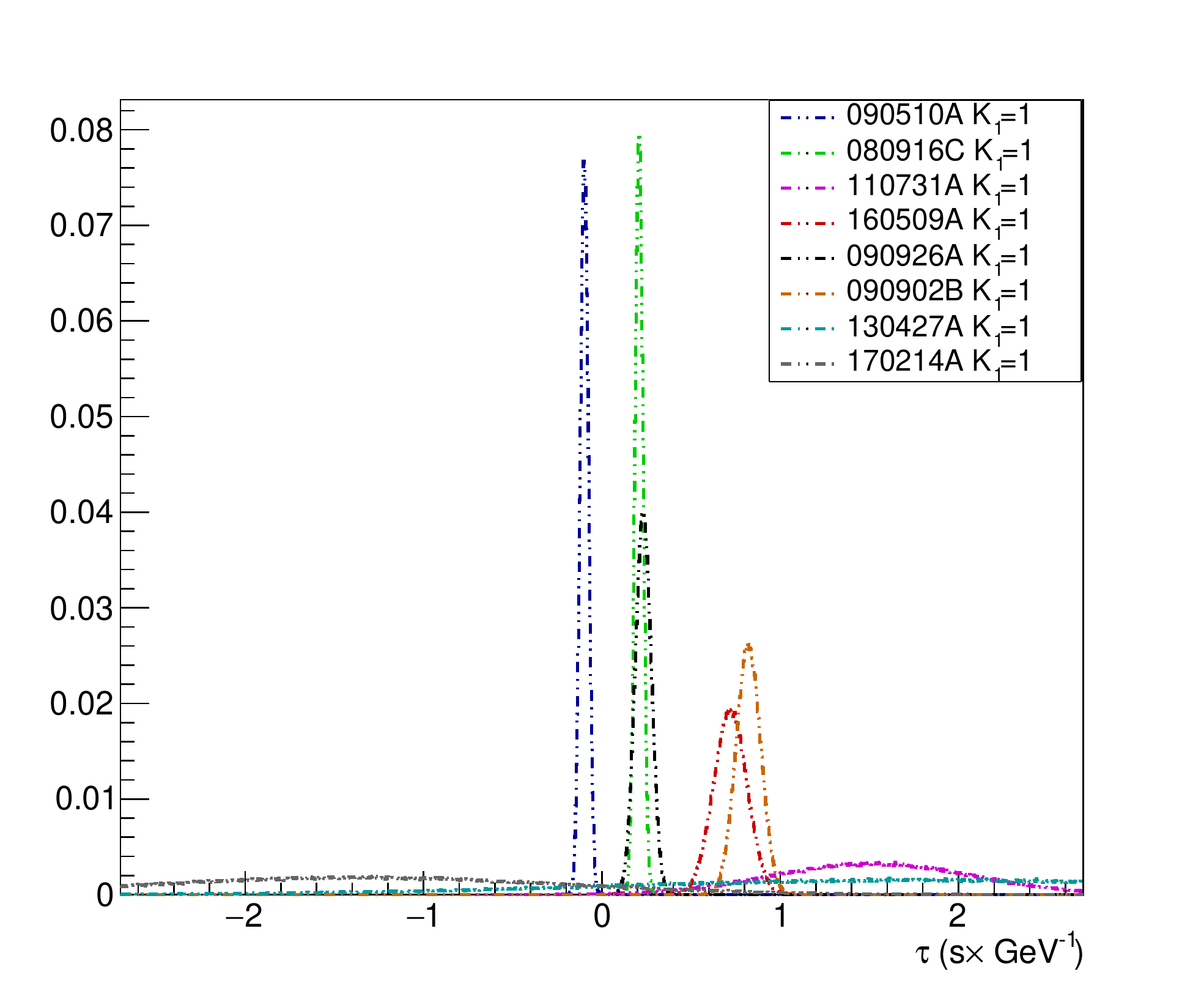}
\vspace{-0.4cm}
\caption{\it Upper row: the overall distributions (left panel) of the corrected values of the compensated parameters for the
eight sources studied (see Table~1), and their K-reduced normalized versions (right panel). Every individual distribution is obtained as an average of
probability distributions obtained using the five estimation procedures described in Section~\ref{sec:stab}. Lower row: the same
distributions as in the top row, corrected for the bias systematic as described in Section~\ref{sec:stab}.
}
\label{fig:COMB1}
\end{figure}

We now address the problem of consolidating our measurements of the compensation parameter obtained for different sources.
In general, we would need to minimize a likelihood function to give an estimate for the distribution of the $\tau (z_0)$ that
combines the information of the individual K-reduced overall probability distribution functions (PDFs). However, given that
our individual K-reduced distributions are very close to Gaussian as seen in the left panel of Fig.~\ref{fig:COMB2}~\footnote{The
Jarque--Berra test~\cite{statBOOK1} has been used as a tool
for study the Gaussianity of the individual overall distributions.},
we can use the minimum $\chi^2$ method to obtain the consolidated PDF.

\begin{figure}
\centering
\includegraphics[width=0.5\textwidth]{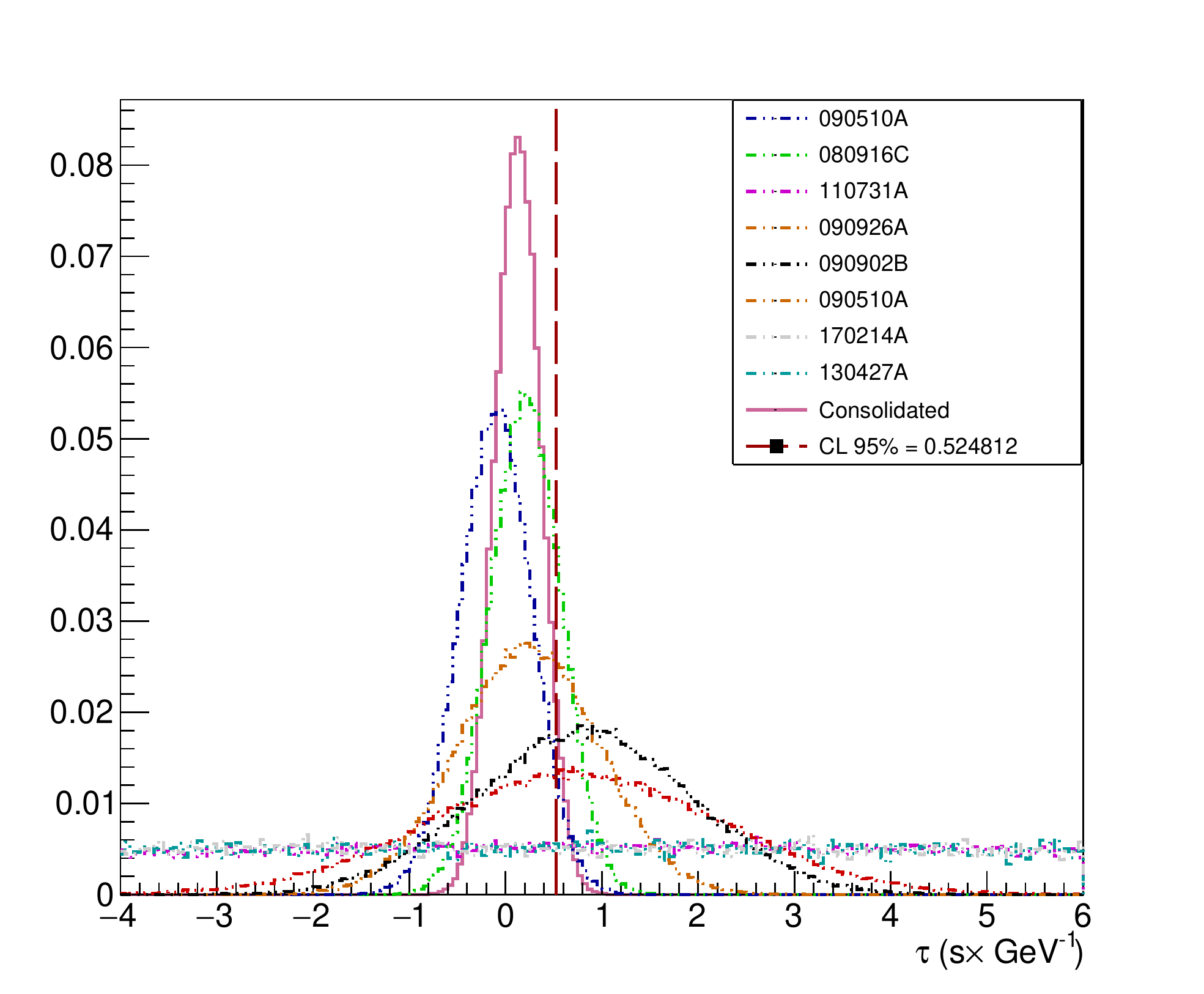}\hspace{0cm}\includegraphics[width=0.5\textwidth]{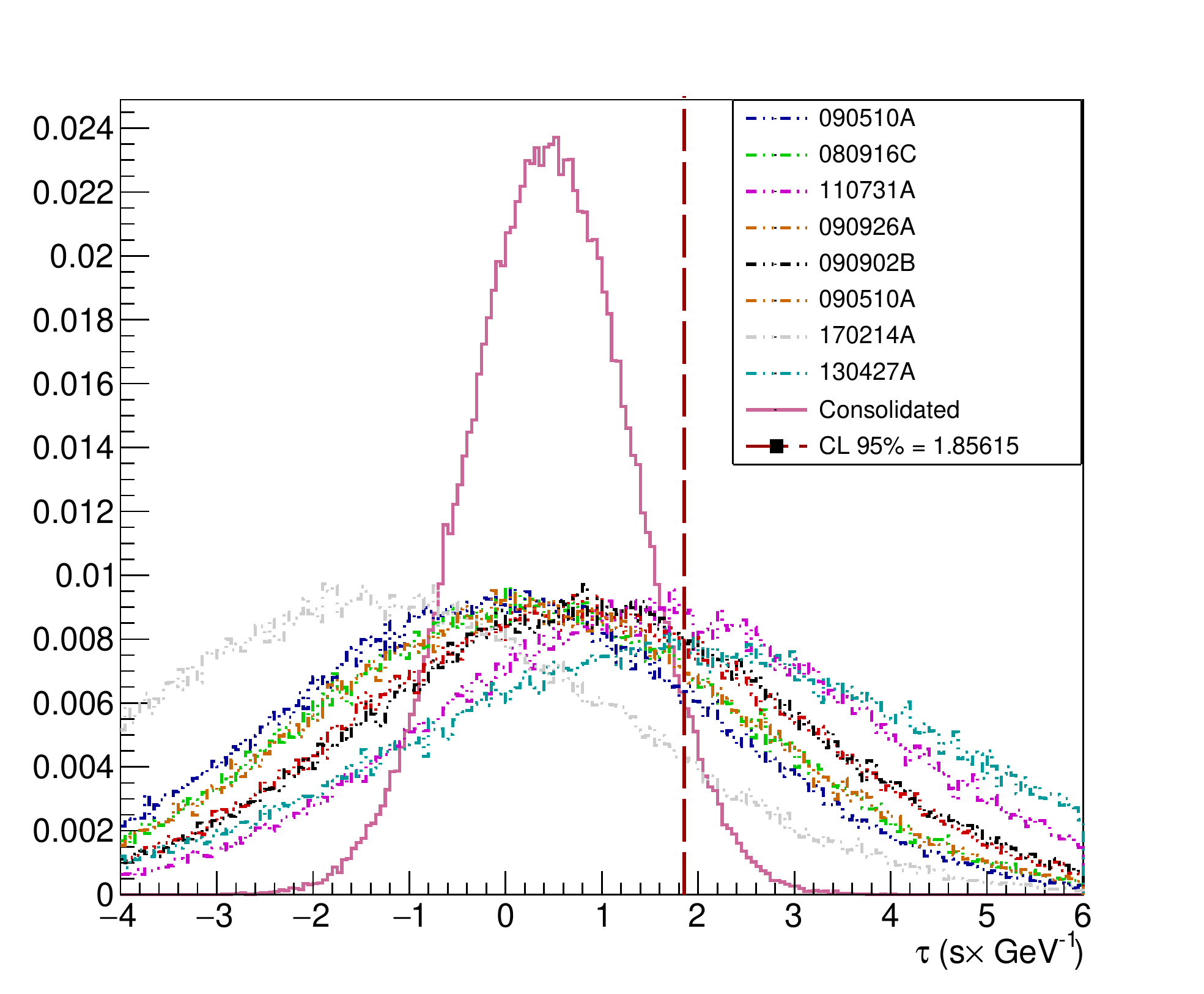}
\vspace{-0.4cm}
\caption{\it Left panel: the normalized K-reduced overall PDFs of the values of the compensated parameters
for all eight sources with standard deviations rescaled by a universal factor $\sqrt{\chi_{\rm raw}^2}$,
together with the consolidated PDF shown s the (red) solid line.
Right panel: the normalized K-reduced overall PDFs with an additional contribution to the standard deviations together with their consolidated PDF,
which is shown as the (red) solid line.
The values of $\tau (z_0)$ to the right of the vertical dashed line are not compatible with zero at the 95\%~CL.
}
\label{fig:COMB2}
\end{figure}

Assuming that there are no correlations between the measurements of different sources, one can construct a common $\chi^2$ function:
\beq
\label{chi2_1}
\chi^2=\sum_{i=1}^{N_{\rm src}}\frac{\overline\tau -\overline\tau_i}{\sigma_{\overline\tau_i}} \, ,
\eeq
where $\overline\tau_i$ and $\sigma_{\overline\tau_i}$ are the means and the standard deviations of the individual Gaussians (see Table~I for details)
and $\overline\tau $ is the mean of the consolidated distribution that minimizes (\ref{chi2_1}).
It is well known that the solution is given by the weighted average
\beq
\label{wav}
\overline\tau =\frac{\sum_{i=1}^{N_{\rm src}}\frac{\overline\tau_i}{\sigma_{\overline\tau_i}^2}}
{\sum_{i=1}^{N_{\rm src}}\frac{1}{\sigma_{\overline\tau_i}^2}} \, ,
\eeq
and the standard deviation of the consolidated distribution is given by
\beq
\label{sav}
\sigma_{\overline\tau } = \sqrt{\frac{1}{\sum_{i=1}^{N_{\rm src}}\frac{1}{\sigma_{\overline\tau_i}^2}}}. \,
\eeq
The results (\ref{wav}) and (\ref{sav}) can be proved as theorems in the framework of conflation~\cite{conflation}, which
provides a recipe for combining the PDFs of different measurements on a point-by-point basis.

Consolidating the K-reduced PDFs of our sources, using the prescriptions (\ref{wav}) and (\ref{sav}), one arrives at a large
raw $\chi^2$ value of 261,
which implies a negligible probability for the individual distributions entering in the combination to be
compatible with each other, implying that the sources are not identical.
Each emission episode is affected by an intrinsic process which might introduce ether stochastic or systematic scatter of the results
for individual sources.

The nature of the radiative processes and energy dissipation
mechanisms of GRBs have not been clearly identified yet, which limits our ability
to model the temporal spectral properties of the emitting region.
Without additional inputs on the physics of the processes
responsible for high energy emission of GRBs~\cite{HE_GRB}, one can
only assume that there are some source-dependent contributions to the spectral evolution of individual sources that
could be responsible for the mistuning we find in the K-reduced distributions of the compensation parameters
obtained for different sources.

In our ignorance, we estimate the possible uncertainties that might be introduced by such unknown effects,
in two ways, namely, using two different scalings of the individual distributions to render them
compatible with a single overall consolidated distribution.
The first possibility is to rescale the standard deviation of the individual distributions by a factor $\sqrt{\chi_{\rm raw}^2}$,
so that the resulting $\chi_{\rm scaled}^2$ becomes unity~\cite{PDG}.
The corresponding re-scaled K-reduced distributions together with the consolidated one are presented in
the left panel of Fig.~\ref{fig:COMB2}.
One can see that the region of 95\% incompatibility with a zero result for the
correct value of the compensation parameter lies beyond the
line $\tau (z_0)[{\rm 95\% CL}] = 0.54\ {\rm s/GeV}$,
which corresponds to the following lower limit on the scale of linear
Lorentz violation:
\beq
\label{mQG95_1}
M_1\ge 8.4\times 10^{17}\ {\rm GeV}.
\eeq
An alternative way of taking into account unknown source-related intrinsic temporal spectral variations would be
to allow for an additional universal
stochastic spread of the PDFs. This may be achieved by adding in quadrature,
for all the PDFs of the sources entered in the analysis, a universal variation in the $\tau (z_0)$ distributions
whose normalization is fixed so that the overall
$\sqrt{\chi^2}\simeq 1$. We estimate this standard deviation to be $2.30\ {\rm s/GeV}$, and
the corresponding re-scaled K-reduced distributions together with the consolidated one are presented in the
right panel of Fig.~\ref{fig:COMB2}. In this case, the region of 95\% incompatibility with zero result exceeds
$\tau (z_0)[{\rm 95\% CL}] = 1.86\ {\rm s/GeV}$, which corresponds to the following lower limit on the
scale of linear Lorentz violation:
\beq
\label{mQG95_2}
M_1\ge 2.4\times 10^{17}\ {\rm GeV} \, ,
\eeq
which is significantly weaker than (\ref{mQG95_1}).

We note that in the case of the $\sqrt{\chi_{\rm raw}^2}$ re-scaling method
the consolidated result is most affected by sources with lower variations
in the K-reduced distributions. On the other hand, in the case of the method
of adding of a universal stochastic spread  the result is depends more equally
on the different sources, since those with narrower distributions are expanded more substantially
than in the re-scaling case.

\section{Discussion of the Results}
\label{sec:concl}

The mean values of the compensation parameter $\tau $ and its
standard deviation $\Delta\tau $ encoded in the K-reduced distributions define the sensitivity of
the source sample to propagation effects due to a quantum-gravity medium.
However, the small statistics of the sources entering in the analysis implies
that another statistical realization of the measurements
of  $\tau $ and $\Delta\tau $, with the same pattern of K-reduced
distributions could have a different sensitivity
for $M_1$. By processing different realizations of the $\tau $ vs $\Delta\tau $
distribution one can assess the robustness of our conclusions about the level of the effects of quantum gravity
allowed by the available measurements of the source sample analyzed. To this
end, we have obtained $\tau $ vs $\Delta\tau $
distributions of the measurements for different realizations using random numbers distributed
according to the cell contents of a two-dimensional histogram with $10\times 10$ cells.
Examples of such realizations are shown in Fig.~\ref{fig:DISTR1}, and the data measurements are
indicated by crosses. In the following, we perform some simple simulation exercises to assess the sensitivity which would be achieved
if the statistical realization of the measurements were different, but assuming the same pattern of $(\tau , \Delta\tau )$
distribution from high-energy GRBs with known red shifts as has been measured by \lat.

\begin{figure}
\centering
\includegraphics[width=0.5\textwidth]{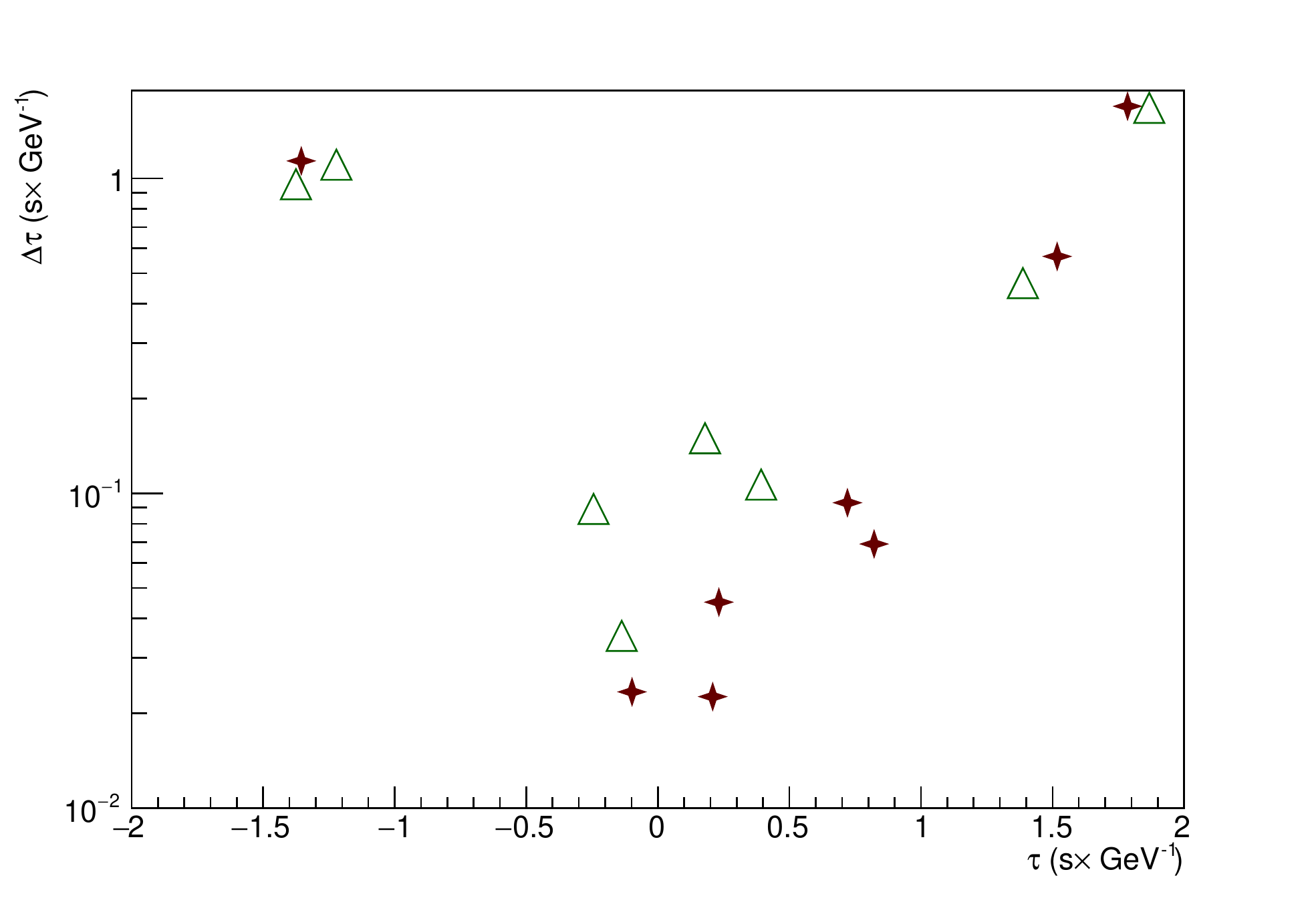}\hspace{0cm}\includegraphics[width=0.5\textwidth]{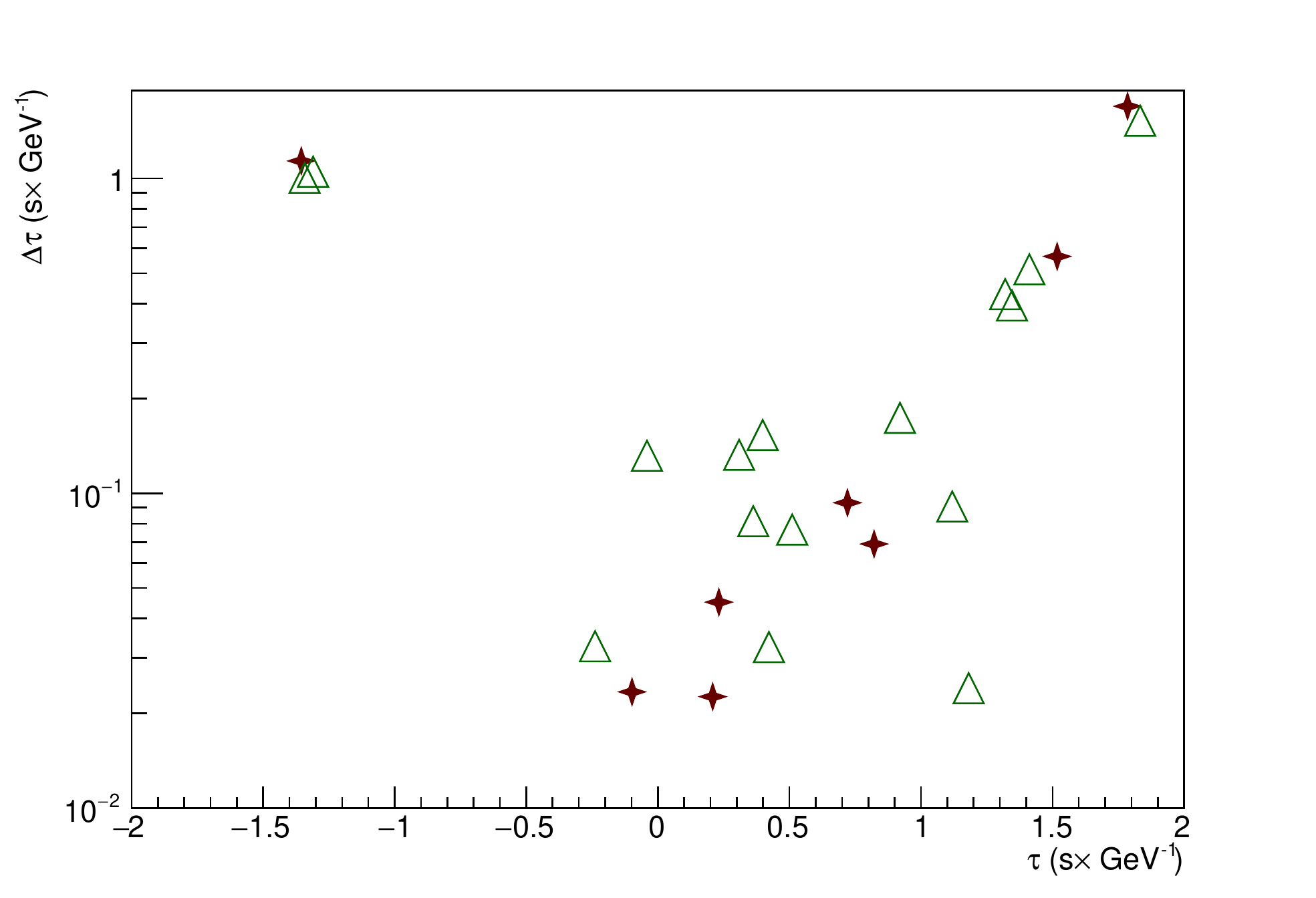}
\vspace{-0.4cm}
\caption{Simulated statistical realizations of the pattern of mean value/variance measurements obtained from the
K-reduced distributions. The pattern of the measured data is shown by the crosses, while the simulated
measurements are marked with triangles. The left and right panels are for 8 and 16 simulated
measurements, respectively.}
\label{fig:DISTR1}
\end{figure}

We first assess what would be expected if we had another realization of the current data set.
We generate 100 thousand realizations of the pattern of  8 sources, as shown by triangles in the left panel of Fig.~\ref{fig:DISTR1}.
Every realization containing only simulated measurements (triangles in the left panel of Fig.~\ref{fig:DISTR1})
has then been processed using the $\sqrt{\chi_{\rm raw}^2}$ rescaling method described in the previous section.
The resulting distribution of 95\% CL limits obtained for realizations
with  $\sqrt{\chi_{\rm raw}^2}$ rescaled measurements is presented in the left panel of Fig.~\ref{fig:limitsDISTR1}.
One can see that in 95\% of the cases the limits fall below $1.4\times 10^{18}$~GeV (indicated by the vertical dashed dotted line), which
we interpret as an effective limit on the sensitivity of the pattern of measurements we have in our disposal.
In the other words, whatever the red-shift distribution, the spectral and temporal content found for 8 emissions
leading to a $(\tau , \Delta\tau )$ distribution similar to the current data, this is
the best sensitivity one could reasonably expect to achieve, which we term the sensitivity end-point.

\begin{figure}
\centering
\includegraphics[width=0.5\textwidth]{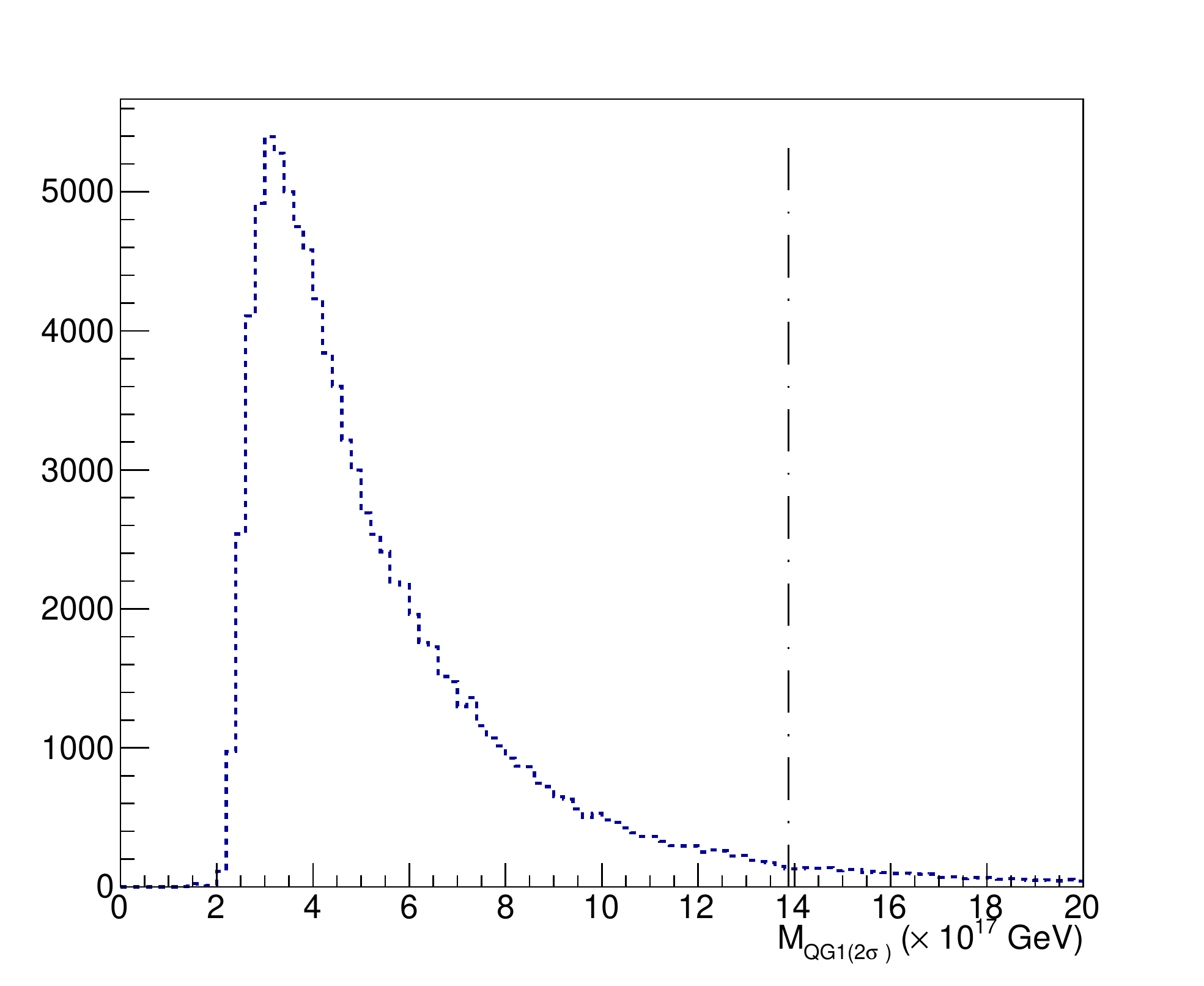}\hspace{0cm}\includegraphics[width=0.5\textwidth]{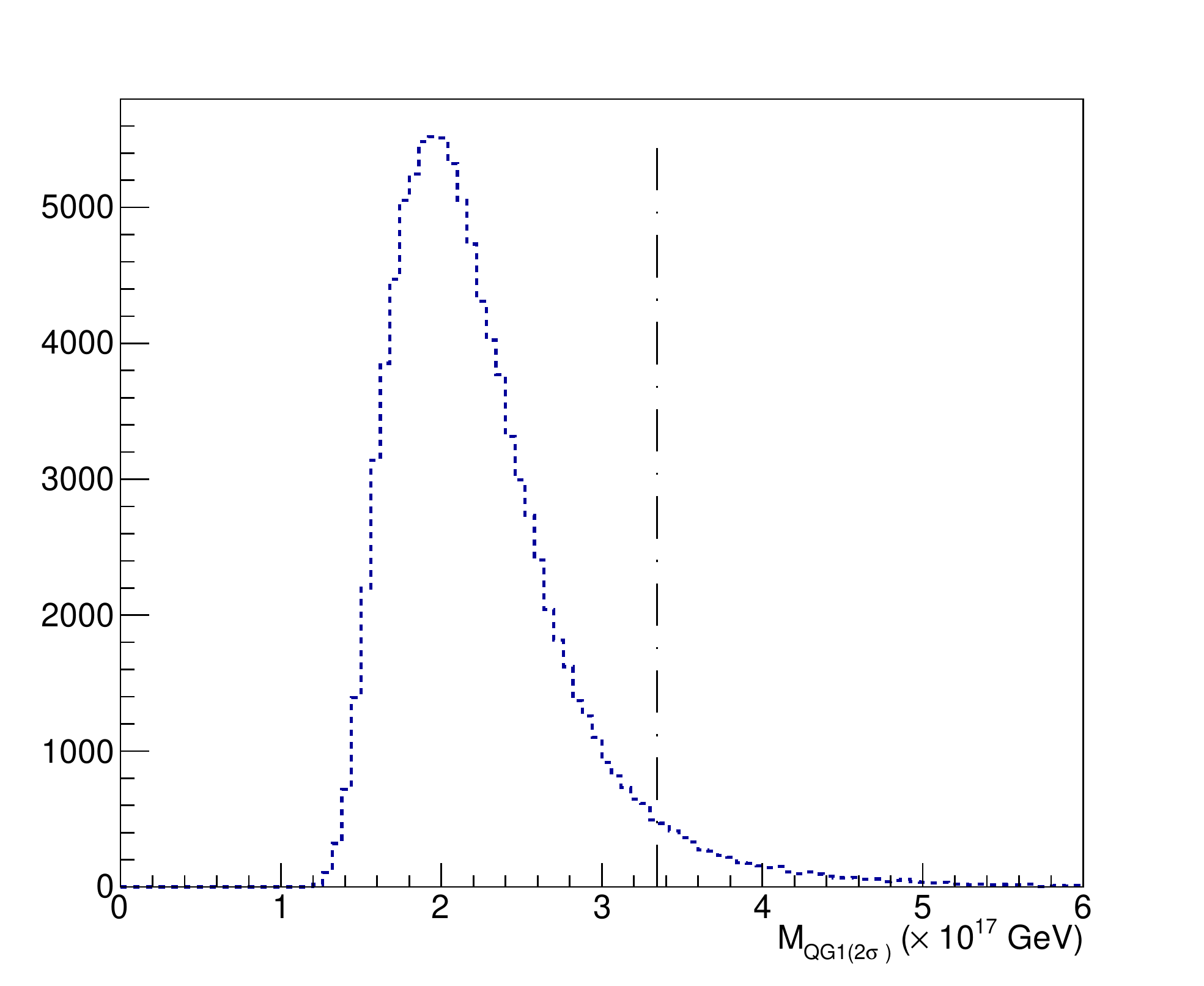}
\caption{Left panel: Distribution of the 95\% CL limits obtained in 100 thousand realizations of eight measurements
processed with the $\sqrt{\chi_{\rm raw}^2}$ standard deviation rescaling method. We find that
95\% of entries do not exceed the value indicated by the vertical dot-dashed line.
Right panel: The same as in the left panel, but for realizations processed by adding a universal stochastic spread.}
\label{fig:limitsDISTR1}
\end{figure}

If instead we process the realizations of eight measurements by adding a universal stochastic spread,
as described previously, we find
that the sensitivity end-point is at $3.3\times 10^{17}$~GeV,
as one can see in the right panel of Fig.~\ref{fig:limitsDISTR1}.

The most probable values of the 95\% CL constraints are quite similar in the two cases, namely
$3.2\times 10^{17}$~GeV in the case of $\sqrt{\chi_{\rm raw}^2}$ rescaled measurements and
$2.1\times 10^{17}$~GeV for processing with the universal stochastic spread.
Doubling the number of simulated measurements in the realizations (see right panel of Fig.~\ref{fig:DISTR1}), however,
we find a sensitivity end-point of $1.0\times 10^{18}$~GeV for $\sqrt{\chi_{\rm raw}^2}$ rescaling
and $2.6 \times 10^{17}$~GeV for universal stochastic spreading, although the most probable values of the 95\% CL
limit stay unchanged.


In practice, working with actual data, it is important
not to underestimate the uncertainties at each step in the analysis flow. In particular, since the temporal distributions of the high-energy emissions
of GRB engines are still poorly understood (see~\cite{HE_GRB}), one has to be careful when cross-correlating directly
the \lat\ multi-GeV events with sub-MeV light curves detected by the the Gamma-ray Burst Monitor (GBM)~\cite{GBM}.
In general, the paucity of multi-GeV photons makes it  difficult to assess the importance of variability and temporal correlations with the
emissions at lower energies.  The latter implies that common features of signals in the sub-MeV and multi-GeV spectral bands could be established
within some time intervals \cite{crossGBMLAT} that exceed substantially the time resolution of the detectors. This ultimately
implies an uncertainty whose neglect can lead to an overstated assessment of the significance of the measurement
obtained on the basis of an analysis~\cite{ch1,ch2,ch3,ch4} cross-correlating sub-MeV and multi-GeV photons.

One can also study the potential impact of accumulating more GRBs with
K-reduced compensation parameter measurements that agree
with the pattern of the eight sources we have analyzed. As a first exercise, we assume that the
existing statistics are doubled so that  eight additional measurements, like
those indicated by the triangles in the left panel of Fig.~\ref{fig:DISTR1}, are available to
be processed along with the current eight measurements indicated
by the crosses in the same plot. Thus we generate 100 thousand realizations each containing 16 measurements,
eight of which are the current measurements as they are, while another eight consist of simulated samples.
The results of processing of the realizations are shown in
the upper row of Fig.~\ref{fig:limitsDISTR3}. The $\sqrt{\chi_{\rm raw}^2}$ re-scaling method (upper left panel of  Fig.~\ref{fig:limitsDISTR3})
leads to a sensitivity end-point at  $8.4\times 10^{17}$~GeV,
while the most probable 95\% CL constraint with this amount of additional statistics is located at $7.0\times 10^{18}$~GeV.
Processing with a universal stochastic spread (upper right panel of  Fig.~\ref{fig:limitsDISTR3}) exhibits a sensitivity end-point
at $2.4\times 10^{17}$~GeV, while the most probable value of the 95\% CL constraint is $2.0\times 10^{17}$~GeV.
Processing the same number of realizations with 16 simulated sources added to the data gives very similar results for the sensitivity
end points (bottom panels of  Fig.~\ref{fig:limitsDISTR3}). However, the distribution of the 95\% CL for the $\sqrt{\chi_{\rm raw}^2}$ re-scaling method
looks rather smooth and symmetric, which slightly decreases the most probable value of the constraint to  $6.0\times 10^{18}$~GeV.

\begin{figure}
\centering
\includegraphics[width=0.5\textwidth]{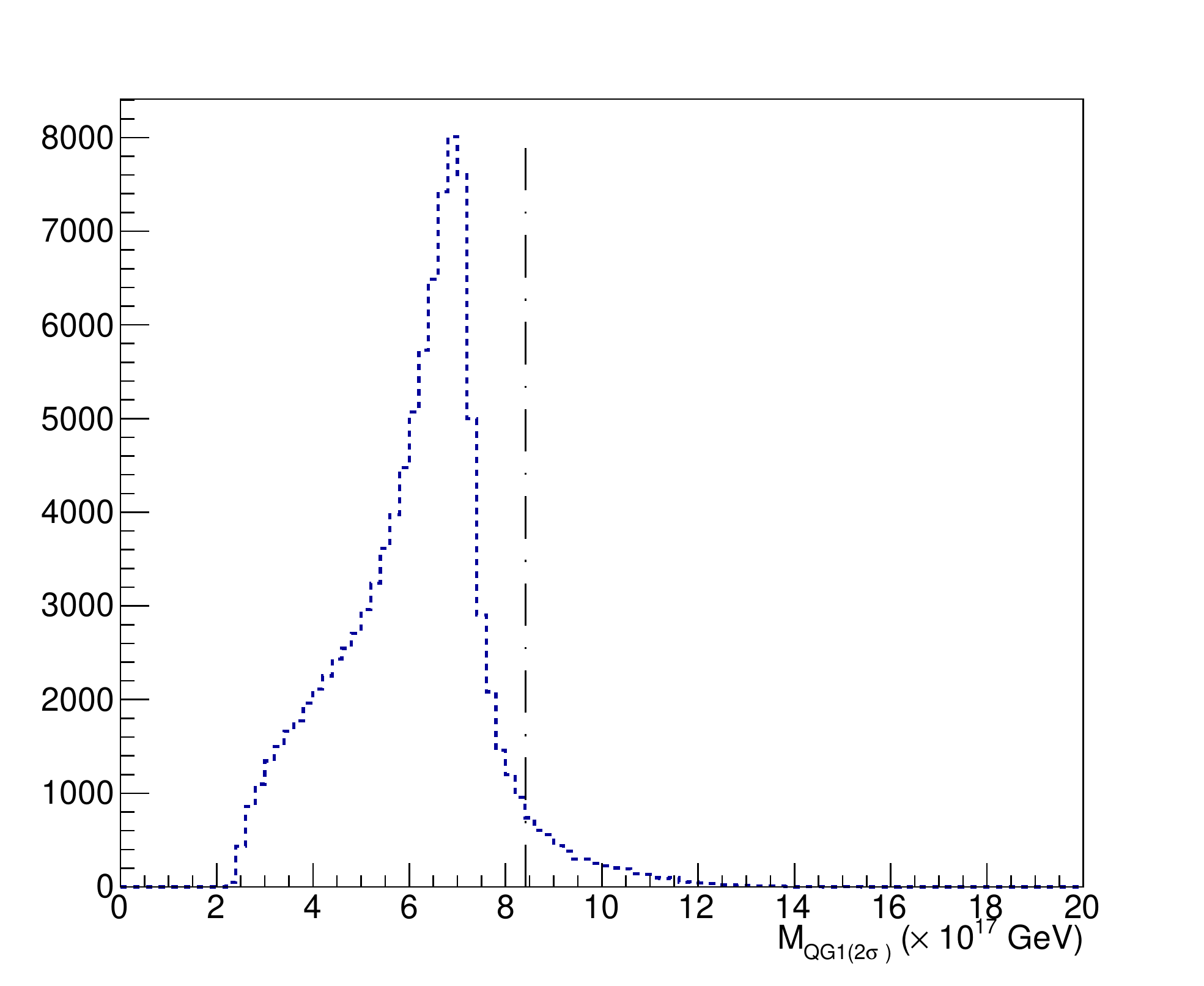}\hspace{0cm}\includegraphics[width=0.5\textwidth]{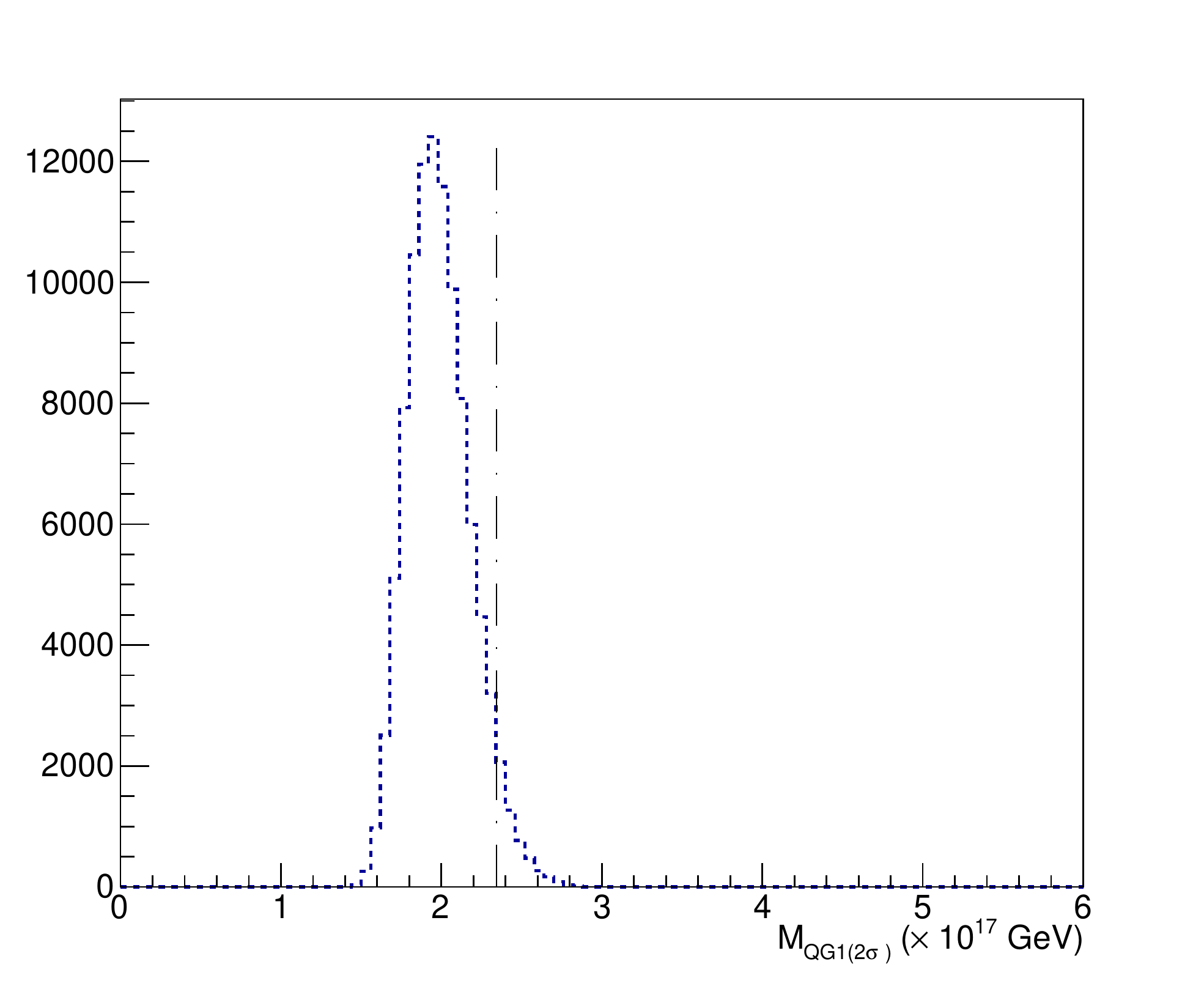}
\includegraphics[width=0.5\textwidth]{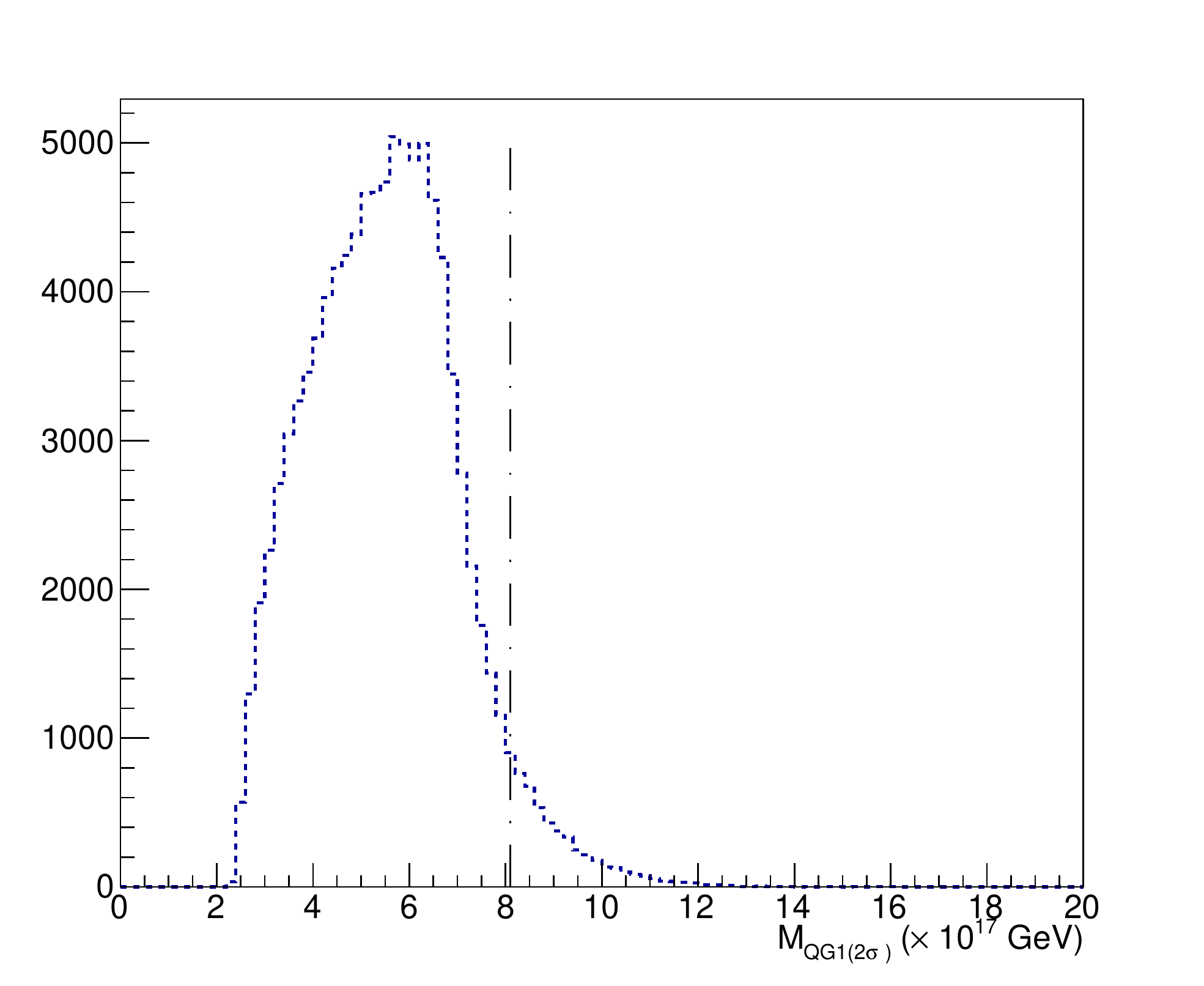}\hspace{0cm}\includegraphics[width=0.5\textwidth]{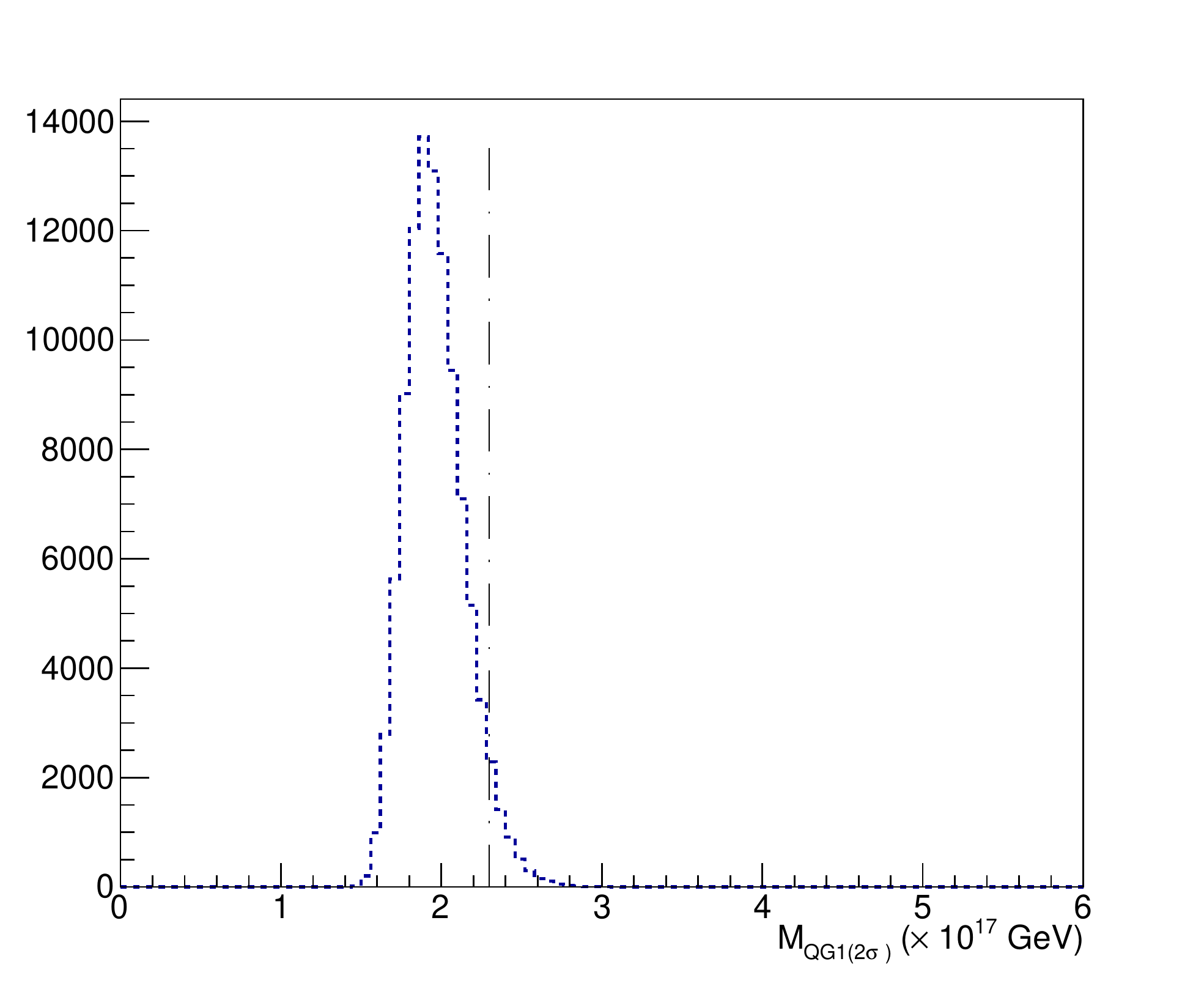}
\vspace{-0.4cm}
\caption{Lefts panels: Distribution of the 95\% CL limits obtained in 100 thousand
realizations with eight (upper) and 16 (lower) simulated measurements added to
the current data, processed with the $\sqrt{\chi_{\rm raw}^2}$ re-scaling method.
We note that 95\% of the entries do not exceed the value indicated by the vertical dotted dashed line.
Right panels: The same as in the left panels, but processing the real and simulated data by adding a universal stochastic spread.}
\label{fig:limitsDISTR3}
\end{figure}

It is also instructive to perform two extreme exercises.
One is to add just one simulated source to the present eight sources, processing the realizations with the
$\sqrt{\chi_{\rm raw}^2}$ rescaling method. The resulting distribution, presented in the left panel of Fig.~\ref{fig:limitsDISTR4},
clearly indicates that the most probable value of  the 2$\sigma$ limit is substantially boosted towards to higher values, namely to
$7.3\times 10^{17}$~GeV, getting quite close to the limit (\ref{mQG95_1}). On the other hand, a substantial increase of statistics,
modelled by adding  28 sources to the eight present sources would lead to a distribution rather
similar to that one shown in the left panel of Fig.~\ref{fig:limitsDISTR1}, with
the most probable value of the limit approaching that one obtained from the statistics of the present data alone.
We recall that  $\sqrt{\chi_{\rm raw}^2}$ re-scaling method of obtaining
limit weighs mostly the measurements with lower variances. Therefore,
simulating one additional source provides, in most of the realizations, one additional measurement with low
variance that increases substantially the constraint obtained. In this sense,
the example with one additional simulated source is evidence of an instability in conclusions about
quantum-gravity effects on photon propagation drawn from analysis of a
single GRB~\cite{Fermi_080916,Fermi_090510,NonFermi_090510,Nemiroff}.

\begin{figure}
\centering
\includegraphics[width=0.5\textwidth]{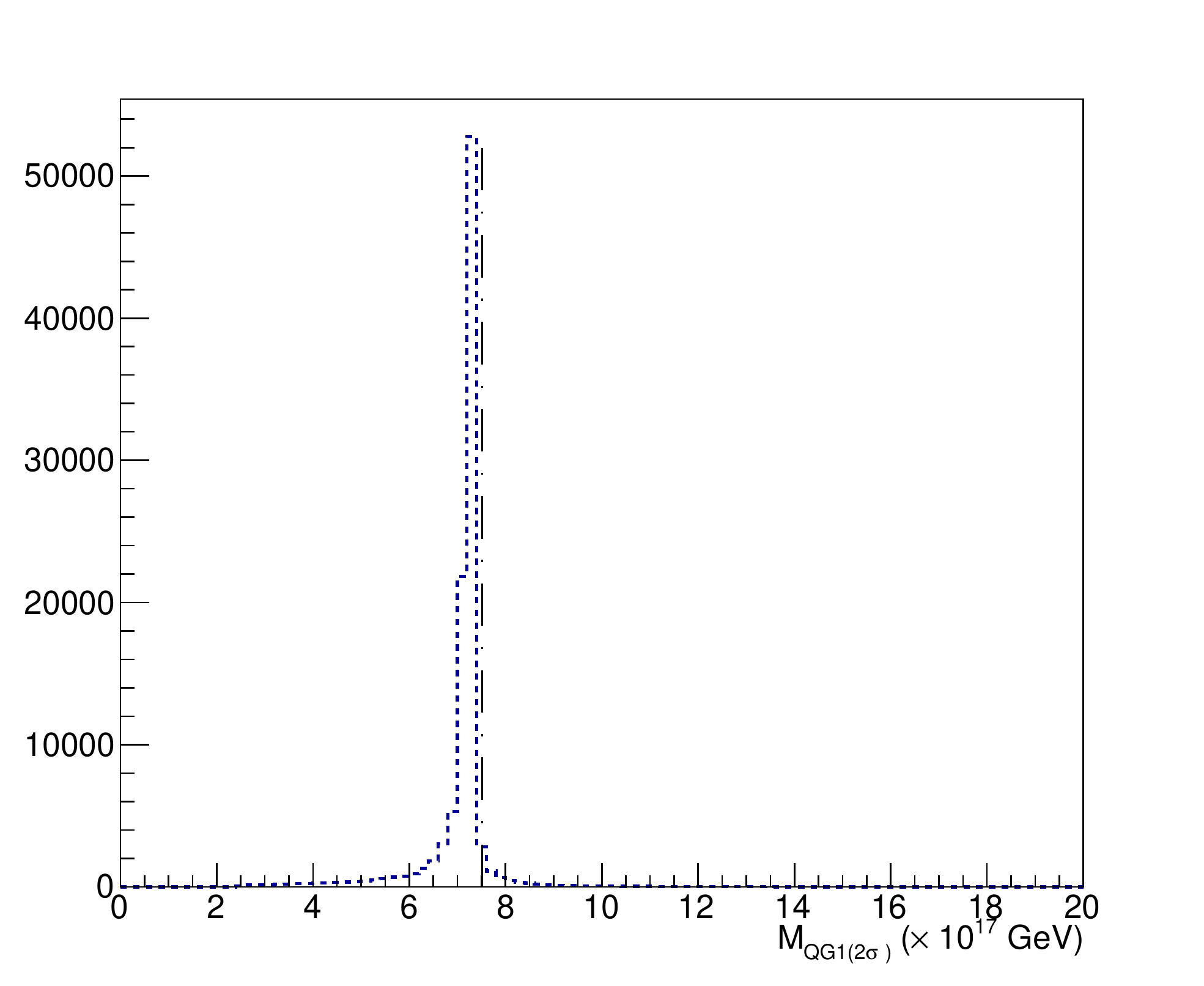}\hspace{0cm}\includegraphics[width=0.5\textwidth]{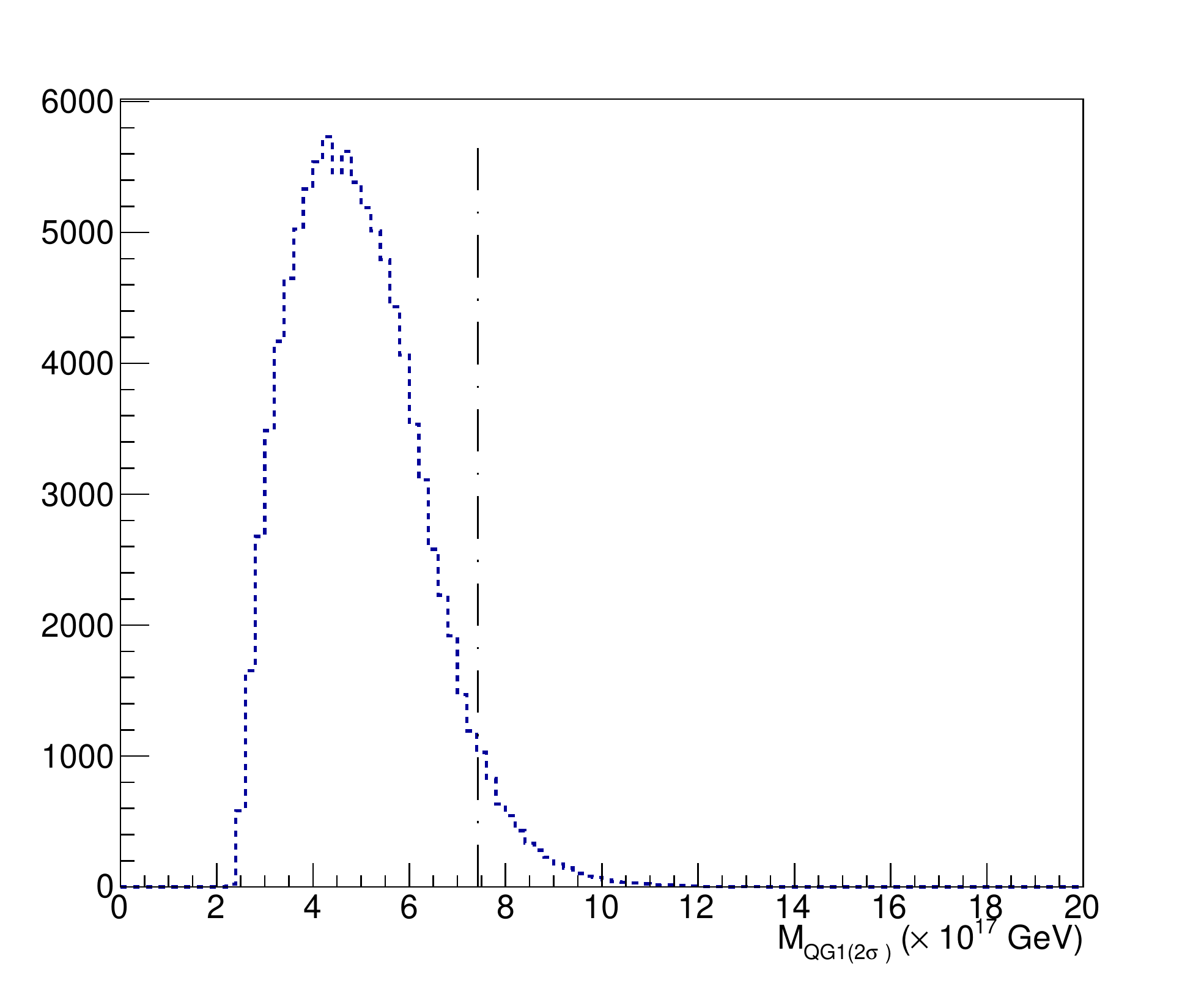}
\vspace{-0.4cm}
\caption{Left panel: Distribution of the 95\% CL limits obtained in 100 thousand realizations of one simulated measurement
added to the eight present sources, processed with the $\sqrt{\chi_{\rm raw}^2}$ rescaling method.
Right panel: The same as in the left panel, but with 28 simulated measurements added to the eight present sources.}
\label{fig:limitsDISTR4}
\end{figure}

In closing this discussion, we emphasize that our analysis was performed in the context of a model
for space-time foam that does not predict birefringence, so that the
strong constraints~\cite{Stec_Polar,INTEGRAL_Polar,strict_Polar,Lin_Pol,Kislat_Polar}
are inapplicable. That said, the statistical techniques developed here could be applied
to a wide class of models for Lorentz violation that predict
anomalous dispersion {\it in vacuo}, providing model-independent constraints that are
complementary to those from searches for birefringence.

\section{Conclusions}
\label{sec:concl2}

We have developed in this paper three distinct statistical non-parametric measures of
GRB emissions, which we have used in an analysis of {\it Fermi}-LAT data to search for the possible effect of
quantum gravity on the propagation of high-energy gamma rays from GRBs.
The measures utilize different types of deformation of the intensity profile of
an envelope of electromagnetic radiation with a burst-like feature that would arise from propagation
through a dispersive quantum-gravity medium. Applying five different estimation procedures developed
on the basis of these statistical measures to the eight observed GRBs that are relatively bright in multi-GeV
energies detected by \lat, we constrain the possibility of a non-trivial
vacuum refractive index for photons. Depending on the method of consolidation
of the results for individual sources, we find that the energy scale $M_1$ characterizing a
linear energy dependence of the refractive index should exceed either $8.4\times 10^{17}\ {\rm GeV}$
or $2.4 \times 10^{17}\ {\rm GeV}$. We have also made simple numerical exercises to explore
the possible sensitivity of the current statistics of~\lat\ sources with measured red shifts together with sources
that might be detected in the future, finding that the sensitivity would probably not exceed significantly $M_{1}\approx 10^{18}\ {\rm GeV}$.

\section*{Acknowledgments:}

We are grateful to N.~Omodei and A.~Moiseev for their advising on {\it Fermi}-LAT GRB data.
The work of JE and NEM was supported in part by the Science and Technology Facilities Council (STFC), UK,
under the research grant ST/P000258/1, and that of JE was also supported in part by the Estonian Research Council
via a Mobilitass Pluss grant.
NEM also wishes to thank the University of Valencia and IFIC for a Distinguished Visiting Professorship held while
this work was initiated. He also currently acknowledges the hospitality of IFIC Valencia through a Scientific Associateship (``Doctor Vinculado'').
The work of RK and LN was supported in part by the US National Science Foundation under Grant No.PHY-1402964.
The work of AS was supported in part by the US National Science Foundation under Grants
No. PHY-1505463 and No. PHY-1402964.


\end{document}